\documentclass[12pt,english]{article}
\usepackage{amsmath}
\usepackage{amsthm}
\usepackage{libertine}
\usepackage[libertine]{newtxmath}
\usepackage[T1]{fontenc}
\usepackage[latin9]{inputenc}
\usepackage{colortbl}
\usepackage{babel}
\usepackage{float}
\usepackage{mathrsfs}
\usepackage{enumitem}
\usepackage{bm}
\usepackage{varwidth}
\usepackage{graphicx}
\usepackage[letterpaper]{geometry}
\usepackage{setspace}
\usepackage[authoryear]{natbib}
\usepackage{bibunits}

\setdisplayskipstretch{0.5}
\usepackage[bookmarks=true,bookmarksnumbered=false,bookmarksopen=false,
 breaklinks=false,pdfborder={0 0 1},backref=false,colorlinks=false]
 {hyperref}
\hypersetup{pdftitle={Coarse Personalization},
 pdfkeywords={Personalization, Targeting, Optimal Transport},
 urlcolor=darkslateblue}

\makeatletter

\floatstyle{ruled}
\newfloat{algorithm}{tbp}{loa}
\providecommand{\algorithmname}{Algorithm}
\floatname{algorithm}{\protect\algorithmname}

\usepackage[utf8]{inputenx}

\usepackage{fancyvrb}
\usepackage{graphicx}
\usepackage{tabularx}
\usepackage{color}
\usepackage{bbm}
\usepackage{soul}

\definecolor{darkslateblue}{rgb}{0.28, 0.24, 0.55}

\usepackage{lscape}
\usepackage{pdflscape}
\usepackage{float}
\usepackage{graphicx}
\usepackage{multicol}
\usepackage{booktabs}
\usepackage{longtable}
\usepackage{multirow}
\usepackage{hhline}
\usepackage{adjustbox}
\usepackage[flushleft]{threeparttable}

\usepackage{algorithm,algpseudocode}

\usepackage{enumitem}
\setlist{nolistsep}

\DeclareMathOperator*{\argmax}{argmax}

\hypersetup{pdfauthor={Walter W. Zhang, Sanjog Misra}}

\ifdefined\showcaptionsetup
 \PassOptionsToPackage{caption=false}{subfig}
\fi
\usepackage{subfig}
\makeatother

\theoremstyle{plain}

\newtheorem{assumption}{\protect\assumptionname}
\newtheorem{prop}{\protect\propositionname}

\newtheorem*{prop*}{\protect\propositionname}
\newtheorem*{cor*}{\protect\corollaryname}
\providecommand{\assumptionname}{Assumption}
\providecommand{\corollaryname}{Corollary}
\providecommand{\propositionname}{Proposition}
\providecommand{\theoremname}{Theorem}

\theoremstyle{definition}
\newtheorem*{example}{Example}
\theoremstyle{plain}

\begin{document}
\begin{bibunit}[ecta]
\begin{spacing}{1.0}

\title{Coarse Personalization\thanks{
    Zhang: \protect\href{mailto:wwz@wharton.upenn.edu}{wwz@wharton.upenn.edu},
    Misra: \protect\href{mailto:sanjog.misra@chicagobooth.edu}{sanjog.misra@chicagobooth.edu};
    We thank the data science and analytics team at the company that is
    the source of our data for their help and numerous insights. We are
    grateful for the discussions, comments, and suggestions of conference
    participants at Marketing Science 2021, TADC 2021, MLESI 2022, QME
    2022, ICSDS 2023, and EC 2024.}}
\author{Walter W. Zhang\\
{\normalsize\emph{The Wharton School, University of Pennsylvania}}\and
Sanjog Misra\\
{\normalsize\emph{University of Chicago Booth School of Business}}}
\date{December 3, 2026}
\maketitle
\begin{abstract}
With advances in estimating heterogeneous treatment effects, firms can
personalize and target individuals at a granular level. However, implementation
costs limit full personalization. In practice, firms assign individuals to
segments and choose a treatment for each segment to maximize profits: We call
this the coarse personalization problem. We propose a two-step solution that
simultaneously makes segmentation and targeting decisions. First, the firm
personalizes by estimating heterogeneous treatment effects. Second, the firm
discretizes using treatment effects to choose which treatments to offer and
their segments. We show that a combination of available machine learning tools
for estimating heterogeneous treatment effects and a novel application of
optimal transport methods provides a viable and efficient solution. With data
from a large-scale field experiment in promotions management, we find our
methodology outperforms extant approaches that segment on consumer
characteristics, consumer preferences, or those that only search over a
prespecified grid. Our framework quantifies the gain in profits from forming
additional segments, allowing firms to weigh these benefits against their
implementation costs. We conclude by discussing how coarse personalization
arises in other domains.\vspace{0.5cm}

\end{abstract}
\noindent\end{spacing}

\thispagestyle{empty}
\newpage{}
\setcounter{page}{1}

\section{Introduction\label{sec:Introduction}}

Improvements in estimating heterogeneous treatment effects enable firms to
target customers on an almost individual level. Leveraging state-of-the-art
machine learning techniques, firms can fully personalize; individual customers
with a unique set of covariates can form their own
segment.\footnote{Personalization using heterogeneous treatment effects
implements one-to-one marketing \citep{Arora2008} when observed covariates
capture relevant individual heterogeneity.} Full personalization involves both
tailoring the marketing mix for each customer and targeting only customers who
generate positive incremental profits for the firm. Consequently, full
personalization is profit-maximizing because individuals are targeted only when
their marginal benefit outweighs their marginal cost.

However, firms face implementation costs when deploying personalized
treatments. Each distinct treatment must be set up before launch and maintained
afterwards, so these costs grow with the number of treatments the firm offers.
For example, on a food delivery platform, each distinct promotion requires
configuring the discount and eligible customers, reviewing the setup, and
monitoring the promotion while it runs. When a promotion fails to apply at
checkout, the failure must also be traced and resolved.\footnote{These issues
were the primary concerns of the firm we partnered with for this application.
Public documentation from other food delivery platforms describes similar steps.
Uber Eats describes how merchants configure individual promotions, monitor their
performance, and end them after launch \citep{uber2026marketing}. DoorDash
reports that diagnosing promotion-code failures at checkout was sufficiently
time-consuming to motivate a dedicated internal tool. Its engineers used the
tool to investigate failures \citep{doordash2026promotion}.} Offering more
distinct promotions therefore increases these implementation costs. Similar
costs arise in email marketing (e.g., Mailchimp), where firms must prepare
message content, configure targeting rules, and verify that customers receive
their intended version. Maintaining a larger number of
distinct emails can therefore increase the work required to manage a campaign.
In direct mail, printing each distinct version is costly. The trade press
similarly emphasizes the high costs of setting up and maintaining the
infrastructure needed to implement personalized
policies.\footnote{\citet{Gilmore1997,HarvardBusinessReviewAnalyticServices2018}
note that implementation costs can limit personalization and customization.}

With these constraints, it is not surprising that firms restrict the
set of unique marketing mix variables, or treatments, to offer. Consequently,
firms must decide to which customers they will assign these treatments.
In essence, firms create smaller sets, or segments, of customers and
optimally choose treatments for each set with the aim of maximizing
profits. We call this problem the \emph{coarse personalization problem}.

The coarse personalization problem is not entirely new. The marketing literature
has long understood the need to segment consumers with the idea of personalizing
the marketing mix for each segment while keeping costs within bounds. The
typical process is sequential: Marketers first group customers into segments and
then personalize marketing mix treatments for the targeted segments. Indeed, the
textbook approach is to implement a \textquotedblleft discretize then
personalize\textquotedblright{} framework where they form customer segments
based on some predefined source of customer heterogeneity, identify which
segments they wish to target, and then tailor marketing mix variables for those
segments \citep{KotlerP.andKeller2014}. Over time, the sources of customer
heterogeneity that underpin the segmentation procedure have evolved. Initially,
marketing researchers proposed forming segments on demographic and psychographic
variables \citep{Smith1956,Wind1978}. More recently, researchers have formed
segments based on estimated consumer preferences and responses
\citep{Kamakura1989a,Bucklin1992,Gupta1994,Krieger1996,Bucklin1998a} via mixture
models \citep{Grover1987,Jain1990}.

In practice, the ``discretize then personalize'' framework is broadly
used in marketing, especially in customer relationship management, 
database marketing, and direct marketing, where marketers first
form segments before making targeting decisions. While the literature
has evolved considerably in terms of the information it uses, the
structure of the framework and the sequence of steps have essentially
remained the same.

In this paper, we show that implementing these steps sequentially
is not necessarily profit maximizing: The initial segmentation procedure
is based on a distance metric that is statistical (e.g., Euclidean)
rather than economic in nature (e.g., profit maximization).
To address this problem, we recharacterize the coarse personalization
problem as a mixed integer optimization problem and derive a theoretically
rigorous yet practically implementable solution using optimal transport theory. 
In doing so, we draw on modern techniques from both the incrementality-based targeting
literature and the optimal transport literature. In particular, we
advocate for a \textquotedblleft personalize then discretize\textquotedblright{}
approach where we \emph{first} construct personalized policies based
on heterogeneous treatment effects obtained using now widely available
machine learning tools, and in a \emph{second }step\emph{,} we use
these policies to construct segments of consumers by minimizing profit
regret---the profits forgone relative to full personalization. 
Crucially, in the second step we \emph{simultaneously} determine
each segment's optimal treatment and customer assignment
to maximize profits in a causally aware segmentation plan. 
Put simply, this approach inverts the \textquotedblleft discretize
then personalize\textquotedblright{} approach by first searching for
what works for each customer and then deciding on a coarsened marketing
mix offering that would not significantly degrade profits relative
to full personalization.

Our two-step framework solves the coarse personalization problem precisely
because it is able to segment consumers based on a profit-relevant
distance metric. As such, we view our key contribution to be the design
of the second step that allows for the construction and implementation
of this solution. The first step's machine learning-based personalization,
done by exploiting heterogeneous treatment effects, is now relatively
straightforward from the literature. On the other hand, the issue
of coarsening these to satisfy the problem is novel and poses challenges
on multiple fronts. The combinatorial nature of the assignment problem
coupled with the optimization of the treatment policy makes the problem
especially difficult.

To give intuition for our framework, we provide a stylized example
comparing our simultaneous segmentation to the classical sequential segmentation. 
We first denote individuals' covariates as $\boldsymbol{x}_{i}$,
the segmentation rule that assigns individuals to segments as $S(\boldsymbol{x}_{i})$,
and the function that maps segments to their assigned treatment as
$T(S(\boldsymbol{x}_{i}),\boldsymbol{x}_{i})$. We define profits from the 
segmentation plan as $\Pi(S(\boldsymbol{x}_{i}),T(\boldsymbol{x}_{i},S))$.

The classical sequential approach forms segments a priori based on
consumer characteristics and preferences and we denote it as $S^{0}(\boldsymbol{x}_{i})$.
After segments are formed, the marketer then solves for the optimal
treatment values $T(S^{0}(\boldsymbol{x}_{i}),\boldsymbol{x}_{i})$ 
while taking the a priori segmentation as given.

In contrast, our approach simultaneously chooses the segmentation rule $S(\mathbf{x}_{i})$
and treatments by $T(S(\boldsymbol{x}_{i}),\boldsymbol{x}_{i})$ to maximize profits. 
Suppressing dependence on $\boldsymbol{x}_i$, we have
\begin{equation}
\underbrace{\max_{S,T}\Pi(S,T(S))}_{\text{Simultaneous profits}}\geq\underbrace{\max_{T}\Pi(S^{0},T(S^{0})).}_{\text{Sequential profits}}\label{eq:Toy-Example}
\end{equation}
The simultaneous approach includes any fixed segmentation $S^0$
among its feasible choices and therefore yields weakly higher
profits. Equality holds if and only if $S^0$ is itself optimal.
Hence, the simultaneous approach will be an upper bound in profits
for \emph{any} sequential segmentation procedure. In our empirical
application, we validate this intuition by evaluating our simultaneous
approach to different sequential benchmarks.

While our simultaneous approach is more profitable analytically, in practice
solving it is quite difficult. 
Even with only 15 customers and 20 possible promotions (ten dollar-off
and ten percentage-off offers), there are ${20 \choose 3}$ ways to
choose three distinct promotions and $3^{15}-3\cdot2^{15}+3$ ways to
assign customers so that each selected promotion serves at least one
customer. Together, these yield over 16 billion possible segmentation plans.
The search space grows combinatorially, and for real world applications with
millions of customers and many possible promotional treatments it only gets more
complex.

To tackle this computational problem, we use optimal transport theory to map 
our mixed integer optimization problem to a semidiscrete optimal transport problem. 
The recharacterization allows a computational solution that adapts 
Lloyd's Algorithm \citep{Lloyd1982}, which has been used to solve other 
semidiscrete optimal transport problems \citep{Galichon2016}. 

To be clear, while one might be able to solve the coarse personalization problem
by brute force, the complexity of the problem is exponential in the number of
people and the treatment configurations grow combinatorially with the
number of treatment dimensions and offered treatments. In contrast, our
algorithm is scalable, provides a transparent visualization of the solution's
procedure, and delivers profits close to the optimum. Although our algorithm is
a heuristic, in our empirical application it delivers profits within
approximately 0.022\% of the upper bound on optimal profits provided by Gurobi
across a tested grid of possible treatments. Our algorithm runs in seconds,
while Gurobi takes up to 28 hours. We also provide theoretical guarantees for
our algorithm.
Practically, its computational efficiency
allows firms to readily solve the coarse personalization problem and implement
the resulting segmentation plan.

In our empirical application, we use data from a food delivery platform that
experimented with a wide range of dollar-off and percentage-off promotions
across 1.2 million customers. The platform seeks to maximize profits by offering
a smaller, optimized set of promotions that accounts for implementation costs.
With implementation costs of \$1,000 per treatment, we find that the firm
maximizes estimated net incremental profits with five carefully formed segments.
Our solution also generates higher estimated profits than sequential
segmentation based on customer covariates or estimated preferences even after
adjusting for the optimizer's curse \citep{Xu2025}.\footnote{After adjusting for
the optimizer's curse, estimated profits net of implementation costs peak at
three treatments instead of five.} The application provides evidence that
our \textquotedblleft personalize then discretize\textquotedblright{} approach
outperforms the traditional \textquotedblleft discretize then
personalize\textquotedblright{} approach.

Substantively, our coarse personalization solution lets us interpolate between
no personalization and full personalization as well as evaluate the surplus at
each step in between. While full personalization is well-studied in the
literature, the gap between full personalization and blanket targeting has
received less attention.\footnote{\citet{Bergemann2011} theoretically show that
more precise advertising targeting increases the social value of advertising.
\citet{Dube2017} find that consumer surplus varies non-monotonically with data
granularity under personalized pricing.} Using our framework, we empirically
document the impact personalization has on consumer and producer surplus.
Coarser personalization can reduce producer surplus due to the extra constraint
on the firm's optimization, but it can boost consumer surplus for some
individuals by assigning them treatment levels that differ from those under full
personalization. In our application, the impact of personalization on consumer
welfare is non-monotonic in the personalization level.

\subsection{Literature Review}

Our paper contributes to the marketing literature on segmentation and
personalization by studying how firms should construct a causally aware
segmentation plan using heterogeneous treatment effects. The plan specifies a
limited set of treatment types and their values and assigns customers to
those treatments. Traditional segmentation procedures first group customers and
then choose the marketing mix for each segment
\citep{Smith1956,Wind1978,Kamakura1989a}. Research on personalization uses
customer heterogeneity for targeted promotions and customized communications
\citep{Rossi1996,Shaffer1995,Ansari2003,Zhang2004,Zhang2009}
and optimizes marketing interventions across customers and over time
\citep{Rust2005,Khan2009}. We study how firms should choose treatments and form
customer segments together when maintaining distinct treatments is costly. This
differs from bundling, where firms design offerings from which customers choose
\citep{Stremersch2002}, and from class pricing, which groups product and service
versions into price classes when pricing is costly \citep{Wernerfelt2008}. In
our setting, the firm groups customers into segments through their assigned
treatments.

Causal targeting and policy learning provide foundations for forming these
assignments using heterogeneous treatment effects
\citep{Swaminathan-Joachims-2015,Athey2016,Kitagawa2018,Wager2018,Farrell2020,Farrell2021}.
This literature considers several restrictions on the resulting policy,
including tree-based assignment rules \citep{Athey2021,Zhou2022}, fairness
constraints \citep{Kallus2021,Lu2025}, and comprehensibility \citep{Zhang2023}.
These functional-form restrictions limit how treatments are assigned and how
much assignments can differ across customers. While our empirical application
includes policy trees as a benchmark, our focus with coarse personalization is
the number of distinct treatments the firm must maintain to implement its
segmentation plan. Our solution using optimal transport allows for more 
flexible segmentation plans than tree-based segmentation.

In marketing, incrementality-based targeting serves as a modern foundation for
personalizing the marketing mix
\citep{Ascarza2018,Yoganarasimhan2020,Ellickson2022}. Related research develops
methods for evaluating targeting policies \citep{Simester2020,Hitsch2024} and
examines the value of targeting information \citep{Rafieian2021}. More recent
work expands the information available for these decisions by pooling evidence
across experiments and outcomes \citep{Yao2022} and learning how incremental
effects vary with both customer and intervention attributes \citep{Huang2024}.
The available treatment choices have also expanded to include novel marketing
content \citep{Ellickson2026}. Policy learning methods have been developed to
accommodate continuous treatment levels \citep{pmlr-v84-kallus18a}, including
learning such policies from experiments with a finite set of treatment levels
\citep{Zhang2026}. We build on these advances to jointly choose a limited
treatment set and its customer assignments using heterogeneous treatment effects.

A closely related literature treats segmentation as part of the firm's problem.
\citet{Cui2022} jointly optimize customer segments and continuous prices for a
specified number of segments using a regression model of customer valuations.
\citet{Aouad2023} integrate segmentation and response modeling through market
segmentation trees. In contrast to these two papers, our approach uses estimated
heterogeneous treatment effects to form segments around incremental profits.
Related work in biostatistics and operations research develops tree-based
treatment rules with continuous doses \citep{Laber2015} and optimizes dose
allocations over a discretized treatment space under resource and fairness
constraints \citep{DeVos2026}. Our formulation directly limits the number of
distinct treatments and optimizes their values and customer assignments without
requiring a tree-based rule or a prespecified treatment grid.

Our contribution is to study how a firm should coarsen a segmentation plan when
customers differ in their causal responses to continuous treatment levels. Given
estimated heterogeneous treatment effects, we address this problem by directly
limiting the number of distinct treatments the firm can offer. We connect this
problem to optimal transport \citep{Peyre2019}, where our transport plan maps
customers to segments and our transport cost reflects the profits
forgone relative to full personalization. This connection motivates our adapted
Lloyd's algorithm \citep{Lloyd1982}, which alternates updates to customer
assignments and segment-level treatments. Comparing segmentation plans across 
different numbers of segments allows the firm to weigh the incremental profits
from greater personalization against the costs of implementation.

Section \ref{sec:Model} formally presents our coarse personalization solution:
The personalization in the first step is described in Section
\ref{subsec:Treatment-effects-estimation} and the discretization in the second
step is detailed in Section \ref{subsec:Discretize-for-segments}. We then
discuss the link to optimal transport and how it motivates our computational solution
in Section \ref{sec:Computational-solution}.
We apply our methodology to an empirical application for promotions management
for a large food delivery firm in Section \ref{sec:Empirical-Application}. The
surplus analysis for coarse personalization is discussed in Section
\ref{sec:Surplus-Analysis} and we conclude with a general discussion in Section
\ref{sec:Conclusion}. 

\section{Model\label{sec:Model}}

In this section, we formulate the coarse personalization problem and our
two-step solution. First, we estimate heterogeneous treatment effects from a
randomized controlled trial (Section \ref{subsec:Treatment-effects-estimation}).
Second, we use these estimates to construct the segmentation plan (Section
\ref{subsec:Discretize-for-segments}). 
We set up the joint problem of choosing a limited treatment set and customer
assignments using heterogeneous treatment effects to solve for the segmentation
plan.

The firm constructs a segmentation plan for $N$ customers and $L$ segments by
jointly choosing which customers belong to each segment and which treatment each
segment receives. Each segment receives a distinct treatment, whose type is
specified by dimension $d\in\{1,\ldots,D\}$ and whose level specifies its
intensity. In our application, dollar-off and percentage-off promotions
correspond to the two treatment dimensions, and the level specifies the
amount or percentage discount. The firm chooses both the promotion type and 
its level for each segment. The firm's objective is to choose the segmentation 
plan that maximizes expected incremental profits.

We now define the notation for our setup.  
For individual $i\in\{1,\ldots,N\}$, we let $\boldsymbol{x}_i$ denote customer
characteristics and $Y_i$ the firm's outcome measure (e.g., sales). The individual's treatment
vector is $\boldsymbol{t}_i=(t_{i,1},\ldots,t_{i,D})$, where $t_{i,d}$ is the
treatment level in dimension $d$. We restrict our attention to \emph{feasible
treatments}, which have at most one nonzero component. Thus, a customer
receiving a treatment in dimension $d$ is assigned
\begin{equation}
\boldsymbol{t}_i=(0,\ldots,t_{i,d},\ldots,0),
\qquad t_{i,d}\neq 0.
\end{equation}
In our application, this restriction means that each customer can receive either
a dollar-off promotion (e.g., \$5 off) or a percentage-off promotion (e.g., 10\%
off), but not both. The vector $\boldsymbol{t}_i=\boldsymbol{0}$ represents no
treatment.

For a feasible treatment in dimension $d$, let $c_d(t_{i,d})$ denote the firm's
per-customer cost of providing treatment level $t_{i,d}$. In our application,
this cost captures the firm's promotional discount cost; for example, 
a customer using a \$5-off promotion costs the firm \$5. This treatment cost is
separate from the implementation cost of maintaining distinct promotions, which
we will introduce below. We normalize the cost of no treatment to zero 
($c_d(0)=0$) and let $R_i$ denote the firm's return from individual $i$. Its expected
return under feasible treatment $\boldsymbol{t}_i$ equals the expected
outcome minus the treatment cost:
\begin{equation}
E[R_i\mid\boldsymbol{x}_i,\boldsymbol{t}_i]
=
E[Y_i\mid\boldsymbol{x}_i,\boldsymbol{t}_i]
-c_d(t_{i,d}).
\end{equation}

\subsection{Step 1: Personalize with heterogeneous treatment effects\label{subsec:Treatment-effects-estimation}}

We first estimate continuous conditional average treatment effect (CATE)
functions, which capture heterogeneous treatment effects across customers with
different characteristics. These functions describe each customer's causal
response to treatment levels. They allow the firm to identify profit-maximizing
treatments under full personalization and provide the inputs for coarsening in
Step 2.

We use an experiment that randomizes customers across treatment dimensions and
levels with a no-treatment group serving as the control. For personalization,
we consider treatment levels $t_{i,d}\in[0,\bar t_d]$, where $\bar t_d$ is the
upper bound of the firm's treatment choices in dimension $d$. In our application,
these bounds are the largest dollar-off and percentage-off promotions tested in
the experiment. Under standard assumptions, the experiment
identifies CATEs relative to the control group at the randomized levels
\citep{Imbens2015}. We use these estimates to construct a continuous CATE
function $\tau_d(\boldsymbol{x}_i,t_{i,d})$ for each dimension, which describes the
expected incremental outcome from assigning treatment level $t_{i,d}$ to a
customer with characteristics $\boldsymbol{x}_i$.

The firm chooses whether to treat each customer and, if so, which treatment to
assign to maximize expected profits. Because feasible treatments have at most
one nonzero component, the firm can optimize the treatment level within each
dimension and then choose the treatment yielding the highest expected profit
across dimensions.

For a feasible treatment $\boldsymbol{t}_i$ that is nonzero in dimension $d$, we
define the CATE as
\begin{equation}
\tau_d(\boldsymbol{x}_i,t_{i,d})
\triangleq
E[Y_i\mid\boldsymbol{x}_i,\boldsymbol{t}_i]
-
E[Y_i\mid\boldsymbol{x}_i,\boldsymbol{0}].
\end{equation}
Since the cost of leaving a customer untreated is zero, assigning this treatment
is more profitable than leaving the customer untreated if and only if
\begin{equation}
E[R_i\mid\boldsymbol{x}_i,\boldsymbol{t}_i]
>
E[R_i\mid\boldsymbol{x}_i,\boldsymbol{0}]
\quad\iff\quad
\tau_d(\boldsymbol{x}_i,t_{i,d})>c_d(t_{i,d}).
\end{equation}
The firm thus targets a customer when the incremental revenue generated by
treatment exceeds its cost, which is analogous to targeting when marginal
revenue exceeds marginal cost.

For each individual $i$, the firm finds the optimal treatment level in dimension
$d$ by solving
\begin{equation}
t^*_{i,d}\in\argmax_{t_{i,d}\in[0,\bar t_d]}
\left\{\tau_d(\boldsymbol{x}_i,t_{i,d})-c_d(t_{i,d})\right\},
\label{eq:Firm-Program-Treatment}
\end{equation}
which maximizes the expected incremental profit from treatment within that
dimension.

The firm then compares the expected incremental profits at these optimal levels
across dimensions and assigns the most profitable treatment to each customer.
This yields the full personalization policy. 

\subsection{Step 2: Discretize to form the segmentation plan\label{subsec:Discretize-for-segments}}

Coarse personalization limits the number of distinct treatments the firm offers.
From Step 1, we have the continuous CATE functions
$\tau_d(\boldsymbol{x}_i,t_{i,d})$ and the individually optimal treatment levels
$t^*_{i,d}$ in each dimension. Using these results, the firm constructs a
segmentation plan by jointly choosing a limited treatment set and its customer
assignments to maximize expected profits.

For a segmentation plan with $L$ distinct treatments, let $\tilde{\mathbf
t}=\{\tilde{\boldsymbol t}_1,\ldots, \tilde{\boldsymbol t}_L\}$ denote the
treatment set. Each $\tilde{\boldsymbol t}_\ell$ is a feasible treatment with
dimension $d_\ell\in\{1,\ldots,D\}$ and level $\tilde t_{\ell,d_\ell}\in[0,\bar
t_{d_\ell}]$. Customers assigned the same treatment form a segment. The firm
therefore chooses each segment's treatment dimension and level jointly with its
customer assignment.

We now lay the groundwork for the problem. We first define the firm's expected
profit from customer $i$ under full personalization as
\begin{equation}
\bar R_i
\triangleq
E[Y_i\mid\boldsymbol{x}_i,\boldsymbol{0}]
+
\max_{d\in\{1,\ldots,D\}}
\left\{
\tau_d(\boldsymbol{x}_i,t^*_{i,d})-c_d(t^*_{i,d})
\right\},
\label{eq:Cost-Max-Return}
\end{equation}
which is the expected baseline (without treatment) outcome plus the highest
incremental profit attainable across treatment dimensions. For each dimension,
incremental profit equals the CATE evaluated at the optimal treatment level from
Step 1 minus the cost of that treatment. Equation \ref{eq:Cost-Max-Return}
represents the maximum expected profit the firm can generate from customer $i$
when it can fully personalize without a limit on the number of segments.

We then define the profits for a given segmentation plan. 
For a given treatment set $\tilde{\mathbf t}$, the firm's expected profit from
assigning treatment $\tilde{\boldsymbol t}_\ell$ in dimension $d_\ell$ to
customer $i$ is
\begin{equation}
\mathcal R_i(\boldsymbol x_i,\tilde{\boldsymbol t}_\ell)
\triangleq
E[Y_i\mid\boldsymbol x_i,\boldsymbol 0]
+\tau_{d_\ell}(\boldsymbol x_i,\tilde t_{\ell,d_\ell})
-c_{d_\ell}(\tilde t_{\ell,d_\ell}).
\label{eq:Cost-Expected-Return}
\end{equation}
To form segments, the firm evaluates each customer's expected profit under every
treatment in $\tilde{\mathbf t}$ and assigns the customer to the treatment
yielding the highest expected profit. Thus, the total expected profit under the
segmentation plan is
\begin{equation}
\mathcal R(\tilde{\mathbf t})
\triangleq
\sum_{i=1}^{N}\max_{\ell\in\{1,\ldots,L\}}
\mathcal R_i(\boldsymbol x_i,\tilde{\boldsymbol t}_\ell).
\label{eq:Segmentation-Plan-Profit}
\end{equation}
For a given $L$, the firm chooses the treatment set $\tilde{\mathbf
t}$ to maximize $\mathcal R(\tilde{\mathbf t})$.

We measure the loss from coarsening as profit regret relative to full
personalization. For customer $i$, the regret from receiving treatment
$\tilde{\boldsymbol t}_\ell$ is
\begin{equation}
C(\boldsymbol x_i,\tilde{\boldsymbol t}_\ell)
\triangleq
\bar R_i-\mathcal R_i(\boldsymbol x_i,\tilde{\boldsymbol t}_\ell).
\label{eq:Assignment-Regret}
\end{equation}

The firm also incurs an implementation cost $\psi(L)$ for maintaining $L$
distinct treatments. We assume $\psi(L)$ is weakly increasing in $L$, so
maintaining additional treatments cannot reduce implementation costs. The
implementation cost applies to the segmentation plan as a whole and is separate
from per-customer treatment costs. The firm chooses a segmentation plan to
maximize expected profits net of implementation costs. Equivalently, it
minimizes aggregate profit regret plus implementation cost by choosing the
number of distinct treatments $L$ and the treatment set $\tilde{\mathbf t}$:
\begin{equation}
\begin{aligned}
\min_{L,\tilde{\mathbf t}}\quad&
\sum_{i=1}^{N}\min_{\ell\in\{1,\ldots,L\}}
C(\boldsymbol x_i,\tilde{\boldsymbol t}_\ell)+\psi(L)\\
\text{subject to}\quad&
|\tilde{\mathbf t}|=L,\quad L\in\{1,\ldots,N\},\quad
d_\ell\in\{1,\ldots,D\},\\
&\tilde{\boldsymbol t}_\ell
=(0,\ldots,\tilde t_{\ell,d_\ell},\ldots,0),\quad
0\leq\tilde t_{\ell,d_\ell}\leq\bar t_{d_\ell},
\quad \ell=1,\ldots,L.
\end{aligned}
\label{eq:CP-Problem}
\end{equation}
Our formulation constructs the coarse personalization problem as a mixed-integer
nonlinear optimization problem in which the inner minimization determines
customer assignments. In Equation \ref{eq:CP-Problem}, the firm chooses the
number of distinct treatments $L$ and the set of offered treatments
$\tilde{\mathbf{t}}$, which includes the treatments' dimensions $d_\ell$ and
their levels $\tilde t_{\ell,d_\ell}$. Offering more treatments allows finer
personalization and can reduce profit regret, but the firm must weigh these
gains against the cost of implementing additional treatments ($\psi(L)$). The
inner minimum assigns each customer to the available treatment with the lowest
regret. The constraints require $L$ distinct treatments, each with at most one
nonzero component and a treatment level $\tilde t_{\ell,d_\ell}\in[0,\bar
t_{d_\ell}]$.
Intuitively, coarse personalization shifts the condition that
marginal revenue equals marginal cost from the individual to the
segment level, where on average equality holds for an interior
solution.\footnote{Online Appendix Section
\ref{sec:First-order-condition-discussion} provides the derivation.}

Solving Equation \ref{eq:CP-Problem} is computationally difficult because
treatment choices and customer assignments depend on one another. Changing a
treatment can change which customers receive it and changing a segment's
customers can change its optimal treatment. A brute-force approach would
enumerate customer assignments and treatment dimensions then optimize treatment
levels for each combination. This search becomes burdensome as the number of
customers and segments grows.

Another approach is to restrict treatment levels to a finite grid and
compute expected profits for each customer-treatment pair.
For a fixed $L$, the resulting problem becomes a mixed-integer linear
program that can be solved using Gurobi. However, finer grids and
larger customer populations increase the complexity of this problem.
We use this approach as a benchmark for our computational solution
which allows treatment levels to vary continuously. 

To simplify treatment optimization within each segment, we impose two
economically motivated assumptions.

\begin{assumption}[Strict concavity of treatment effects]
\label{assu:TE-Concavity}
For every customer $i$ and treatment dimension $d$, $\tau_d(\boldsymbol{x}_i,t)$
is continuous and strictly concave in $t$ on $[0,\bar t_d]$. For distinct $t,t'\in[0,\bar t_d]$ and
$\alpha\in(0,1)$, $\tau_d(\boldsymbol{x}_i,\alpha t+(1-\alpha)t') >
\alpha\tau_d(\boldsymbol{x}_i,t) +(1-\alpha)\tau_d(\boldsymbol{x}_i,t').$
\end{assumption}

\begin{assumption}[Convexity of treatment costs]
\label{assu:Cost-Convexity}
For every treatment dimension $d$, $c_d(t)$ is continuous and convex in $t$ on $[0,\bar t_d]$. For
$t,t'\in[0,\bar t_d]$ and $\alpha\in[0,1]$, $c_d(\alpha t+(1-\alpha)t')
\leq
\alpha c_d(t)+(1-\alpha)c_d(t').$
\end{assumption}

In our promotions setting, Assumption \ref{assu:TE-Concavity} implies
diminishing sensitivity to promotional levels. As the discount increases, each
additional dollar off has a smaller effect on the expected outcome. Assumption
\ref{assu:Cost-Convexity} rules out economies of scale in treatment costs and
allows linear costs, such as dollar-off promotions whose cost equals the
discount amount. Together, the assumptions make a segment's aggregate profit
strictly concave in its treatment level within each dimension. Holding customer
assignments fixed, the firm can therefore optimize the treatment level
separately in each dimension and compare the resulting profits. The joint
problem remains nonconvex because customer assignments and treatment dimensions
also vary.

We next establish two properties of the coarse personalization problem that
guide our computational approach: First, Assumptions \ref{assu:TE-Concavity} and
\ref{assu:Cost-Convexity} simplify treatment optimization for a given segment.
Second, because segment sizes are unrestricted, optimal customer assignments to  
segments can be deterministic.

\begin{prop}[Optimal treatment levels within a segment]
\label{prop:Conditional-Treatment-Optimization}
Let $S_\ell$ be a fixed, nonempty segment and fix its treatment dimension
$d_\ell$. Under Assumptions \ref{assu:TE-Concavity} and
\ref{assu:Cost-Convexity},
\[
\max_{t\in[0,\bar t_{d_\ell}]}
\sum_{i\in S_\ell}
\left\{\tau_{d_\ell}(\boldsymbol x_i,t)-c_{d_\ell}(t)\right\}
\]
has a unique solution.\footnote{The individual treatment problem in Equation
\ref{eq:Firm-Program-Treatment} is the special case with $S_\ell=\{i\}$.} The
segment's aggregate profit regret is strictly convex in its treatment level.
\end{prop}

\begin{prop}[Deterministic segmentation]
\label{prop:Deterministic-Segmentation}
For any fixed treatment set $\tilde{\mathbf t}$, when segment sizes
are unrestricted, a profit-maximizing customer assignment can be
chosen deterministically. Specifically, assigning each customer $i$
to a treatment satisfying
\[
\ell_i\in\arg\min_{\ell\in\{1,\ldots,L\}}
C(\boldsymbol x_i,\tilde{\boldsymbol t}_\ell)
\]
minimizes aggregate profit regret.
\end{prop}

Propositions \ref{prop:Conditional-Treatment-Optimization} and
\ref{prop:Deterministic-Segmentation} establish two properties that support our
computational approach. For a given treatment set, a deterministic assignment of
each customer to a profit-maximizing treatment is optimal. For a given segment,
the optimal treatment level is unique within each dimension, and the firm can
compare the resulting profits across dimensions to find the segment's optimal
feasible treatment. Online Appendix Section \ref{sec:Model-Proofs} provides the
proofs.

Section \ref{subsec:Optimal-transport-connection} connects these assignment and treatment decisions
to semidiscrete optimal transport and optimal quantization. Together
with Assumptions \ref{assu:TE-Concavity} and \ref{assu:Cost-Convexity},
this connection motivates our adaptation of Lloyd's algorithm,
which alternates between assigning customers to treatments and
optimizing each segment's treatment.

\section{Computational solution and optimal transport\label{sec:Computational-solution}}

In this section, we connect the coarse personalization problem to optimal
transport to motivate our computational solution that adapts Lloyd's algorithm. We
then present the algorithm, establish its improvement and convergence
properties, and compare its performance and computational cost with alternative
approaches. Finally, we discuss why the optimal transport perspective is
fruitful for coarse personalization.

\subsection{Connection to optimal transport\label{subsec:Optimal-transport-connection}}

The coarse personalization problem requires the firm to jointly choose
treatments and assign customers to them. We now express the joint decision in
Equation \ref{eq:CP-Problem} through two nested problems. The outer problem
chooses the treatments and segment sizes, while the inner problem determines
customer assignments given those choices. Optimal transport provides a
theoretically-rigorous framework for the inner problem. We use profit regret
$C(\boldsymbol{x}_i,\tilde{\boldsymbol t}_\ell)$ as the transport cost of
assigning customer $i$ to treatment $\tilde{\boldsymbol t}_\ell$, so minimizing
aggregate assignment regret maximizes expected profit for given treatments and
segment sizes.

\begin{figure}[!htbp]
\begin{spacing}{1.0}
\begin{centering}
\includegraphics[width=0.8\textwidth]{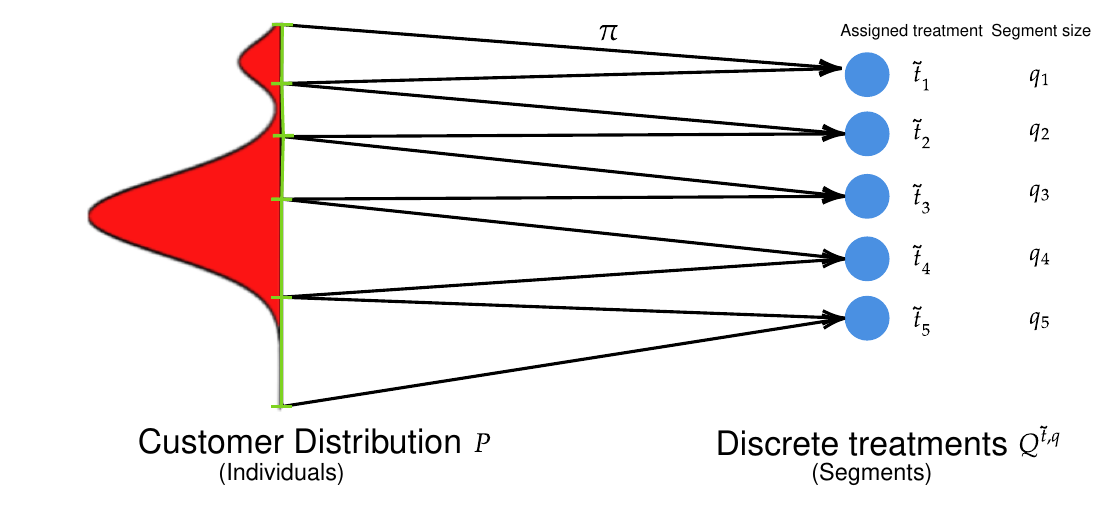}
\caption{Semidiscrete optimal transport \label{fig:OT-Semidiscrete}}
\medskip{}
\par\end{centering} {\footnotesize \singlespacing Note: The figure illustrates the assignment
of customers, represented by their continuous CATE functions, to five treatments
($L=5$). $P$ represents the customer distribution. Customers assigned treatment
$\tilde{\boldsymbol t}_\ell$ form segment $\ell$, whose normalized size $q_\ell$
is the fraction of customers receiving that treatment. The treatments and their
normalized segment sizes define the discrete treatment distribution
$Q^{\tilde{\mathbf t},\mathbf q}$. The arrows represent customer assignments
under the transport plan $\pi$, which minimizes aggregate profit regret for
given treatments and segment sizes.\par}
\end{spacing}
\end{figure}

Figure \ref{fig:OT-Semidiscrete} illustrates the assignment problem for a
one-dimensional treatment. Customers are characterized by their continuous CATE
functions, and customers assigned to the same treatment form a segment. The
transport plan $\pi$ describes these assignments. When the population
distribution of customer characteristics is continuous, the assignment problem
maps this continuous distribution to a discrete set of treatments, which leads to a
semidiscrete optimal transport problem.

The firm also chooses the treatments and segment sizes in the outer problem.
This choice connects coarse personalization to optimal quantization:
The firm discretizes a heterogeneous customer population into a limited set
of treatments while minimizing profit regret. The connection between
assignment and treatment optimization motivates the alternating computational
approach developed in the next subsection. Online Appendix Section
\ref{sec:Mathematics-of-Coarse-Personalization} provides further mathematical
background on optimal transport.

We now describe the link to optimal transport more formally. For a given number
of distinct treatments $L$, we let $P$ denote the distribution of customer
characteristics $\boldsymbol{x}$, which index the continuous CATE functions
$\tau_d(\boldsymbol{x},t)$. This distribution describes the customer population
and is provided from the first step.

We let $Q^{\tilde{\mathbf t},\mathbf q}$ denote the discrete distribution of assigned
treatments. Each treatment $\tilde{\boldsymbol t}_\ell$ receives a normalized
segment size $q_\ell$, defined as the fraction of customers receiving that
treatment. For example, $q_\ell=0.2$ means that 20 percent of customers receive
$\tilde{\boldsymbol t}_\ell$. These sizes satisfy $q_\ell\geq0$ and
$\sum_{\ell=1}^{L}q_\ell=1$, to account for
the entire customer population. By choosing $\tilde{\mathbf t}$ and $\mathbf q$,
the firm determines which treatments to offer and how many customers receive
each. The assignment problem then determines which customers receive them.

For a given treatment set $\tilde{\mathbf t}$ and normalized segment sizes
$\mathbf q$, we let $\pi$ describe the customer assignments through a joint
distribution of customer characteristics and assigned treatments. This joint
distribution specifies how customers on the left of Figure
\ref{fig:OT-Semidiscrete} are assigned to treatments on the right. Its two
marginal distributions must equal $P$ and $Q^{\tilde{\mathbf t},\mathbf q}$,
preserving the customer population and the specified segment sizes. A joint
distribution satisfying these requirements is called a \emph{coupling} in
optimal transport terminology. Different couplings can assign different
customers to each treatment while maintaining the same segment sizes. Let
$\mathcal M(P,Q^{\tilde{\mathbf t},\mathbf q})$ denote the set of these
assignment plans or couplings. The optimal transport problem selects a plan from
this set that minimizes expected profit regret.

More formally, for given treatments and segment sizes, the firm chooses the
transport plan $\pi$ linking the customer distribution $P$ to the treatment
distribution $Q^{\tilde{\mathbf t},\mathbf q}$ by solving
\begin{equation}
\mathcal W(\tilde{\mathbf t},\mathbf q)
\triangleq
\min_{\pi\in\mathcal M(P,Q^{\tilde{\mathbf t},\mathbf q})}
E_\pi[C(X,T)],
\label{eq:MK-Problem}
\end{equation}
where $X$ denotes customer characteristics and $T$ the assigned
treatment.\footnote{Equation \ref{eq:MK-Problem} is known as the
Monge-Kantorovich problem in the optimal transport literature (Online Appendix
Section \ref{sec:Mathematics-of-Coarse-Personalization}).} The expectation
averages profit regret (Equation \ref{eq:Assignment-Regret}) over
customer-treatment assignments under $\pi$. Economically, the firm determines
which customers receive each treatment while maintaining the specified segment
sizes. Thus, $\mathcal W(\tilde{\mathbf t},\mathbf q)$ is the lowest average
profit regret attainable for those treatments and sizes. In our formulation, $P$
places weight $1/N$ on each observed customer, so total profit regret is
$N\mathcal W(\tilde{\mathbf t},\mathbf q)$.

Altogether, the firm chooses the number of treatments $L$, the treatment set
$\tilde{\mathbf t}$, and segment sizes $\mathbf q$ to solve
\begin{equation}
\begin{aligned}
\min_{L,\tilde{\mathbf t},\mathbf q}\quad&
N\mathcal W(\tilde{\mathbf t},\mathbf q)+\psi(L)\\
\text{subject to}\quad&
|\tilde{\mathbf t}|=L,\qquad L\in\{1,\ldots,N\},\\
&q_\ell\geq0,\quad \ell=1,\ldots,L,\qquad
\sum_{\ell=1}^{L}q_\ell=1,
\end{aligned}
\label{eq:OT-Problem}
\end{equation}
where the treatments in $\tilde{\mathbf t}$ are feasible and their levels lie
within the treatment domains specified in Equation \ref{eq:CP-Problem}. The firm
weighs the reduction in profit regret from offering additional treatments
against their implementation cost, accounting for optimal customer assignments
through $\mathcal W$. Optimizing over $\mathbf q$ allows segment sizes to adjust
to customer assignments as in the general coarse personalization problem.

To see this, we can fix the treatment set ($\tilde{\mathbf t}$). Proposition
\ref{prop:Deterministic-Segmentation} establishes that assigning each customer
to a lowest-regret treatment is optimal when segment sizes are unrestricted.
Allowing these assignments to determine $\mathbf q$ therefore gives, for the
observed customer sample,
\begin{equation}
N\left(\min_{\substack{\mathbf q\geq0,\sum_{\ell=1}^{L}q_\ell=1}}
\mathcal W(\tilde{\mathbf t},\mathbf q)\right)
=
\sum_{i=1}^{N}\min_{\ell\in\{1,\ldots,L\}}
C(\boldsymbol x_i,\tilde{\boldsymbol t}_\ell).
\label{eq:OT-CP-Equivalence}
\end{equation}
Substituting this equality into Equation \ref{eq:OT-Problem} recovers the coarse
personalization problem in Equation \ref{eq:CP-Problem}.

Because the firm chooses segment sizes freely, it can minimize profit regret by
deterministically assigning each customer to an available treatment yielding the
highest expected profit. A coupling induced by such an assignment is called a
Monge coupling. 
These
assignments determine both the segments and their sizes, so the firm does not
need to optimize assignments or segment sizes separately.
This connects our formulation to the Monge problem in the
optimal transport literature, whose solution assigns each customer to a single
treatment.
We formalize this
connection in Proposition \ref{prop:Optimal-Monge-Coupling} and Online Appendix
Section \ref{sec:OT-Proofs} provides the proof.

\begin{prop}[Existence of an optimal Monge coupling]
\label{prop:Optimal-Monge-Coupling}
For the empirical customer distribution $P$ and any fixed treatment set
$\tilde{\mathbf t}$, optimizing over segment sizes $\mathbf{q}$ and
transport plans $\pi$ admits an optimal Monge coupling. 
\end{prop}

The proposition shows that randomizing across treatments cannot improve on
assigning each customer to a lowest regret treatment when segment sizes are
unrestricted. Thus, for any candidate treatment set, the assignment problem
reduces to comparing the profits each treatment generates for each customer. The
remaining challenge is to then choose the treatments themselves.

Choosing treatments together with customer assignments connects coarse
personalization to optimal quantization \citep{Pollard1982}. In our setting,
coarsening is guided by profit regret derived from continuous CATEs,
so customers are grouped according to the profitability of assigning them a
common treatment. This connection motivates our adaptation of Lloyd's algorithm
\citep{Lloyd1982}, which alternates between assigning customers to treatments
and optimizing each segment's treatment. Proposition
\ref{prop:Optimal-Monge-Coupling} supports the assignment step, while Proposition
\ref{prop:Conditional-Treatment-Optimization} ensures a unique optimal treatment
level within each dimension for a fixed, nonempty segment. The next subsection
develops our computational solution and establishes its theoretical properties.

\subsection{Adapted Lloyd's algorithm\label{subsec:Adapted-Lloyd-Algorithm}}

Building on the connection to optimal transport, we adapt Lloyd's algorithm to
construct a segmentation plan using the continuous CATE functions
($\tau_d(\boldsymbol{x}_i,t)$) and individually optimal treatment levels
($t^*_{i,d}$) from Step 1. For a given treatment budget $L$, the algorithm
alternates between assigning customers to treatments and updating each segment's
treatment dimension and level to improve expected profits. We run the algorithm
for different values of $L$ and compare the resulting plans' expected profits
net of implementation costs. We establish the algorithm's improvement and
convergence properties and evaluate its performance against a treatment-grid
MILP benchmark solved using Gurobi in our empirical application.

\begin{algorithm}[!htb]
\begin{spacing}{1.0}
    \caption{Adapted Lloyd's Algorithm}
\label{alg:Adapted-Lloyd's-Algorithm}

\textbf{Notation:}
$L$ is the number of treatments and $D$ the number of treatment dimensions.
Treatment $\ell$ at iteration $k$ is $\tilde{\boldsymbol t}_\ell^{\,k}$ and
$S_\ell^{\,k}$ is its customer segment. The tolerance is $\delta>0$.

\textbf{Step 0:}
Initialize $L$ feasible treatments
$\tilde{\mathbf t}^{\,0}
=(\tilde{\boldsymbol t}_1^{\,0},\ldots,
\tilde{\boldsymbol t}_L^{\,0})$
within their treatment domains.

\textbf{Step $\boldsymbol{k}$} ($k=0,1,\ldots$):
\begin{itemize}
\item (\emph{Form segments})
Assign each customer to the lowest-regret treatment and resolve ties
by the smallest treatment index:
\[
\ell_i^{\,k}\in
\arg\min_{j\in\{1,\ldots,L\}}
C(\boldsymbol x_i,\tilde{\boldsymbol t}_j^{\,k}),
\qquad
S_\ell^{\,k}=\{i:\ell_i^{\,k}=\ell\}.
\]
These segments are generalized Voronoi cells under profit regret.

\item (\emph{Optimize treatment levels}) For each nonempty segment
$S_\ell^{\,k}$ and dimension $d$, find the treatment level that maximizes the segment's
expected incremental profit:
\[
t'_{\ell,d}
\in
\arg\max_{t\in[0,\bar t_d]}
\sum_{i\in S_\ell^{\,k}}
\left\{\tau_d(\boldsymbol x_i,t)-c_d(t)\right\},
\qquad d=1,\ldots,D.
\]

\item (\emph{Choose treatment dimensions})
For each nonempty segment, compare the maximized profits across
dimensions and resolve ties by the smallest dimension index:
\[
d_\ell^{\,k+1}
\in
\arg\max_{d\in\{1,\ldots,D\}}
\sum_{i\in S_\ell^{\,k}}
\left\{
\tau_d(\boldsymbol x_i,t'_{\ell,d})
-c_d(t'_{\ell,d})
\right\}.
\]
Set $\tilde{\boldsymbol t}_\ell^{\,k+1}$ to the feasible vector whose component
in dimension $d_\ell^{\,k+1}$ is $t'_{\ell,d_\ell^{\,k+1}}$ and whose other
components are zero. If $S_\ell^{\,k}$ is empty, retain $\tilde{\boldsymbol
t}_\ell^{\,k+1} =\tilde{\boldsymbol t}_\ell^{\,k}$.

\end{itemize}

\textbf{Terminate} when
$\max_{\ell=1,\ldots,L}
\|\tilde{\boldsymbol t}_\ell^{\,k+1}
-\tilde{\boldsymbol t}_\ell^{\,k}\|_\infty<\delta$
or the iteration limit is reached.

\textbf{Output:}
Reassign customers to the final treatments and return the segmentation plan.

\end{spacing}
\end{algorithm}

Starting from an initial set of feasible treatments, we let $\tilde{\boldsymbol
t}_\ell^{\,k}$ denote treatment $\ell$ at iteration $k$, and let $S_\ell^{\,k}$
denote the customers assigned to it. Each iteration proceeds in two steps:
First, the algorithm assigns each customer to an available treatment that yields
the highest expected profit. The resulting segments form generalized Voronoi
cells under profit regret, which contain customers assigned to the same
lowest-regret treatment. Second, holding customer assignments fixed, the
algorithm finds the profit-maximizing treatment level in each dimension for each
nonempty segment. For an interior solution, this equates average marginal
revenue and average marginal cost within the segment. It then compares profits
across dimensions and assigns the segment its most profitable treatment. The
updated treatments become the candidates for the next iteration. Algorithm
\ref{alg:Adapted-Lloyd's-Algorithm} summarizes the procedure. For a fixed number
of treatments $L$, implementation cost $\psi(L)$ is constant and thus affects
neither assignment nor treatment updates. 

\begin{figure}[!htb]
\begin{centering}
\includegraphics[width=0.85\textwidth,trim={25pt 25pt 25pt
15pt},clip]{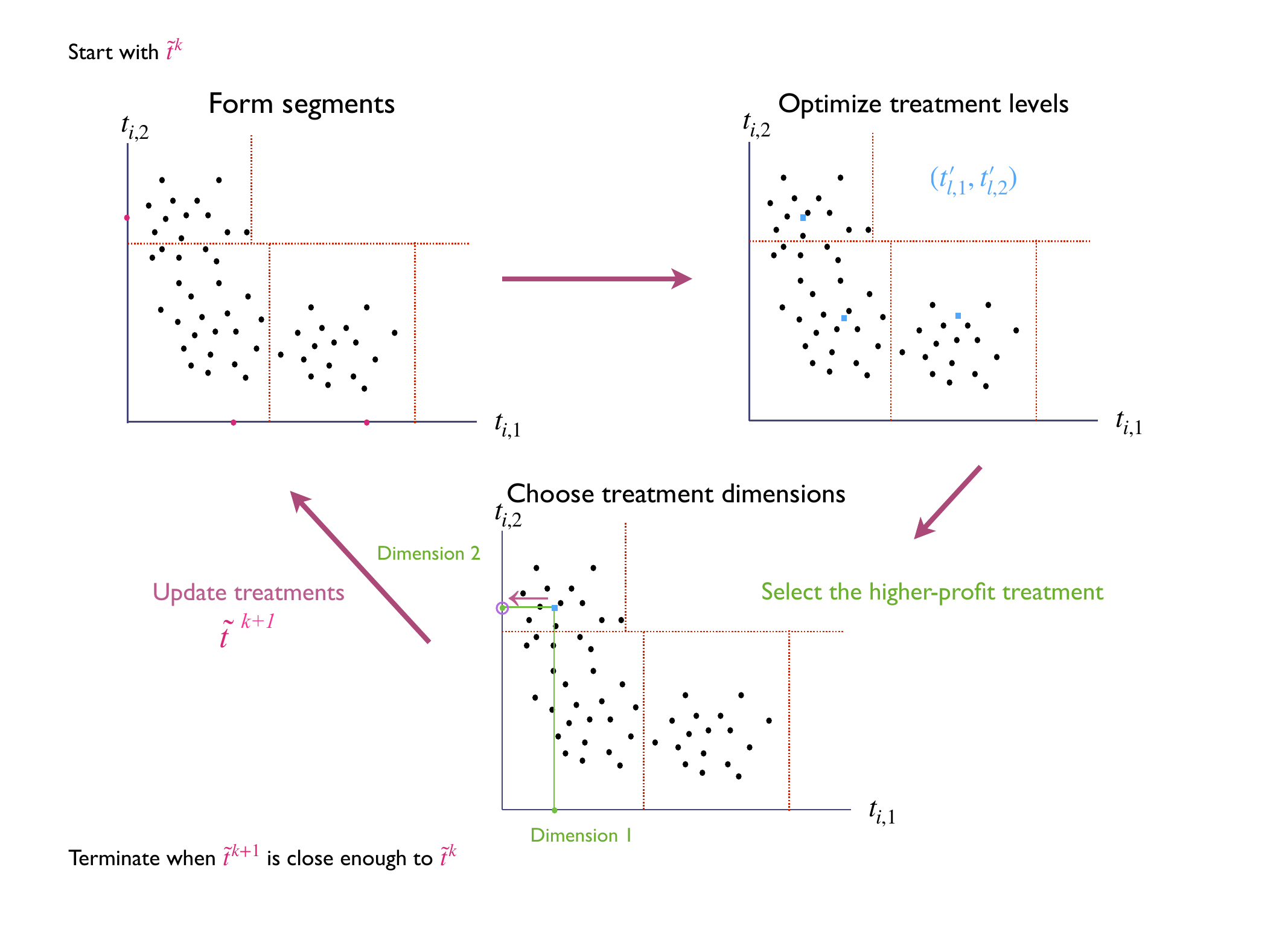}
\caption{Adapted Lloyd's Algorithm ($D=2$)\label{fig:Adapted-Lloyd's-Algorithm}}
\medskip{}
\par\end{centering} {\footnotesize \singlespacing Note: The figure illustrates
one iteration of Algorithm \ref{alg:Adapted-Lloyd's-Algorithm} with two
treatment dimensions ($D=2$) and three segments ($L=3$). Each panel corresponds
to one step of the algorithm. Black points show customers' individually optimal
treatment levels $(t^*_{i,1},t^*_{i,2})$. Purple points in the first panel (top
left) show the current treatments. Blue squares in the second panel (top right) show each
segment's optimal candidate levels in both dimensions. The third panel (bottom)
illustrates the dimension choice for one segment: Green points show the feasible
alternatives, and the purple ring and arrow mark the treatment yielding higher
total segment profit. The boundaries visualize the segments whose membership is
determined by profit regret. \par}
\end{figure}

Figure \ref{fig:Adapted-Lloyd's-Algorithm} illustrates the procedure for two
treatment dimensions. To explain the steps using our dollar-off and
percentage-off promotions example, suppose the firm initially offers \$1 off,
\$2 off, and 5\% off. Customers for whom the \$1-off promotion yields the
highest expected profit form its segment. Holding this group of customers fixed,
the firm finds the profit-maximizing promotion in each dimension. Suppose these
candidates are \$1.50 off and 2\% off. Because each customer receives only one
type of promotion, the firm compares the segment's total expected profits under
these two offers and retains the more profitable one. After updating all
segments' offers, the firm reassigns customers to their most profitable
available promotion. Thus, both the treatment offers and segment membership can
change across iterations.

Although treatment optimization within a fixed segment and dimension is a convex
optimization problem, jointly choosing customer assignments and treatments
remains nonconvex. Different starting treatments can therefore lead to different
segmentation plans. In our application, we sample individually optimal
treatments from Step 1, favoring treatments for customers whose profit regret
remains high under the offers already selected, to initialize our algorithm. We
then run the algorithm from multiple initial treatment sets and retain
the plan yielding the highest expected profit. We also consider a greedy
initialization procedure, described below, that provides a theoretical bound on
solution quality.

\subsubsection{Theoretical properties\label{subsubsec:Lloyd-Theoretical-Properties}}

In this section, we establish theoretical guarantees for our algorithm's
performance. Each iteration weakly increases expected profits, and the algorithm
converges under our stated conditions. We also establish a bound on solution
quality relative to the global optimum for a greedy initialization procedure.
These results provide a theoretical basis for assessing the algorithm's
performance and the role of initialization. We present the main results here and
provide the proofs in Online Appendix Section \ref{sec:Algo-Proofs}.

Each iteration improves the segmentation plan in two steps. Holding assignments
fixed, the treatment update maximizes each segment's expected profit.
Reassigning customers to their most profitable available treatment can then only
increase total expected profit.

\begin{prop}[Profit improvement and convergence]
\label{prop:Lloyd-Convergence}
Under fixed $L$ and Assumptions \ref{assu:TE-Concavity} and
\ref{assu:Cost-Convexity}, exact updates in Algorithm
\ref{alg:Adapted-Lloyd's-Algorithm} satisfy
\[
\mathcal R(\tilde{\mathbf t}^{\,k+1})
\geq \mathcal R(\tilde{\mathbf t}^{\,k}),
\]
and expected profits converge to a value no greater than $\sum_{i=1}^{N}\bar
R_i$. The algorithm reaches a fixed point after finitely many
iterations.
\end{prop}

This result applies to any initialization, but does not guarantee a globally
optimal segmentation plan. To bound solution quality, we consider a greedy
procedure that chooses the initial treatments. The procedure starts with the
most profitable blanket treatment and adds the treatment generating the largest
increase in total expected profit until $L$ treatments have been selected.
Adding an offer benefits only customers for whom it improves on the available
treatments, so the incremental gain from adding a given offer weakly decreases
as the treatment set expands. This property yields the following bound.

\begin{prop}[Greedy initialization]
\label{prop:Greedy-Initialization}
Let $\mathcal R_m^*$ denote the highest expected profit attainable
with at most $m$ feasible treatments from the continuous treatment
domain. The greedy procedure first selects the most profitable
blanket treatment and then adds the feasible treatment generating
the largest increase in total expected profit until $L$ treatments
have been selected. For $L\geq2$, the resulting treatment set
$\tilde{\mathbf t}^{\,0}$ satisfies
\[
\mathcal R_L^*-\mathcal R(\tilde{\mathbf t}^{\,0})
\leq
\left(1-\frac1L\right)^{L-1}
\left(\mathcal R_L^*-\mathcal R_1^*\right).
\]
For $L=1$, the procedure selects the optimal blanket treatment.
\end{prop}

Proposition \ref{prop:Greedy-Initialization} bounds the profit loss from greedy
initialization relative to the globally optimal segmentation plan with at most
$L$ treatments. By Proposition \ref{prop:Lloyd-Convergence}, subsequent
iterations weakly increase expected profits. These updates therefore preserve
the initial guarantee, and each profit gain reduces the bound on the remaining
gap by the same amount. In practice, a greedy start can also be combined with
randomized starts. Retaining the highest-profit result preserves the guarantee
while allowing other starts to improve the final segmentation plan.

Together, these results establish that the algorithm weakly improves expected
profits from any starting treatment set and provides a bound on solution quality
under greedy initialization. We next evaluate its profits and runtime
in our application.

\subsubsection{Computational performance\label{subsec:Computational-performance}}

Algorithm \ref{alg:Adapted-Lloyd's-Algorithm} reduces computational burden by
solving customer assignments and segment-level treatment updates separately,
avoiding an exhaustive search over segmentation plans. In our application,
assignment requires $O(NL)$ operations and the closed-form treatment updates
require $O(ND+LD)$ operations per iteration. Thus, $I$ iterations require
$O(I(NL+ND+LD))$ operations per start, excluding initialization, or $O(INL)$ for fixed dimension $D$. By
comparison, a grid containing $K$ feasible treatments produces $\binom{K}{L}$
possible treatment sets. Evaluating customer assignments for each set requires
$O(NL)$ operations, giving an exhaustive-search runtime of
$O\!\left(\binom{K}{L}NL\right)$ once each customer's expected profit under each
treatment has been computed. Our algorithm avoids this combinatorial enumeration,
and its computation per iteration grows linearly in the number of segments and
treatment dimensions.

We first benchmark Algorithm \ref{alg:Adapted-Lloyd's-Algorithm} against
Gurobi's solution to a treatment grid MILP. Both methods use data from
approximately 1.2 million customers and a common grid of 102 feasible treatments
spaced in increments of 0.25 across two dimensions, dollar-off and
percentage-off. Algorithm \ref{alg:Adapted-Lloyd's-Algorithm} achieves expected
incremental profits within approximately $0.022\%$ of Gurobi's proven upper
bound across $L=5,\ldots,10$. This bound, obtained by relaxing the integer
constraints, limits the profit that any feasible solution on the same grid can
achieve. For example, for $L=9$, Algorithm \ref{alg:Adapted-Lloyd's-Algorithm} takes
approximately 5.17 seconds and uses 0.90 GB RAM while Gurobi uses 28.04 hours
and 440 GB RAM. Online Appendix Table \ref{tab:Solver-Performance} reports the
full comparisons.\footnote{As the number of dimensions or $L$ increases, the
number of possible treatment sets can grow rapidly, and we expect our
algorithm's runtime advantage over Gurobi to widen.} On the tested grid,
Algorithm 1 achieves near-optimal profits using less than 1\% of Gurobi's
runtime and peak memory. These computational savings make our approach practical
for firms that need to deploy promotions quickly and at scale.

We also compare our algorithm with continuous-treatment optimizers, including
multi-start L-BFGS-B, simulated annealing, differential evolution, and a genetic
algorithm. Our algorithm attains the highest expected profits in these
comparisons. L-BFGS-B matches these profits but takes substantially longer. The
other optimizers achieve lower profits despite much greater computational
effort. Online Appendix Table \ref{tab:Solver-Heuristics} reports these
comparisons.

\subsection{Why optimal transport for coarse personalization?\label{subsec:Why-OT-for-CP}}

The optimal transport connection arises naturally from the customer assignment
problem. The transport plan specifies which treatment each customer receives and
therefore defines the firm's segments. Profit regret guides these assignments
using estimated causal effects and treatment costs, so our causally aware
segments reflect the profitability of assigning customers a common treatment.
Traditional segmentation approaches first form customer segments and then choose
a treatment for each segment. Our ``personalize then discretize'' approach
jointly determines segments and their treatments using estimated continuous
CATEs and treatment costs. The firm chooses each segment's treatment by
balancing average marginal revenue against average marginal cost across the
segment's customers.

The connection to optimal transport and optimal quantization also guided our
computational approach. It led us to adapt Lloyd's algorithm, which alternates
between assigning customers to available treatments and updating each segment's
treatment. The individually optimal treatments from Step 1 also provide informed
starting values for the randomized runs, which use the estimated continuous
CATEs to guide initialization. Assumptions \ref{assu:TE-Concavity} and
\ref{assu:Cost-Convexity} make treatment optimization within each segment and
dimension tractable.\footnote{The optimal transport formulation remains
applicable without these assumptions. They simplify treatment optimization by
imposing economically-motivated restrictions on treatment responses and costs.}
Although general-purpose optimizers can also solve for segmentation plans, the
connection to optimal transport suggests a procedure that exploits the structure
of the firm's problem. The preceding theoretical results and computational
comparisons establish the properties and practical performance of this
procedure. This connection also places the coarse personalization problem within
the broader literature on optimal transport and quantization.

For marketing applications, the framework connects heterogeneous causal
responses to an implementable segmentation plan. CATE estimates describe how
customers respond to treatment and the coarsening step determines which
treatments to offer and which customers should receive them. Comparing plans
across treatment budgets allows the firm to directly evaluate profit gains from
finer personalization against implementation costs. We next illustrate these
decisions in our empirical application for promotions management.

\section{Empirical application\label{sec:Empirical-Application}}

In this section, we illustrate our methodology with data from a large-scale
field experiment conducted by a food delivery platform. The firm chooses a
limited set of dollar-off and percentage-off promotions and assigns customers to
them to maximize expected incremental profits. We examine how much profit coarse
personalization retains relative to full personalization and how implementation
costs affect the choice of the number of treatments. We show that our
simultaneous approach generates higher estimated profits than sequential
segmentation procedures and other targeting benchmarks.

\subsection{Setting}

We use data from a large-scale randomized controlled trial conducted by a food
delivery platform across three core-based statistical areas in the United
States. Maintaining distinct promotions requires configuring offers, assigning
eligible customers, and managing campaigns. While the firm experiments with
several offers to learn customer responses, their implementation costs motivate
choosing a smaller set of promotions for subsequent deployment. The data contain
over 1.2 million customers and $126$ pretreatment characteristics describing
their demographics and recency, frequency, and monetary (RFM) behavior.
Customers were randomly assigned a dollar-off promotion (\$2, \$3, \$4, or \$5),
a percentage-off promotion (5\%, 10\%, 15\%, or 20\%), or no promotion.
Promotions were sent by email and could be used only on the customer's next
order. Customers receiving no promotion form the control group. We assume
customers do not engage in coupon gaming or stockpiling. Online Appendix Section
\ref{sec:Randomization-check} reports the randomization checks.

\begin{figure}[!htb]
\begin{centering}
\includegraphics[width=\textwidth]{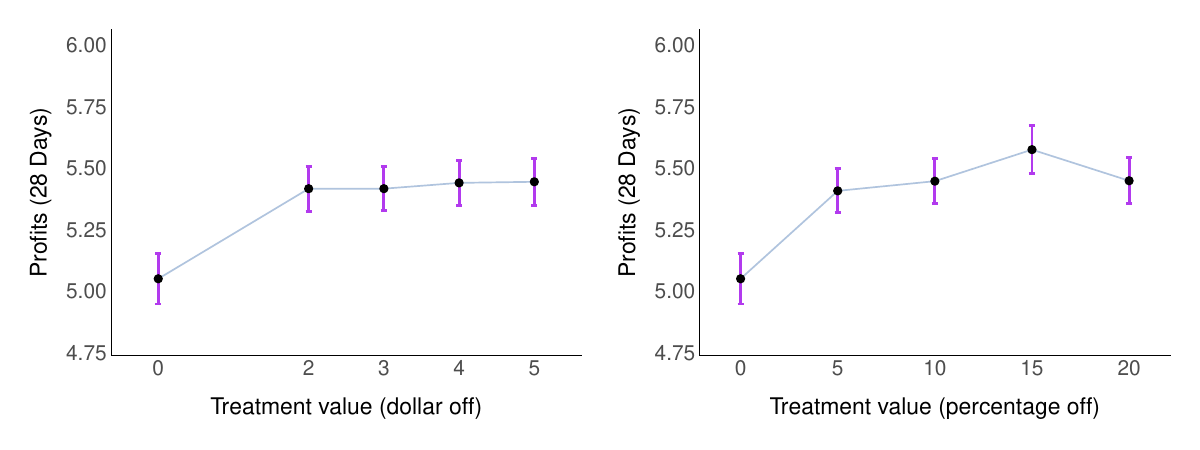}
\par\end{centering} \vspace{-0.5cm}
{\footnotesize \singlespacing
Note: Profits are measured per customer over the 28 days after the promotion
was issued, as defined in Equation \ref{eq:Profit-Equation-Setting}.
The error bars show 95\% confidence intervals for mean profits.
The points are raw means from the data. Table \ref{tab:Experiment-ATEs}
reports doubly robust estimates of the average treatment effects.
\par}
\caption{Profits by treatment\label{fig:Profits-by-Treatment}}
\end{figure}

Our outcome is the firm's total profit from each customer in the 28 days after
the promotion was sent. We measure customer spending net of payments to
restaurants and delivery drivers, then deduct the promotional cost:
\begin{equation}
\text{Profits after 28 days}_{i}
=
\text{Spending after 28 days}_{i}
-
\text{Promotion cost}_{i}.
\label{eq:Profit-Equation-Setting}
\end{equation}
The promotional cost depends on the type and level of the assigned offer and is
described below. Figure \ref{fig:Profits-by-Treatment} plots average observed
profits across the experimental treatments. These profits exclude the
implementation costs of maintaining distinct promotions, which we introduce
separately.

Table \ref{tab:Experiment-ATEs} reports the estimated average treatment effect
of each promotion on profits relative to no promotion. All point estimates are
positive, with the effects of the \$4-off, 15\%-off, and 20\%-off promotions
statistically significant at the 5\% level in two-sided $t$-tests. These positive
effects suggest an opportunity for the firm to personalize promotions to
individual customers.

\begin{table}[!htb]
\begin{centering}
\caption{Average treatment effects on profits \label{tab:Experiment-ATEs}}
\vspace{-0.7cm}
\begingroup
\singlespacing
\renewcommand{\arraystretch}{1.0}
\begin{tabular}{rrr|rl}
\hline
Treatment value & ATE (\$) & SE & $t$-stat & $p$-value\\
\hline
\multicolumn{3}{l|}{Dollar off} & & \\
\hspace{1em}2 & 0.033 & 0.045 & 0.735 & 0.462\\
\hspace{1em}3 & 0.073 & 0.046 & 1.600 & 0.110\\
\hspace{1em}4 & 0.127 & 0.046 & 2.762 & 0.006\\
\hspace{1em}5 & 0.076 & 0.049 & 1.558 & 0.119\\
\multicolumn{3}{l|}{Percentage off} & & \\
\hspace{1em}5 & 0.064 & 0.045 & 1.417 & 0.156\\
\hspace{1em}10 & 0.079 & 0.047 & 1.689 & 0.091\\
\hspace{1em}15 & 0.166 & 0.047 & 3.506 & $\mathord{<}0.001$\\
\hspace{1em}20 & 0.099 & 0.045 & 2.190 & 0.029\\
\hline
\end{tabular}
\endgroup
 \par\end{centering}
\noindent\raggedright{} {\footnotesize \singlespacing Note: The outcome is profit
per customer over the 28 days after the promotion was issued. We report doubly
robust estimates of each promotion's average treatment effect on profits
relative to no promotion, using causal forests. SE denotes the standard error.
The $t$-statistic equals ATE/SE. The $p$-value tests whether the ATE differs
from zero using a two-sided $t$-test.
\par}
\end{table}

Following Section \ref{sec:Model}, we let $R_i$ denote profits from customer $i$
over the 28-day period and $Y_i$ denote spending net of payments to restaurants
and delivery drivers before deducting promotion costs. The treatment vector
is $\boldsymbol{t}_i=(t_{i,1},t_{i,2})$, where $t_{i,1}$ is the dollar-off
amount and $t_{i,2}$ is the percentage-off level, measured in percentage
points. Each customer receives at most one promotion type, so at most one
component is nonzero. The vector $(0,0)$ represents no promotion.

We specify the dollar-off promotion cost as the discount amount. For a
percentage-off promotion, we use the customer's average spending per promotional
order over the preceding year:
\begin{equation}\label{eq:promo-costs}
c_1(t_{i,1})=t_{i,1},
\qquad
c_2(t_{i,2})=(t_{i,2}/100)\times\text{average past promo spend}_i.
\end{equation}
For customers with no prior promotional orders, we impute this spending using
the sample median. Both costs are zero when no promotion is assigned. 

In addition to the discounts themselves, the firm incurs costs to implement and
maintain distinct promotions. As a food delivery platform, the firm must
configure offers, verify customer eligibility, and manage promotions after
launch. Customer service is particularly costly for this firm because each offer
is handled separately and requires support to resolve redemption problems and
other issues. These support requirements grow with the number of distinct
promotions deployed, $L$. The firm faces implementation costs of approximately
\$1,000 per distinct promotion for this period, which we represent as
$\psi(L)=1{,}000L$.

The continuous CATE $\tau_d(\boldsymbol{x}_i,t_{i,d})$ describes the effect of
promotion type $d$ and level $t_{i,d}$ on expected spending relative to no
promotion. We assume these functions are strictly concave in the promotion
level, capturing diminishing marginal responses to larger discounts and
satisfying Assumption \ref{assu:TE-Concavity}. The promotion costs are linear in
the treatment level (Equation \ref{eq:promo-costs}) and satisfy Assumption
\ref{assu:Cost-Convexity}. Our specification also allows responses to differ
across promotion types since the distinction between absolute and percentage
discounts matters for modeling promotional responses \citep{Briesch1997}. We
thus retain separate continuous CATE response functions for dollar-off and
percentage-off promotions.

\subsection{Estimation procedure}

Our estimation procedure has three steps. We estimate customer CATEs for
spending and identify individually optimal promotions, construct coarsened
segmentation policies with a limited set of promotions, and apply an
$m$-out-of-$n$ bootstrap to quantify uncertainty and correct for possible upward
bias in estimated profits due to the optimizer's curse.

We estimate the effect of each experimental promotion relative to no promotion
using causal forests \citep{Wager2018}. We fit eight forests, one for each
offer-versus-control comparison, using the pretreatment customer
characteristics. We estimate the forests using generalized random forests
\citep{Athey2019}, which estimate propensity scores to account for possible
imbalances between each promotion arm and the control group. The forests
estimate heterogeneous effects on profits at the experimental treatment
levels.\footnote{We add back promotion costs (Equation \ref{eq:promo-costs}) to
the estimated treatment effects on profits to obtain estimated treatment effects
on spending before fitting the logarithmic functions in Equation
\ref{eq:CATE-Log-Form}. This logarithmic functional form can change the implied
profit levels relative to the estimates in Table \ref{tab:Experiment-ATEs}.}

We then construct continuous CATE functions from the estimated effects at the
experimental treatment levels. For each customer and promotion type, we
approximate the estimated spending responses using the logarithmic function
\begin{equation}
\hat{\tau}_d(\boldsymbol{x}_i,t_{i,d})
=\hat{\alpha}_{i,d}+\hat{\beta}_{i,d}\log(1+t_{i,d}),
\label{eq:CATE-Log-Form}
\end{equation}
with $\hat{\beta}_{i,d}>0$ to ensure strict concavity.\footnote{Other strictly
concave functions can also be used. Logarithmic functions have been used to
capture diminishing sensitivity in CRM models \citep{Rust2005}.} This
specification allows us to evaluate responses at promotion levels not directly
tested in the experiment. Using the coefficient estimates
$(\hat{\alpha}_{i,d},\hat{\beta}_{i,d})$, we solve Equation
\ref{eq:Firm-Program-Treatment} to obtain each customer's optimal promotion
level $t^*_{i,d}$ within each type. Comparing the resulting expected profits
across types yields the fully personalized policy. These estimates provide the
continuous CATE functions and fully personalized benchmark used for coarsening.

In the coarsening step, we use the estimated continuous CATE functions to
jointly choose a limited set of promotions and assign customers to them. For
each $L$, we apply Algorithm \ref{alg:Adapted-Lloyd's-Algorithm}, alternating
between assigning customers to their most profitable available promotion and
updating each segment's promotion type and level.\footnote{Online Appendix
Algorithm \ref{alg:Empirical-Application} provides the exact updating rules for
our specification.} The procedure seeks to minimize aggregate profit regret
relative to full personalization. We repeat this procedure for $L=1,\ldots,10$
and compare the resulting profits before and after deducting implementation
costs, $\psi(L)=1{,}000L$. This comparison shows how much profit additional
promotions generate and whether those gains justify their implementation costs.
We compare our coarse personalization solution with full personalization,
sequential segmentation procedures, and policy tree benchmarks.

To account for uncertainty in both CATE estimation and coarsening, we use an
$m$-out-of-$n$ bootstrap \citep{Bickel1997,Politis1999}. Optimization over
treatments and assignments can introduce non-smoothness and motivates us to use
this bootstrap approach. In each replication, we draw $m<n$ customers with
replacement, reestimate the causal forests and continuous CATE functions,
and reconstruct the segmentation plans. We use these bootstrap replications to
estimate standard errors for the uncorrected profit estimates.

We also address the optimizer's curse \citep{Smith2006}, where using the same
estimates to select segmentation plans and evaluate their profits can favor
plans whose profitability is overestimated.
This concern is particularly relevant here because optimized promotion levels
can fall outside the experimental offers and cannot be validated directly using
observed outcomes in a holdout sample. We use out-of-bag causal forest
predictions to limit overfitting and then adapt the bootstrap bias correction of
\citet{Xu2025} to adjust estimated profits.\footnote{Specifically, we use 2000
trees for each forest and 500 bootstrap iterations for the $m$-out-of-$n$
bootstrap.} We adopt their recommended resample size, $m=\lfloor
n^{0.95}\rfloor$, and fix this tuning parameter for our analysis. Our
implementation differs in two respects. First, we focus on the population
performance of the different segmentation methods, rather than the performance
of a segmentation policy selected from the observed sample. Second, we scale the
estimated bias correction by $\sqrt{m/n}$ to adjust from the bootstrap sample
size $m$ to the original sample size $n$. The Online Appendix
\ref{sec:Bootstrap-Bias-Correction-Details} provides details of the bootstrap
bias correction for our procedure.

\subsection{Results\label{subsec:Empirical-Results}}

\begin{figure}[!htb]
\begin{centering}
\includegraphics[width=\textwidth]{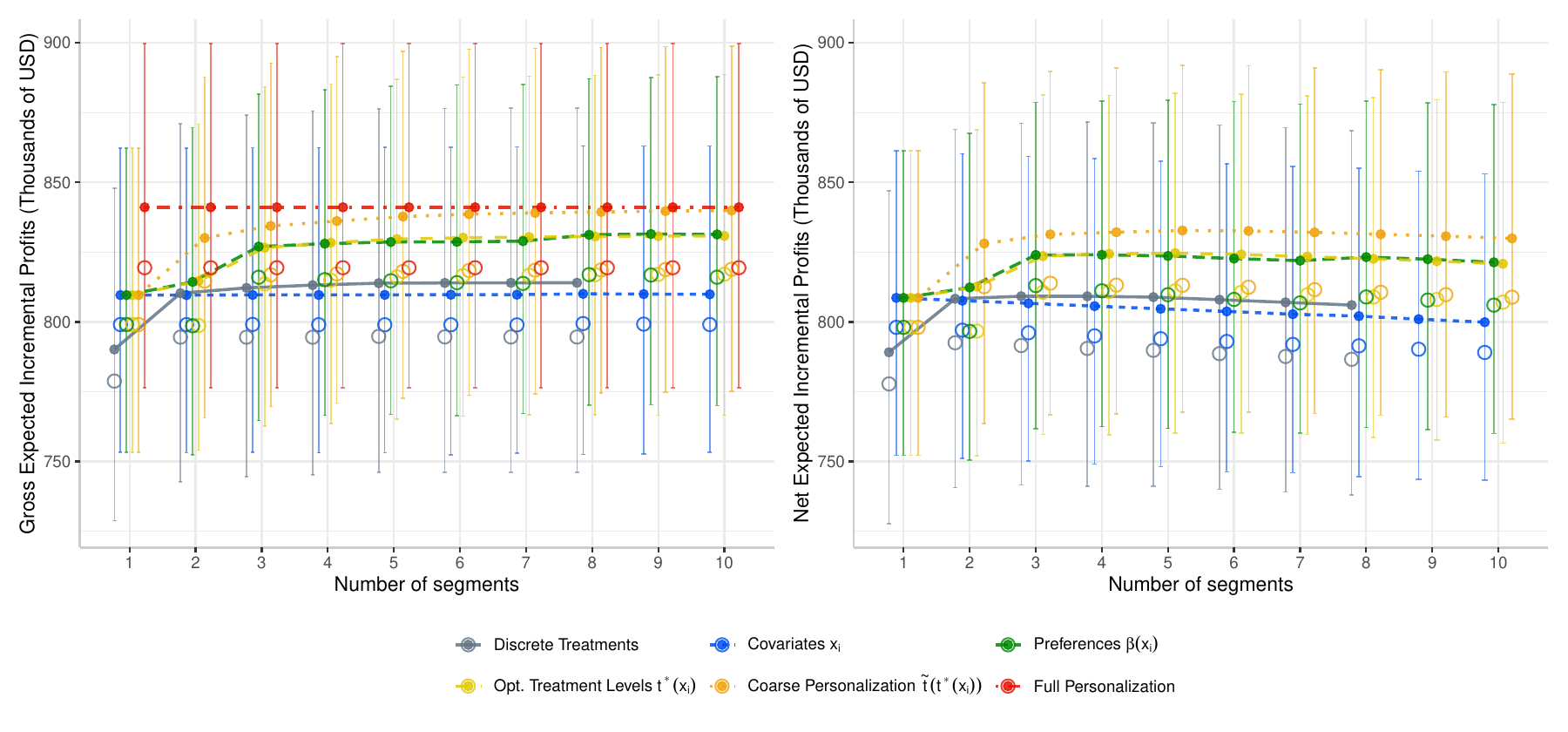}
\par\end{centering} {\footnotesize Note: The left panel shows expected
incremental profits before implementation costs. The right panel incorporates
implementation costs of \$1,000 per treatment. Filled points and lines show
uncorrected estimates, and hollow points show bootstrap bias-corrected
estimates. Error bars show 95\% bootstrap intervals for the uncorrected
estimates, based on 500 $m$-out-of-$n$ bootstrap replications. Full
personalization encompasses 2,084 distinct promotions when rounded to three
significant figures and appears only in the left panel. Online Appendix
Table \ref{tab:Profits-Uncertainty-Correction} reports uncorrected profits,
standard errors, and bootstrap corrections for all methods.\par} \caption{Profit
comparisons\label{fig:Profit-to-granular}}
\end{figure}

In our application, our coarse personalization solution recovers nearly all
full-personalization profits with only a few treatments and before accounting
for implementation costs. Adding implementation costs of \$1,000 per treatment,
coarse personalization maximizes estimated net incremental profits at five
treatments using uncorrected estimates and three after bootstrap bias
correction. Figure \ref{fig:Profit-to-granular} summarizes these profit
comparisons with and without implementation costs.

The first-stage causal forests use all 126 covariates, and the continuous
response parameterization produces average $R^2$ of $0.926$ and $0.779$ for
dollar-off and percentage-off promotions, respectively.\footnote{We pre-process
the CATE estimates and exclude 0.68\% of customers whose estimated slope in
either promotion dimension is at the lower bound of $10^{-6}$, which still leaves
over 1.2 million customers for the coarsening step.} We also validate the
continuous CATE functional form by predicting responses at held-out treatment
levels (Online Appendix Section \ref{subsec:Honest-validation}). 

We now compare expected incremental  profits across methods and numbers of
segments using the continuous CATE estimates. Figure
\ref{fig:Profit-to-granular} shows profits before and after implementation costs
and Table \ref{tab:Profits-table-by-treatments} reports the corresponding
estimates before implementation costs. Filled points show uncorrected estimates,
while hollow points show bootstrap bias-corrected estimates. The intervals
around the uncorrected estimates reflect uncertainty from both response
estimation and policy construction. The correction is generally larger for full
personalization than for covariate-based segmentation. This pattern is
consistent with greater flexibility in treatment choice and customer assignment
leading to a larger optimizer's curse.

The left panel of Figure \ref{fig:Profit-to-granular} shows that, before
implementation costs, our coarse personalization solution recovers 99.2\% of
full-personalization profits with three treatments and 99.6\% with five
treatments. After the bootstrap bias correction, these shares rise to 99.7\% and
99.8\%, respectively. Thus, restricting the firm to a small treatment set
sacrifices little estimated profit relative to full personalization. Coarse
personalization also generates higher estimated profits than the sequential
segmentation benchmarks, both before and after correction. Coarse
personalization with two treatments also generates higher estimated profits than
the depth-1 and depth-2 policy trees, both of which use two distinct offers
(Table \ref{tab:Profits-table-by-treatments}). We examine these benchmark
comparisons in more detail below.

The right panel of Figure \ref{fig:Profit-to-granular} incorporates
implementation costs of \$1,000 per treatment. Additional treatments raise
expected profits, but their diminishing incremental gains must be weighed
against implementation costs. For coarse personalization, estimated net profits
peak at five treatments using uncorrected estimates and three after bootstrap
bias correction. The correction reduces the estimated gains from additional
treatments, so fewer treatments justify their implementation costs. After the 
correction, each additional treatment beyond three generates less than \$1,000
in estimated incremental profits over the range examined. Implementation costs
of \$1,000 or more per treatment therefore favor at most three treatments, which
gives the firm a break-even comparison.

Although the uncorrected profit estimates have substantial sampling uncertainty,
comparisons between methods are more precise because each replication evaluates
them on the same resampled data. Paired comparisons across bootstrap iterations
show that coarse personalization generates significantly higher uncorrected
profits than all three sequential segmentation benchmarks at $L=10$. At $L=3$
and $L=5$, the differences remain significant against two of the three
sequential segmentation benchmarks at the 5\% level (Online Appendix Table
\ref{tab:Profits-Paired-Comparisons}). The standard errors and significance
tests only apply to the uncorrected profit estimates as we report
bias-corrected profits as point estimates. In a separate subsampling exercise
drawing 80\% of customers without replacement, coarse personalization achieves
the highest in-sample profit among the segmentation methods in all 100
replications at each $L=2,\ldots,10$ (Online Appendix Figure
\ref{fig:Segmentation-Ranks}). Together, these results provide supporting
evidence that coarse personalization performs well relative to the sequential
benchmarks, despite substantial sampling uncertainty in estimated profits.

\begin{figure}[!htb]
\begin{centering}
\includegraphics[width=\textwidth]{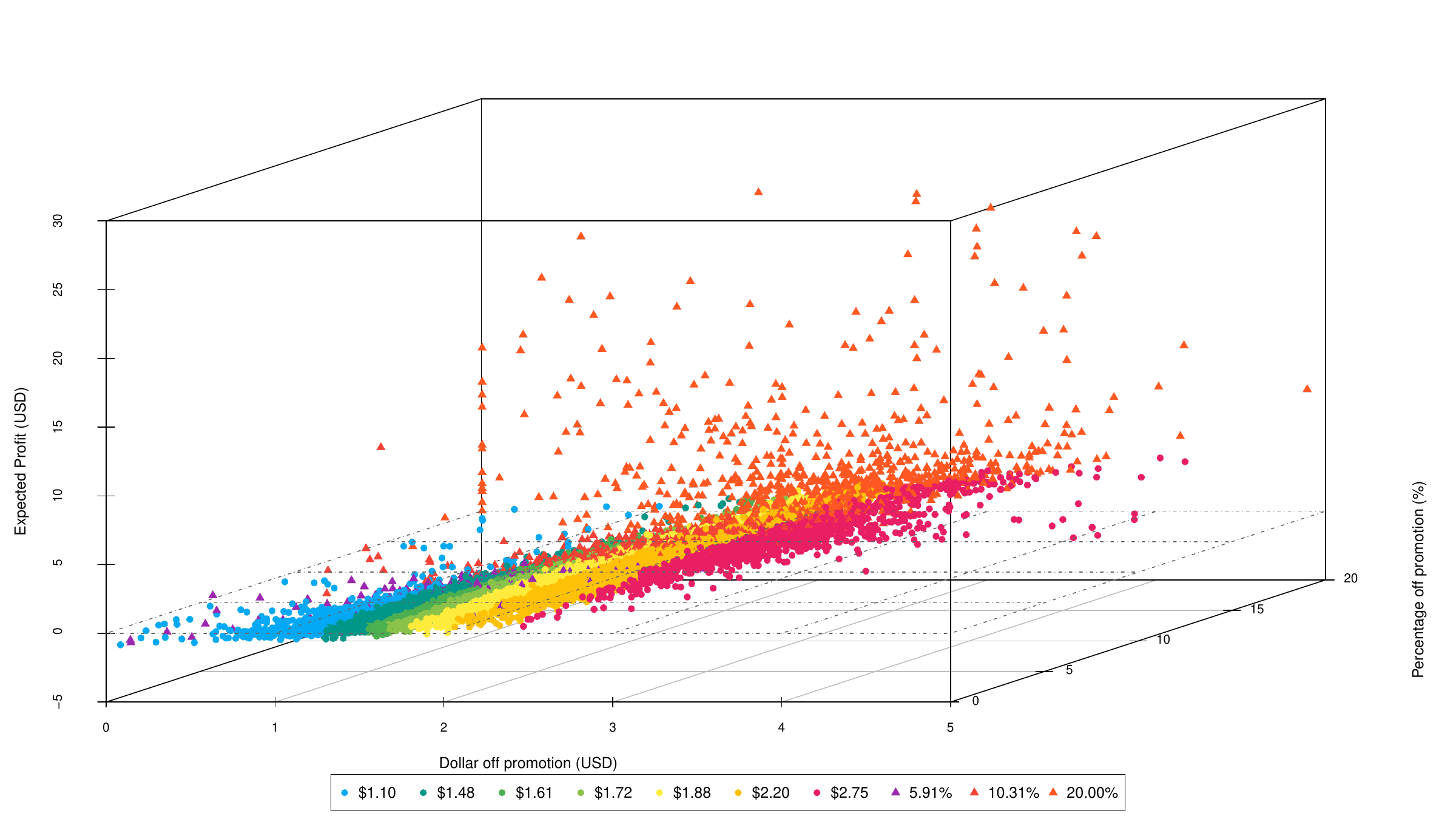}
\par\end{centering}
\begin{raggedright}
{\footnotesize Note:
The horizontal axes show the customer's individually optimal dollar-off and
percentage-off treatment values. The vertical axis shows expected incremental
profit from the treatment assigned under coarse personalization with ten
treatments, before implementation costs. Colors identify assigned treatments.
Circles indicate dollar-off promotions and triangles indicate percentage-off
promotions. The horizontal plane marks zero profit. Online Appendix Table
\ref{tab:Treatment-assignments-10} reports treatment values and segment sizes,
and Table \ref{tab:Profits-table-by-treatments} reports aggregate expected
profits.\par}
\par\end{raggedright}
\raggedright{}{\footnotesize{} }\caption{Treatment assignments ($L=10$ treatments)\label{fig:Treatment-assignments-10}}
\end{figure}

Figure \ref{fig:Treatment-assignments-10} visualizes the segmentation produced
by our optimal transport procedure with $L=10$ treatments. 
It plots customers' individually optimal dollar-off and percentage-off treatment
levels on the horizontal axes and the expected incremental profits from their
assigned treatments on the vertical axis.
We highlight three takeaways from the segmentation plot.
First, assignments form discrete groups within each promotion type,
with each customer receiving the most profitable treatment available in the
chosen set. The solution concentrates treatments where most customers are
assigned: Five dollar-off promotions between \$1 and \$2 serve approximately
93\% of customers (Online Appendix Table \ref{tab:Treatment-assignments-10}).

Second, expected incremental profits vary substantially across customers and
generally increase with their individually optimal treatment levels. Most
customers generate between \$0 and \$1, while a small number generate negative
profits, primarily under the lowest dollar-off and percentage-off promotions.
Third, a small group of customers has an optimal percentage-off level at the
20\% upper bound. The fitted responses suggest potential gains from larger
promotions for these customers, but we retain the largest experimental treatment
levels as upper bounds. This finding points to an opportunity for the firm to
test higher promotion levels in future experiments.

\subsubsection{Comparison to sequential segmentation procedures
\label{subsec:Classical-segmentation-procedures}}

We compare our coarse personalization solution (``personalize then discretize'')
with three sequential segmentation procedures (``discretize then personalize'').
Each segmentation procedure uses $k$-means to group customers by their
covariates, estimated preferences, or individually optimal treatment levels,
then selects the profit-maximizing promotion for each segment. These comparisons
illustrate the value of incorporating profits directly into segment formation,
with results reported in Figure \ref{fig:Profit-to-granular} and Table
\ref{tab:Profits-table-by-treatments}.

\begin{table}[!htb]
\centering
\caption{Expected incremental profits by segmentation method}
\label{tab:Profits-table-by-treatments}
\begin{singlespace}
\small
\setlength{\tabcolsep}{6pt}
\renewcommand{\arraystretch}{1.02}
\begin{tabular}{@{}rrrrrr@{}}
\toprule
$L$ & \shortstack{Discrete\\treatments} & Covariates & Preferences & \shortstack{Optimal\\treatment levels} & \shortstack{Coarse\\personalization} \\
\midrule
\multicolumn{6}{@{}l}{\textit{A. Uncorrected profits}} \\[2pt]
1 & 790.1 & 809.6 & 809.6 & 809.6 & 809.6 \\
2 & 810.3 & 809.6 & 814.4 & 814.7 & 830.1 \\
3 & 812.2 & 809.7 & 827.0 & 826.6 & 834.4 \\
4 & 813.2 & 809.7 & 828.1 & 828.4 & 836.1 \\
5 & 813.9 & 809.7 & 828.6 & 829.7 & 837.8 \\
6 & 814.0 & 809.8 & 828.7 & 830.2 & 838.6 \\
7 & 814.0 & 809.8 & 829.0 & 830.4 & 839.1 \\
8 & 814.0 & 810.1 & 831.2 & 830.6 & 839.4 \\
9 & \textemdash & 810.0 & 831.5 & 830.8 & 839.7 \\
10 & \textemdash & 809.9 & 831.4 & 830.8 & 839.9 \\
\addlinespace[3pt]
\multicolumn{6}{@{}l}{\footnotesize Full personalization: 841.1; policy trees (depth 1 / 2): 804.1 / 804.9.} \\
\midrule
\multicolumn{6}{@{}l}{\textit{B. Bias-corrected profits}} \\[2pt]
1 & 778.8 & 799.1 & 799.1 & 799.1 & 799.1 \\
2 & 794.5 & 799.0 & 798.6 & 798.7 & 814.6 \\
3 & 794.5 & 799.1 & 816.0 & 813.6 & 816.9 \\
4 & 794.5 & 799.0 & 815.2 & 814.9 & 817.2 \\
5 & 794.8 & 799.0 & 814.7 & 816.2 & 818.1 \\
6 & 794.6 & 799.0 & 814.1 & 816.6 & 818.4 \\
7 & 794.6 & 798.9 & 813.8 & 816.7 & 818.6 \\
8 & 794.6 & 799.4 & 816.9 & 816.9 & 818.6 \\
9 & \textemdash & 799.2 & 816.8 & 817.0 & 818.8 \\
10 & \textemdash & 799.1 & 816.0 & 817.1 & 818.9 \\
\addlinespace[3pt]
\multicolumn{6}{@{}l}{\footnotesize Full personalization: 819.4; policy trees (depth 1 / 2): 791.7 / 791.8.} \\
\bottomrule
\end{tabular}
\par\medskip
\begin{minipage}{\textwidth}
\footnotesize
\raggedright
Note: Profits are in thousands of dollars over 1.2 million customers net of
promotion costs and before implementation costs. $L$ is the number of segments
for the segmentation methods and the number of offers for the discrete treatment
benchmark. Panel B subtracts the bootstrap bias correction from Panel A, using
500 full-pipeline replications with $m=\lfloor n^{0.95}\rfloor$.
Discrete treatments are restricted to the eight experimental offers and dashes
denote unavailable treatment counts. We report policy trees of depths 1 and 2,
each using two distinct offers. Full personalization is unrestricted by $L$ and
entails 2,084 distinct promotions when rounded to three significant figures.
Online Appendix Table \ref{tab:Profits-Uncertainty-Correction} reports
uncorrected profits, standard errors, and bootstrap corrections for all methods.
\end{minipage}
\end{singlespace}
\end{table}

\emph{Covariate segmentation}. We form segments using customer characteristics
that describe demographics, platform behavior and recency, frequency, and
monetary value (RFM). We retain $59$ covariates selected by Lasso models
predicting subsequent purchase incidence or purchase amount.\footnote{These
covariates capture past spending, tipping, delivery versus pickup, order
cancellations, promotional spending, location, and ordering device.} This
benchmark represents \emph{classical ex ante segmentation}, which groups
customers with similar observed characteristics and purchase histories.

\emph{Preference segmentation}. We segment customers based on their estimated
preference coefficients, $\{\hat{\beta}_{i,1},\hat{\beta}_{i,2}\}_{i=1}^{N}$
from Equation \ref{eq:CATE-Log-Form}. These coefficients capture sensitivity to
dollar-off and percentage-off promotions. This benchmark groups customers with
similar promotional sensitivity. 

\emph{Optimal treatment levels segmentation}. We form segments using customers'
individually optimal dollar-off and percentage-off promotion levels,
$\{t^{*}_{i,1},t^{*}_{i,2}\}_{i=1}^{N}$. These levels combine information about
customers' promotional sensitivity with treatment costs. This benchmark
therefore groups customers for whom similar promotional levels are individually
optimal.

Figure \ref{fig:Profit-to-granular} and Table
\ref{tab:Profits-table-by-treatments} show that covariate segmentation generates
little additional profit as the number of segments increases. Customers with
similar observed characteristics do not necessarily respond similarly to the
same promotion. Segmenting on preferences or optimal treatment levels generates
higher estimated profits by incorporating information about promotional
responses, consistent with the benefits of preference-based targeting
discussed in \citet{Gupta1994,Rossi1996}. Optimal treatment levels also account
for treatment costs, and customers with similar promotional sensitivities may
have different optimal promotions and thus fall into different segments. Coarse
personalization generates higher estimated profits than all three sequential
benchmarks at each $L=2,\ldots,10$, both before and after the bootstrap bias
correction.

The performance differences across procedures reflect the distance metrics they
use for segmentation. Optimal-level segmentation assigns customers to the
nearest cluster center in Euclidean distance, then optimizes the treatment for
each segment. Coarse personalization instead uses expected profit loss as its
distance metric, assigning each customer the most profitable treatment in the
chosen set. This distinction matters because customers with similar individually
optimal treatment levels can incur different profit losses from receiving the
same promotion. Online Appendix Figures \ref{fig:Treatment-assignments-10-2D}
and \ref{fig:Treatment-assignments-10-2D-opt-t-levels} illustrate this
difference. Under coarse personalization, a customer can receive a promotion
farther from their individually optimal treatment level, or in another treatment
dimension, if it generates higher expected profits. Optimal treatment level
segmentation instead fixes membership by proximity to cluster centers and does
not reassign customers when another segment's promotion would generate higher
expected profits for the firm. By \emph{directly} targeting profits when forming
segments and choosing treatments, coarse personalization aligns both decisions
with the firm's objective.

Finally, our finding that coarse personalization recovers nearly all profits
from full personalization before implementation costs is specific to this
empirical application. With implementation costs of \$1,000 per treatment,
estimated net profits with coarse personalization peak at five treatments before
bootstrap bias correction and three after correction. Both the profit gap and
the optimal number of treatments may differ across applications and depend on
heterogeneity in treatment effects as well as the treatment and implementation
cost functions.

\subsubsection{Discrete treatments
benchmark\label{subsec:A/B-testing-comparison}}

In this section, we evaluate the cost of restricting dollar-off and
percentage-off promotions to the discrete levels used in the experiment. We
first examine how much profit the firm forgoes by not fully optimizing its
blanket promotion, then assess the cost of restricting personalized promotions
to the experimental treatment levels. These comparisons show the value of
optimizing treatment levels themselves. We then discuss how coarse
personalization can accommodate discrete actions, such as free shipping,
alongside continuous treatments.

We first ask how much the firm can gain by optimizing its blanket promotion,
before introducing any personalization. We compare the firm's historical $10\%$
off offer with the best experimental offer and the optimal promotion when
treatment levels can vary continuously. The historical offer generates expected
incremental profits of \$332,420, or 39.5\% of full personalization profits.
Under the estimated continuous CATEs, choosing the best experimental offer, \$2
off, raises profits to \$790,135. Allowing continuous treatment levels yields an
optimal blanket promotion of approximately \$1.69 off and profits of \$809,619
(Table \ref{tab:Profits-table-by-treatments}).\footnote{These are uncorrected
profit estimates and exclude implementation costs. They use the fitted continuous CATE functions in Equation
\ref{eq:CATE-Log-Form}.} Thus, optimizing the blanket
offer alone goes a long way toward recovering full personalization profits,
while allowing its level to vary beyond the experimental offers generates
further gains.

We next ask how much profit the firm can generate by personalizing assignments
while restricting promotions to the eight experimental offers. This restriction
connects our comparison to policy learning methods that assign customers to a
fixed set of discrete treatments.\footnote{\citet{pmlr-v84-kallus18a} adapt
policy learning for continuous treatments, but their reliance on kernel
estimation faces the curse of dimensionality as the number of treatment
dimensions increases.} For each $L$, we search over all combinations of $L$
experimental offers, assign each customer the most profitable offer in each
combination, and finally select the combination with the highest aggregate
profit. This discrete treatment benchmark provides an upper bound on fitted
profits for policy learning methods restricted to the same experimental offers
and treatment count. Coarse personalization generates higher estimated profits
at every $L$, both before and after bootstrap bias correction (Table
\ref{tab:Profits-table-by-treatments}), which demonstrates the gains from optimizing
treatment levels beyond the experimental offers.

We also examine policy trees of depths one and two \citep{Athey2021,Zhou2022}
and find that both generate lower estimated profits than the optimized two-offer
discrete policy, before and after bootstrap bias correction. This comparison
illustrates the cost of restricting customer assignments to a shallow tree in
this application. We train the trees using pretreatment customer characteristics
and the same fitted-profit objective as the discrete-treatment benchmark, with
the eight experimental promotions as available treatments. Online Appendix
Figure \ref{fig:Policy-Trees} visualizes the policy trees. Both trees select two
offers, \$2 off and $20\%$ off and generate uncorrected profits of \$804,052 and
\$804,904, respectively, compared with \$810,296 for the optimized two-offer
discrete policy (Table \ref{tab:Profits-table-by-treatments}). Thus,
more flexible customer assignments than what tree-based models allow can improve
estimated profits.

Although the preceding comparisons illustrate the value of treatment continuity,
some promotions, such as free expedited shipping, are inherently discrete.
Coarse personalization can accommodate these discrete treatments alongside
continuous treatments by including them as candidates when updating each
segment's promotion. For each segment, we compare the aggregate expected profit
from each discrete action with that from the best promotion in each continuous
dimension and select the most profitable candidate. Customers continue to
receive the most profitable offer in the chosen treatment set. Thus, whether a
discrete action appears in the set is determined by its profitability. Online
Appendix Algorithm \ref{alg:Mixed-Treatments} details the extension.
The next subsection considers restricting continuous treatment levels to rounded
increments.

\subsubsection{Rounded treatment levels\label{subsec:Rounded-treatment-levels}}

\begin{figure}[!htb]
\centering
\includegraphics[width=\textwidth]{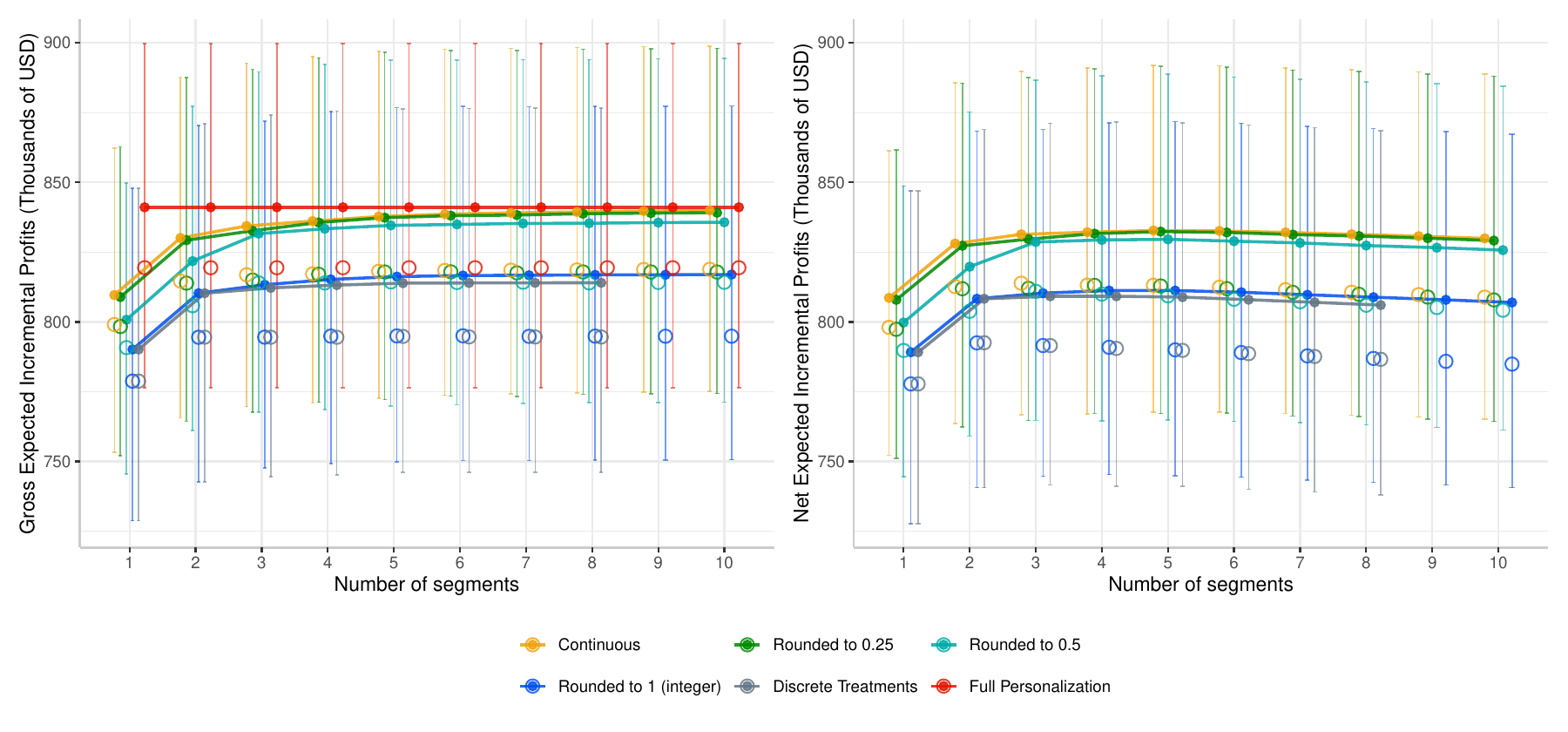}
\par\medskip {\footnotesize\raggedright Note: Rounded increments are in dollars
for dollar-off promotions and percentage points for percentage-off promotions
with treatment levels optimized on the grid within each update. The left panel
excludes implementation costs and the right panel subtracts \$1,000 per
treatment for implementation costs. Filled points show uncorrected profits,
hollow points show bias-corrected profits, and error bars show 95\% bootstrap
intervals for the uncorrected estimates from 500 replications. Full
personalization appears only on the left panel. Online Appendix Table
\ref{tab:Profits-table-by-treatments-discrete} reports the integer-grid profits.
\par}
\caption{Profit comparisons for coarse personalization with rounded treatments}
\label{fig:Profit-to-granular-rounded}
\end{figure}

Promotions with fractional dollar or percentage amounts may be impractical, so
we examine how much profit the firm forgoes when treatment levels must be
rounded. We restrict dollar-off and percentage-off promotions to increments of
$0.25$, $0.5$, and $1$, measured in dollars and percentage points, respectively.
At each iteration, we select the profit-maximizing permitted level for each
segment and treatment dimension, then update customer assignments. This approach
directly incorporates rounding into the optimization routine.
Online Appendix Algorithm \ref{alg:Rounded-Treatments} details the procedure and
Figure \ref{fig:Profit-to-granular-rounded} reports the resulting profits.

The left panel of Figure \ref{fig:Profit-to-granular-rounded} shows that modest
rounding preserves most of the profits from continuous treatment optimization.
At $L=10$, increments of $0.25$ and $0.5$ reduce bias-corrected profits by
approximately $0.1\%$ and $0.6\%$, respectively, relative to continuous coarse
personalization, while integer increments reduce profits by $2.9\%$. Allowing
promotions in smaller increments generates higher estimated profits than using
only the experimental offers. 
In contrast, integer treatments offer a modest improvement over the experimental
promotions after bias correction, even though ten segments recover nearly all
the profits from integer-rounded full personalization (Online Appendix Table
\ref{tab:Profits-table-by-treatments-discrete}). Thus, the main profit loss
comes from restricting the available treatment levels rather than limiting
personalization to a small number of segments.

The right panel of Figure \ref{fig:Profit-to-granular-rounded} shows that
implementation costs further limit the value of additional treatments. At
\$1,000 per treatment, uncorrected net profits peak at five treatments for all
three rounding grids. After bootstrap bias correction, net profits peak at four
treatments for the $0.25$ grid, three for the $0.5$ grid, and two for the
integer grid. The preferred rounding level ultimately depends on the firm's
application and how much profit it is willing to forgo for simpler treatment
values. Even when the firm commits to rounded treatments, coarse personalization
continues to form profitable segments.

\section{Surplus analysis\label{sec:Surplus-Analysis}}

We now use our coarse personalization framework to study how the degree of
personalization affects consumer, producer, and total surplus. While
personalization allows the firm to offer more profitable promotions, these
changes may increase or decrease the value consumers receive \citep{Dube2017}.
Implementation costs further reduce the gains from offering additional
treatments. We empirically examine how expanding the treatment set beyond an
optimized blanket promotion changes the surplus accruing to the firm and its
customers in our application.

To complete the surplus calculation, we impose assumptions on consumers'
valuation of the promotions. We assume that consumers value dollar-off
promotions at face value and percentage-off promotions at the discount rate
times their average past promotional order value. Thus, $v_i(t_{i,1})=t_{i,1}$
and $v_i(t_{i,2})=(t_{i,2}/100)\times\text{average past promo spend}_i$, where
percentage-off treatments are expressed in percentage points. Under these
assumptions, the monetary value of the promotion is a transfer from the firm to
the consumer. Implementation costs reduce producer and total surplus but leave
consumer surplus unchanged. We defer the complete surplus calculation details to Online
Appendix Section \ref{sec:Surplus-Appendix}.

\begin{figure}[!htb]
\begin{centering}
\includegraphics[width=0.8\textwidth]{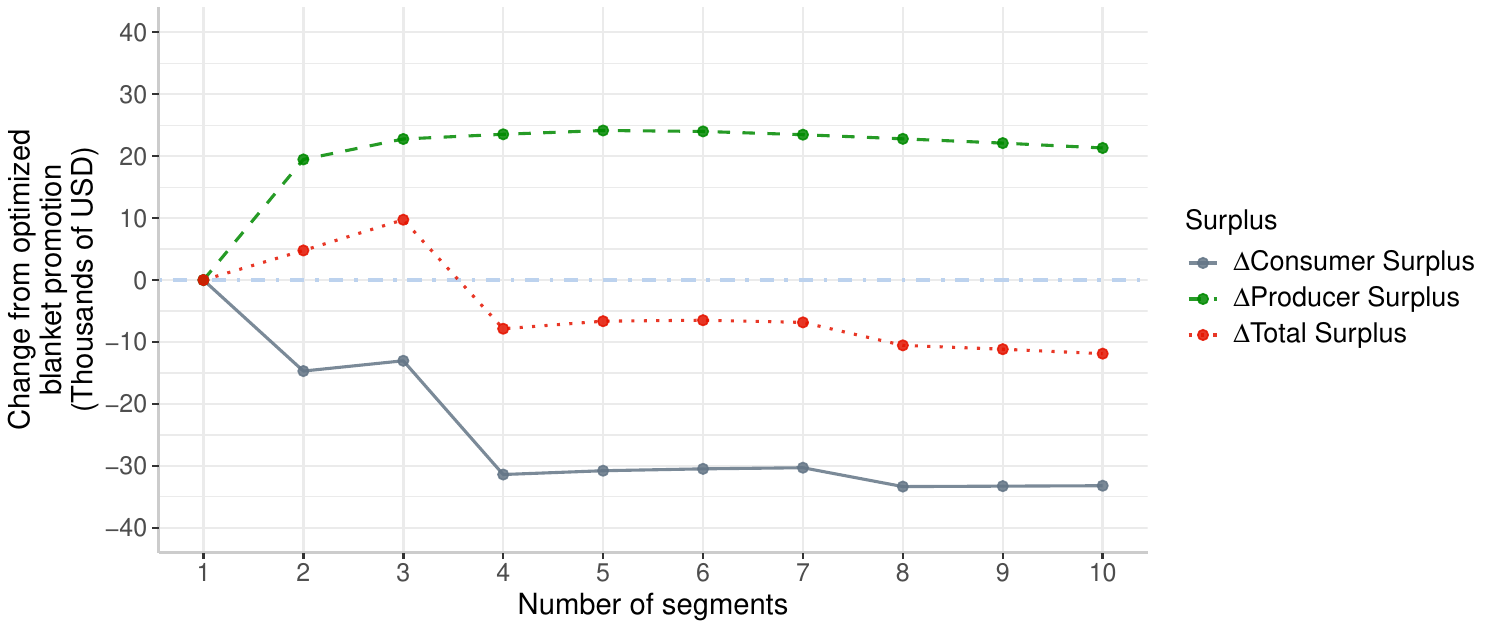}
\par\end{centering}
\begin{singlespace}
{\footnotesize Note: Uncorrected estimates of changes in surplus relative to the
optimized blanket promotion ($L=1$). Producer and total surplus deduct
implementation costs of \$1,000 per treatment for both policies.
} \caption{Surplus decomposition across treatments
\label{fig:Welfare-decomposition-bootstrap}}
\end{singlespace}
\end{figure}

Figure \ref{fig:Welfare-decomposition-bootstrap} shows changes in surplus
relative to the optimized blanket promotion, after implementation costs of
\$1,000 per treatment. Producer surplus increases initially and peaks at $L=5$,
while consumer surplus peaks at $L=1$, but it does not decline monotonically.
Estimated total surplus is highest at $L=3$, where the producer surplus gain of
\$22,768 exceeds the consumer surplus loss of \$13,032, yielding a net gain of
\$9,736 relative to the blanket promotion. For $L=4,\ldots,10$, the consumer
surplus losses exceed the net producer surplus gains, leaving total surplus
below the blanket treatment benchmark. Online Appendix Table
\ref{tab:Welfare-decomposition-treatments} reports the full decomposition.
Estimated total surplus also peaks at $L=3$ after correcting producer surplus
for the optimizer's curse (Online Appendix Figure
\ref{fig:Welfare-decomposition-corrected}).

These results illustrate why the gains from personalization should be evaluated
from both the firm's and its customers' perspectives. The firm's most profitable
segmentation plan does not necessarily generate the greatest total surplus
because the firm chooses promotions to maximize profits rather than consumer
surplus.
In our application, a small treatment set captures much of the benefit
of personalization. Three treatments retain most of the attainable net
profit gains and yield the highest estimated total surplus among the
segmentation plans examined.
While the firm chooses $L$ to maximize profits net of implementation
costs, the surplus analysis makes explicit the consumer gains or losses
associated with that choice.

\section{Conclusion \label{sec:Conclusion}}

Recent advances in estimating heterogeneous treatment effects have made
granular, incrementality-based targeting and personalization feasible. However,
maintaining many distinct treatments can be costly, so firms must weigh the
additional profits from personalization against its implementation costs. We
refer to this as the coarse personalization problem, where firms need to choose
how many distinct treatments to maintain, what marketing mix each treatment
represents, and which segments of customers to offer each treatment to.

In this paper, we propose a two-step approach to the coarse personalization
problem. The first step uses existing methods to estimate conditional average
treatment effects and identify individually optimal treatments. The second step
uses these estimates to jointly form causally aware customer segments and choose their
treatment types and continuous levels, subject to a limit on the number of
distinct treatments. We formulate customer assignment as an optimal transport
problem, mapping the distribution of customers, characterized by their
heterogeneous treatment responses, to a limited set of treatments using profit
regret to full personalization as the transport cost. This formulation connects
coarse personalization to optimal quantization and enables our adaptation of
Lloyd's algorithm, which provides a scalable computational approach.
We therefore advocate for a ``personalize then discretize'' approach that
forms segments and chooses their assigned marketing mix simultaneously,
instead of the traditional ``discretize then personalize'' approach that
first forms segments and then chooses their marketing mix.

To illustrate the practical relevance of our methodology, we present an
empirical application for promotions management using field experiment data from
a food delivery platform. For the platform, implementation costs include
configuring each promotion and providing customer support. With our approach, we
find that five optimized treatments recover 99.6\% of estimated
full-personalization profits before implementation costs. With implementation
costs of \$1,000 per treatment, estimated profits after implementation costs
peak at five treatments before adjusting for the optimizer's curse and three
afterward. Our approach generates higher estimated profits than sequential
procedures that first segment on covariates, preferences, or individually
optimal treatment levels, both before and after adjusting for the optimizer's
curse. These results have direct managerial implications. We offer a practical
way for firms and marketers to jointly choose customer segments and their
marketing mix, while weighing the gains from additional segments against their
implementation costs.

Our framework can be adapted to other marketing applications where maintaining
distinct options is costly. For example, firms may choose a limited set of
advertising variants for customers or incentive plans for salespeople. In each
setting, the firm must jointly choose these options and decide who receives
each. The appropriate degree of personalization will depend on heterogeneity in
responses as well as treatment and implementation costs. Future research can
apply the framework in these settings and use field experiments to evaluate the
resulting segmentation plans. We hope this framework encourages further research
that bridges advances in personalization and marketing practice.

\newpage
\begin{spacing}{1.0}
\small
\putbib[bib]
\end{spacing}

\end{bibunit}

\begin{bibunit}[ecta]
\makeatletter
\def\@extra@b@citeb{@appendix}
\def\@extra@binfo{@appendix}
\makeatother

\appendix

\section*{Online Appendix\pdfbookmark{Online Appendix}{sec:Online-Appendix}}

\section{Mathematics of Coarse
Personalization\label{sec:Mathematics-of-Coarse-Personalization}}

This section introduces the optimal transport concepts used in Section
\ref{subsec:Optimal-transport-connection}.
We provide a canonical example used in the optimal transport literature to
introduce the general framework. Then we link it to our promotions management
setting from our empirical application. We direct readers to
\citet{Galichon2016}, \citet{Santambrogio2015}, and \citet{Villani2009} for a
more formal introduction to optimal transport.

In essence, optimal transport is the study of moving mass from one distribution
to another while minimizing an established cost function. These two
distributions can be continuous or discrete. The solution to the optimal
transport problem is an \emph{optimal transport plan} that specifies how mass is
allocated between the two distributions.

For a concrete example, consider transporting ore from mines to factories. The
optimal transport problem determines the least costly way to move the ore, given
each mine's supply and each factory's required quantity. The optimal transport
plan specifies how much ore is moved from each mine and where it is sent. A
\emph{transport map} is a special case of a transport plan in which \emph{all}
the ore from each mine is sent to a single factory. In other words, a map cannot
\emph{split mass} by sending half of a mine's ore to one factory and the other
half to another. A transport plan allows such splitting, while a transport map
requires a deterministic assignment of each mine to one factory.

The transportation example extends to continuous distributions of supply and
demand. When the source distribution is continuous and the receiving
distribution has finitely many points, the problem is called semidiscrete
optimal transport.

In our promotions setting, customers take the place of mines and offered
treatments take the place of factories. Customer characteristics index the
treatment responses estimated in Step 1, and customers assigned the same
treatment form a segment. A continuous distribution of customer characteristics
and a finite set of treatments, each defining a segment, therefore yield a
semidiscrete formulation. In our empirical implementation, we approximate the
customer population using the observed sample, assigning probability $1/N$ to
each customer.

Our coarse personalization problem in Equation \ref{eq:OT-Problem}
has two nested problems. The outer problem chooses the number of
treatments, their dimensions and levels, and segment sizes. Given
these choices, the inner optimal transport problem assigns customers
to treatments.

We focus on the inner optimal transport problem (Equation \ref{eq:MK-Problem}).
We define $P$ as the distribution of customer characteristics $\boldsymbol{x}$,
which index the continuous CATE functions $\tau_d(\boldsymbol{x},t)$. We let
$\tilde{\mathbf t}=\{\tilde{\boldsymbol t}_1,\ldots, \tilde{\boldsymbol t}_L\}$
denote the $L$ offered treatments and $\mathbf q=(q_1,\ldots,q_L)$ their
normalized segment sizes. The treatment distribution $Q^{\tilde{\mathbf
t},\mathbf q}$ assigns probability $q_\ell$ to treatment $\tilde{\boldsymbol
t}_\ell$, where $q_\ell\geq0$ and $\sum_{\ell=1}^{L}q_\ell=1$. Economically,
$q_\ell$ is the fraction of customers receiving treatment $\ell$.

In this setup, the transport plan assigns customers represented by
$P$ to the $L$ offered treatments. Customers receiving the same
treatment form a segment, so $L$ also denotes the number of segments
in the plan. The normalized size $q_\ell$ describes the fraction
of customers in segment $\ell$, while the transport plan determines
which customers belong to it.

A \emph{coupling} is a joint distribution of customer characteristics and
assigned treatments whose marginal distributions equal $P$ and
$Q^{\tilde{\mathbf t},\mathbf q}$. In our setting, it describes customer
assignments while preserving the customer population and the specified segment
sizes. A coupling can allow probabilistic assignments, whereas a transport map
assigns each customer to a single treatment.

More formally, following \citet{Galichon2016}, let $X\sim P$
denote customer characteristics in $\mathcal X$ and
$Y\sim Q^{\tilde{\mathbf t},\mathbf q}$ denote assigned treatments
in $\mathcal Y$. We write
$\mathcal M(P,Q^{\tilde{\mathbf t},\mathbf q})$ for the set of
joint probability distributions $\pi$ on $\mathcal X\times\mathcal Y$
satisfying
\[
\pi(A\times\mathcal Y)=P(A),
\qquad
\pi(\mathcal X\times B)=Q^{\tilde{\mathbf t},\mathbf q}(B)
\]
for all measurable sets $A\subseteq\mathcal X$ and
$B\subseteq\mathcal Y$. These marginal conditions ensure that the assignment
plan preserves the customer distribution and the specified normalized segment
sizes.

When $Y=T(X)$ with probability one for a measurable function $T$,
the coupling is a \emph{Monge coupling} and $T$ is a
\emph{transport map}. In our setting, this map assigns each
customer to a single treatment and therefore a single segment.

For given treatments and normalized segment sizes, the \emph{Monge
problem} chooses a transport map that minimizes expected transport cost:
\begin{equation}
\begin{aligned}
\min_T\quad & E_P[C(X,T(X))]\\
\text{subject to}\quad & T_{\#}P=Q^{\tilde{\mathbf t},\mathbf q}.
\end{aligned}
\end{equation}
Here, $T_{\#}P$ is the distribution of assigned treatments induced
by applying $T$ to customers drawn from $P$. The constraint requires
these assignments to match the specified normalized segment sizes.
In our setting, the transport cost $C(X,T(X))$ is the profit regret
from the assigned treatment relative to full personalization.

In general, the Monge problem may not have a solution. Allowing
probabilistic assignments gives the \emph{Monge--Kantorovich problem},
\begin{equation}
\min_{\pi\in\mathcal M(P,Q^{\tilde{\mathbf t},\mathbf q})}
E_\pi[C(X,Y)].
\end{equation}
This problem optimizes over couplings rather than only deterministic
transport maps. It allows customers to be assigned probabilistically
across treatments while preserving the specified normalized segment sizes.

In our setting, the firm chooses segment sizes freely. For any fixed treatment
set, it can therefore minimize profit regret by assigning each customer to a
treatment that yields the highest expected profit. These assignments determine
the segment sizes and form an optimal Monge coupling, as established in
Proposition \ref{prop:Optimal-Monge-Coupling}. Thus, allowing probabilistic
assignments does not improve the optimal value. Online Appendix Section
\ref{sec:OT-Proofs} provides the proof.

\section{Proofs}

\subsection{Proofs for Section \ref{sec:Model} \label{sec:Model-Proofs}}

We provide proofs for Propositions
\ref{prop:Conditional-Treatment-Optimization} and
\ref{prop:Deterministic-Segmentation} in Section
\ref{subsec:Discretize-for-segments}.

\begin{prop*}[Optimal treatment levels within a segment]
Let $S_\ell$ be a fixed, nonempty segment and fix its treatment
dimension $d_\ell$. Under Assumptions \ref{assu:TE-Concavity} and
\ref{assu:Cost-Convexity},
\[
\max_{t\in[0,\bar t_{d_\ell}]}
\sum_{i\in S_\ell}
\left\{\tau_{d_\ell}(\boldsymbol x_i,t)-c_{d_\ell}(t)\right\}
\]
has a unique solution. The segment's aggregate profit regret is
strictly convex in its treatment level.
\end{prop*}

\begin{proof}
Under Assumption \ref{assu:TE-Concavity}, each customer's treatment
effect is continuous and strictly concave in $t$. Under Assumption
\ref{assu:Cost-Convexity}, the treatment cost is continuous and convex.
Subtracting the treatment cost therefore gives a continuous,
strictly concave incremental profit function. Summing across a
nonempty segment preserves these properties. Continuity on the
closed, bounded treatment interval ensures that a maximum exists,
and strict concavity makes it unique.

Aggregate profit regret equals the negative of this objective plus
the constant
\[
\sum_{i\in S_\ell}
\left\{\bar R_i-E[Y_i\mid\boldsymbol x_i,\boldsymbol 0]\right\}.
\]
It is therefore strictly convex and has the same unique optimal
treatment level.
\end{proof}

\begin{prop*}[Deterministic segmentation]
For any fixed treatment set $\tilde{\mathbf t}$, when segment sizes
are unrestricted, a profit-maximizing customer assignment can be
chosen deterministically. Specifically, assigning each customer $i$
to a treatment satisfying
\[
\ell_i\in\arg\min_{\ell\in\{1,\ldots,L\}}
C(\boldsymbol x_i,\tilde{\boldsymbol t}_\ell)
\]
minimizes aggregate profit regret.
\end{prop*}

\begin{proof}
Fix the treatment set and consider any randomized assignment.
Let $p_{i\ell}$ be the probability that customer $i$ receives
treatment $\tilde{\boldsymbol t}_\ell$, with $p_{i\ell}\geq0$ and
$\sum_{\ell=1}^{L}p_{i\ell}=1$. The customer's expected regret satisfies
\[
\sum_{\ell=1}^{L}p_{i\ell}
C(\boldsymbol x_i,\tilde{\boldsymbol t}_\ell)
\geq
\min_{\ell\in\{1,\ldots,L\}}
C(\boldsymbol x_i,\tilde{\boldsymbol t}_\ell),
\]
because a weighted average cannot be smaller than its smallest component.
Assigning the customer entirely to a treatment attaining the minimum achieves
equality. Since segment sizes are unrestricted, this choice can be made
separately for every customer. Summing across customers establishes optimality.
If several treatments attain the minimum, selecting any one gives a
deterministic optimal assignment. We select the one with the lowest treatment
index to break the tie.
\end{proof}

\subsection{Proofs for Section \ref{subsec:Optimal-transport-connection}
\label{sec:OT-Proofs}}

We provide the proof of Proposition \ref{prop:Optimal-Monge-Coupling} in Section
\ref{subsec:Optimal-transport-connection}.

\begin{prop*}[Existence of an optimal Monge coupling]
For the empirical customer distribution $P$ and any fixed treatment
set $\tilde{\mathbf t}$, optimizing over segment sizes $\mathbf q$
and transport plans $\pi$ admits an optimal Monge coupling.
\end{prop*}

\begin{proof}
For each customer characteristic vector $\boldsymbol x$ in the sample, select a
treatment attaining the lowest regret, resolving ties by the smallest treatment
index. Denote this assignment by $T^*(\boldsymbol x)$. Let $q_\ell^*$ be the
fraction of customers assigned treatment $\tilde{\boldsymbol t}_\ell$, and let
$\pi^*$ be the joint distribution of $(X,T^*(X))$ when $X$ follows $P$.

The customer marginal of $\pi^*$ is $P$, and its treatment marginal is
$Q^{\tilde{\mathbf t},\mathbf q^*}$. Moreover, $T=T^*(X)$ with probability one
under $\pi^*$, so it is a Monge coupling \citep{Galichon2016}. Since segment
sizes are chosen freely, $\mathbf q^*$ can be chosen.

For any segment sizes $\mathbf q$ and coupling $\pi\in\mathcal
M(P,Q^{\tilde{\mathbf t},\mathbf q})$, each assigned treatment yields regret no
smaller than the customer's lowest available regret. Because every admissible
coupling has customer marginal $P$, it follows that 
\[
E_\pi[C(X,T)]
\geq
E_P\!\left[
\min_{\ell\in\{1,\ldots,L\}}
C(X,\tilde{\boldsymbol t}_\ell)
\right]
=
E_{\pi^*}[C(X,T)].
\]
Thus, $\pi^*$ attains a lower bound on the expected regret of every admissible
coupling and is optimal.
\end{proof}

\subsection{Proofs for Section \ref{subsubsec:Lloyd-Theoretical-Properties}
\label{sec:Algo-Proofs}}

We restate and prove Propositions \ref{prop:Lloyd-Convergence} and
\ref{prop:Greedy-Initialization}. The first establishes profit improvement and
finite termination under exact updates. The second bounds the profit loss under
greedy initialization relative to the globally optimal segmentation plan.

\begin{prop*}[Profit improvement and convergence]
Under fixed $L$ and Assumptions \ref{assu:TE-Concavity} and
\ref{assu:Cost-Convexity}, exact updates in Algorithm
\ref{alg:Adapted-Lloyd's-Algorithm} satisfy
\[
\mathcal R(\tilde{\mathbf t}^{\,k+1})
\geq \mathcal R(\tilde{\mathbf t}^{\,k}),
\]
and expected profits converge to a value no greater than $\sum_{i=1}^{N}\bar
R_i$. The algorithm reaches a fixed point after finitely many
iterations.
\end{prop*}

\begin{proof}[Proof of Proposition \ref{prop:Lloyd-Convergence}]
Let $\boldsymbol z=(z_1,\ldots,z_N)$ denote customer assignments
to the $L$ indexed treatments, and define expected profit holding
these assignments fixed as
\[
J(\tilde{\mathbf t},\boldsymbol z)
\triangleq
\sum_{i=1}^{N}
\mathcal R_i(\boldsymbol x_i,\tilde{\boldsymbol t}_{z_i}).
\]
Because the assignment step selects the most profitable treatment
for each customer,
\[
J(\tilde{\mathbf t}^{\,k},\boldsymbol z^{\,k})
=
\mathcal R(\tilde{\mathbf t}^{\,k}).
\]
The treatment update maximizes each nonempty segment's profit
across feasible treatment dimensions and levels. Empty segments
contribute zero and retain their treatments. Reassignment then
maximizes profits under the updated treatments. Consequently,
\[
\mathcal R(\tilde{\mathbf t}^{\,k})
\leq J(\tilde{\mathbf t}^{\,k+1},\boldsymbol z^{\,k})
\leq \mathcal R(\tilde{\mathbf t}^{\,k+1}).
\]
Each customer's expected profit is bounded above by their
full-personalization profit $\bar R_i$. Thus, the sequence of total
expected profits is weakly increasing and bounded above by
$\sum_i\bar R_i$, and therefore converges.

To establish finite termination, define the highest profit
attainable for a fixed assignment:
\[
V(\boldsymbol z)
=
\sum_{\ell:S_\ell(\boldsymbol z)\neq\varnothing}
\max_{\substack{d\in\{1,\ldots,D\}\\t\in[0,\bar t_d]}}
\sum_{i\in S_\ell(\boldsymbol z)}
\left\{
E[Y_i\mid\boldsymbol x_i,\boldsymbol0]
+\tau_d(\boldsymbol x_i,t)-c_d(t)
\right\},
\]
where $S_\ell(\boldsymbol z)=\{i:z_i=\ell\}$. The exact treatment
update and subsequent reassignment imply
\[
V(\boldsymbol z^{\,k+1})
\geq J(\tilde{\mathbf t}^{\,k+1},\boldsymbol z^{\,k+1})
\geq J(\tilde{\mathbf t}^{\,k+1},\boldsymbol z^{\,k})
=V(\boldsymbol z^{\,k}).
\]

If $V(\boldsymbol z^{\,k+1})=V(\boldsymbol z^{\,k})$, every
inequality is an equality. Each customer's reassignment gain is
nonnegative, so every customer's gain must be zero. Their previous
treatment therefore remains a most profitable option. Because
assignment ties are resolved by the smallest treatment index,
\[
z_i^{\,k+1}\leq z_i^{\,k}
\quad\text{for every }i.
\]
If any assignment changes across equally profitable options, $\sum_i z_i$
strictly decreases. Thus, each assignment change either strictly increases
$V(\boldsymbol z)$, or leaves it unchanged and strictly decreases $\sum_i z_i$.
Since there are finitely many assignments, assignments can change only finitely
many times.

Once consecutive assignments agree, the treatment update repeats
the same treatments: Assumptions \ref{assu:TE-Concavity} and
\ref{assu:Cost-Convexity} give a unique optimal level within each
dimension, dimension ties follow a fixed rule, and empty segments
retain their treatments. The next reassignment is therefore
unchanged as well. Thus the algorithm reaches a fixed point
after finitely many iterations.
\end{proof}

\begin{prop*}[Greedy initialization]
Let $\mathcal R_m^*$ denote the highest expected profit attainable
with at most $m$ feasible treatments from the continuous treatment
domain. The greedy procedure first selects the most profitable
blanket treatment and then adds the feasible treatment generating
the largest increase in total expected profit until $L$ treatments
have been selected. For $L\geq2$, the resulting treatment set
$\tilde{\mathbf t}^{\,0}$ satisfies
\[
\mathcal R_L^*-\mathcal R(\tilde{\mathbf t}^{\,0})
\leq
\left(1-\frac1L\right)^{L-1}
\left(\mathcal R_L^*-\mathcal R_1^*\right).
\]
For $L=1$, the procedure selects the optimal blanket treatment.
\end{prop*}

\begin{proof}[Proof of Proposition \ref{prop:Greedy-Initialization}]
We adapt the greedy approximation argument of \citet[Proposition
4.1]{Nemhauser1978} to our profit objective, starting from the optimal blanket
treatment and allowing continuous treatment levels.

Let $\mathcal T$ denote the continuous domain of feasible treatments.
It is the union of the $D$ closed, bounded treatment intervals,
each embedded in its corresponding coordinate axis. Under
Assumptions \ref{assu:TE-Concavity} and \ref{assu:Cost-Convexity},
each customer's expected profit is continuous in the treatment
level within each dimension. For any finite, nonempty treatment
set $A\subseteq\mathcal T$, we define
\[
r_i(A)
\triangleq
\max_{\boldsymbol t\in A}
\mathcal R_i(\boldsymbol x_i,\boldsymbol t).
\]
The increase in total expected profit from adding treatment
$\boldsymbol t$ to the set $A$ is
\[
\Delta(\boldsymbol t\mid A)
\triangleq
\mathcal R(A\cup\{\boldsymbol t\})-\mathcal R(A)\\
=
\sum_i
\max\left\{
\mathcal R_i(\boldsymbol x_i,\boldsymbol t)-r_i(A),\,0
\right\}.
\]
If $A\subseteq B\subseteq\mathcal T$ are finite and nonempty,
then $r_i(A)\leq r_i(B)$. Therefore,
\[
\Delta(\boldsymbol t\mid A)
\geq
\Delta(\boldsymbol t\mid B).
\]
This result is a diminishing returns property where adding a given treatment
produces a weakly smaller gain when more treatments are already
available.

The blanket profit objective and $\Delta(\boldsymbol t\mid A)$
are continuous in the candidate treatment level within each
dimension. Since there are finitely many compact treatment
intervals, both attain global maxima. An optimal treatment set
with at most $L$ treatments also exists. 
We represent a treatment set with at most $L$ treatments by $L$
treatment vectors and allow repetitions. For each choice of
dimensions, total expected profit is a continuous function on
a compact product of intervals, and there are finitely many
dimension choices. Thus, the continuous greedy selections and
the optimal comparison plan are well defined.

Let $A_j$ be the greedy treatment set after $j$ selections, and
let $O\subseteq\mathcal T$ be an optimal treatment set containing at most $L$
treatments. Adding all treatments in $O$ to $A_j$ yields at least
the profit of $O$. Adding those treatments one at a time and
applying the diminishing-returns property gives
\[
\mathcal R^*_L-\mathcal R(A_j)
\leq
\mathcal R(A_j\cup O)-\mathcal R(A_j)\\
\leq
\sum_{\boldsymbol t\in O}\Delta(\boldsymbol t\mid A_j).
\]
The greedy rule selects a treatment with the globally largest
gain over $\mathcal T$, comparing all dimensions and levels.
Each treatment in $O$ therefore yields a marginal gain no greater
than the gain from the greedy addition,
$\mathcal R(A_{j+1})-\mathcal R(A_j)$.
Since $O$ contains at most $L$ treatments, we obtain a bound
on the profit gain from each greedy addition, for $j=1,\ldots,L-1$,
\[
\mathcal R^*_L-\mathcal R(A_j)
\leq
L\left\{\mathcal R(A_{j+1})-\mathcal R(A_j)\right\},
\]
which we rearrange to get the lower bound,
\[
\mathcal R(A_{j+1})-\mathcal R(A_j)
\geq
\frac{1}{L}\bigl[\mathcal R_L^*-\mathcal R(A_j)\bigr].
\]

This bound shows that the greedy addition increases profits
by at least $1/L$ of the current profit gap. We now write the new gap
as the current gap minus this profit gain,
\[
\mathcal R_L^*-\mathcal R(A_{j+1})
=
\mathcal R_L^*-\mathcal R(A_j)
-\bigl[\mathcal R(A_{j+1})-\mathcal R(A_j)\bigr].
\]
Substituting this lower bound on the profit gain then gives
\[
\mathcal R_L^*-\mathcal R(A_{j+1})
\leq
\left(1-\frac{1}{L}\right)
\bigl[\mathcal R_L^*-\mathcal R(A_j)\bigr].
\]
The first selected treatment is the optimal blanket treatment,
so $\mathcal R(A_1)=\mathcal R^*_1$. Applying this
inequality over the remaining $L-1$ selections, with
$A_L=\tilde{\mathbf t}^{\,0}$, yields
\[
\mathcal R^*_L
-\mathcal R(\tilde{\mathbf t}^{\,0})
\leq
\left(1-\frac1L\right)^{L-1}
\left(
\mathcal R^*_L-\mathcal R^*_1
\right).
\]
If the largest marginal gain is zero before $L$ selections,
the preceding inequalities imply that the current treatment
set already attains $\mathcal R_L^*$. Any further additions
leave its profit unchanged, so the bound still holds. For $L=1$,
selecting the optimal blanket treatment gives the result directly.
\end{proof}

The connection between the propositions follows from the identity
\[
\mathcal R^*_L
-\mathcal R(\tilde{\mathbf t}^{\,k})
=
\mathcal R^*_L
-\mathcal R(\tilde{\mathbf t}^{\,0})\\
\quad-
\left\{
\mathcal R(\tilde{\mathbf t}^{\,k})
-\mathcal R(\tilde{\mathbf t}^{\,0})
\right\}.
\]
Proposition \ref{prop:Lloyd-Convergence} makes the last bracket
nonnegative. Thus, every profit increase reduces the bound on
the remaining profit gap by exactly that amount.

\section{First-order condition for coarse personalization
\label{sec:First-order-condition-discussion}}

At an interior solution to the coarse personalization problem, the average marginal effect of
treatment equals its average marginal cost within each segment. We derive this
condition from the firm's problem in Equation \ref{eq:CP-Problem}.

We fix a nonempty segment $S_\ell$ and its treatment dimension $d_\ell$. The firm
chooses the segment's treatment level to solve
\begin{equation}
\tilde t_{\ell,d_\ell}^{*}
\in \arg\max_{t\in[0,\bar t_{d_\ell}]}
\sum_{i\in S_\ell}
\left\{
\tau_{d_\ell}(\boldsymbol x_i,t)-c_{d_\ell}(t)
\right\}.
\end{equation}
Since baseline outcomes do not depend on the treatment level and implementation cost
$\psi(L)$ is constant for a fixed number of treatments, neither will enter this
optimization problem.

We suppose the treatment effect and treatment cost functions are differentiable
and the optimal level is in the interior. Differentiating the objective and
dividing by the number of customers in the segment gives
\begin{equation}
\underbrace{
\frac{1}{|S_\ell|}
\sum_{i\in S_\ell}
\left.
\frac{\partial\tau_{d_\ell}(\boldsymbol x_i,t)}{\partial t}
\right|_{t=\tilde t_{\ell,d_\ell}^{*}}
}_{\text{Average marginal effect}}
=
\underbrace{
c_{d_\ell}'(\tilde t_{\ell,d_\ell}^{*}).
}_{\text{Average marginal cost}}
\end{equation}
Under Assumptions \ref{assu:TE-Concavity} and \ref{assu:Cost-Convexity}, the
segment's objective is strictly concave, so an interior level satisfying this
condition is its unique maximizer. If the optimal level lies on a boundary, the
equality may not hold.

The first-order condition implies that the firm balances the average marginal
effect of treatment against its average marginal cost across the customers
assigned to a segment. When the outcome is revenue, this equates average
marginal revenue and marginal cost. Under full personalization, the firm
balances these quantities separately for each customer. Thus, coarse
personalization extends the same profit-maximizing principle from the individual
to the segment level. This condition determines the treatment level within each
dimension, and the firm then compares expected profits across dimensions to choose
the segment's treatment.

\section{Bootstrap Bias Correction\label{sec:Bootstrap-Bias-Correction-Details}}

This appendix explains our bootstrap bias correction and its application to the
population performance of the different personalization methods. Our comparison
focuses on personalization methods that share the same estimated customer
continuous CATE functions but differ in how they form segments and choose
treatments. We assess each method's population performance by the profit it can
attain under the underlying customer continuous CATE functions, subject to its
restrictions on promotions and assignments. The bootstrap reestimates these
continuous CATE functions and repeats each method to account for how first-stage
estimation noise affects estimated profits. We apply the bootstrap bias
correction to coarse personalization and all benchmark methods to mitigate the
optimizer's curse from using the same data to learn policies and evaluate their
profits.\footnote{Even a blanket treatment can exhibit this bias because its
treatment dimension and level are chosen using estimated customer responses.}
Methods with greater flexibility to optimize treatments and assignments have
more opportunities to select policies that benefit from estimation noise. Full
personalization may therefore exhibit greater upward bias than segmentation
based on customer characteristics. Consistent with this intuition, our main
results show larger bootstrap corrections for full personalization than for
segmentation based on customer characteristics.

We compare the population performance of each personalization method instead of
selecting a single method with the highest estimated profit. However, these
methods use the data to choose treatments and assign customers to segments.
These choices can favor policies whose estimated profits benefit from estimation
noise. \citet{Xu2025} address this problem by correcting the estimated value of
the particular policy selected from the data. Our target is instead the
population profit attainable under each method's segmentation and treatment
selection procedure using the underlying customer CATE functions.

To illustrate the distinction, suppose we run a segmentation method from two
random initializations, producing two candidate segmentation plans, A and B.
Each plan specifies customer assignments and segment treatments. Using the
estimated customer CATE functions, Plan A yields an estimated profit of \$110
and Plan B yields \$105, so we select Plan A and report \$110. Holding these
plans fixed, their profits under the true customer CATE functions are \$100 and
\$102, respectively.

\citet{Xu2025} seek to correct the estimated value of the plan selected using
the observed data. Here, that is Plan A, whose true population profit is \$100,
so the reported profit overstates its value by \$10. Our question instead
concerns the population performance of the segmentation method. In this
simplified example, choosing between the two plans using the true customer CATE
functions would select Plan B and yield \$102. Relative to this target, the
reported profit is overstated by \$8. The remaining \$2 reflects the loss from
selecting the inferior plan because of estimation error.

We construct the correction using the bootstrap estimate of bias. For a given
method and number of promotions $L$, let $\hat V_n$ denote estimated profit per
customer in the original sample and $\hat V_m^{*(b)}$ the corresponding estimate
in bootstrap replication $b$. In each replication, we reestimate customer
responses and repeat the method's segmentation and treatment selection
procedures. We calculate the unscaled correction as
\begin{equation}
\hat b_m
=
\frac{1}{B}\sum_{b=1}^{B}
\left(\hat V_m^{*(b)}-\hat V_n\right)
\end{equation}
and use $B=500$ replications. This correction measures how much the method's
estimated profit changes on average when we repeat estimation and optimization
in the bootstrap samples. The reference value is the method's original-sample
profit estimate \citep{Efron1994}. In contrast, \citet{Xu2025} evaluate each
bootstrap-selected policy using the original estimates and use that
policy-specific value as the reference. We then adjust $\hat b_m$ for the
difference in sample sizes and subtract it from $\hat V_n$, as described below.

Following \citet{Xu2025}, we draw bootstrap samples of size $m=\lfloor
n^{0.95}\rfloor$ with replacement. The correction calculated from these smaller
samples reflects estimation noise at sample size $m$, whereas our reported
profits are estimated using $n$ observations. Motivated by the subsampling
bias-reduction construction of \citet[Section 3.7 and Remark
3.7.1]{Politis1999}, we adjust the correction using the ratio of rates at the
two sample sizes. We adopt a square-root rate specification, yielding the factor
$\sqrt{m/n}$, and examine sensitivity to alternative scaling
exponents.\footnote{\citet{Xu2026} use the same $\sqrt{m/n}$ factor to convert
their bootstrap estimate of bias to the full-sample scale in their
setting.}The corrected profit estimate is
\begin{equation}
\hat V_n^{\mathrm{BC}}
=
\hat V_n-\sqrt{m/n}\,\hat b_m.
\end{equation}
Intuitively, the correction is estimated using smaller samples, so we adjust it
to reflect the amount of bias expected in the full sample. 

To illustrate how the correction works, we consider two segmentation plans that
generate the same expected profit. In this setting, any estimated gain from
choosing one plan over the other arises from sampling variation. The example
shows how repeating that choice in the bootstrap and scaling the resulting
correction by $\sqrt{m/n}$ reduces the upward bias in estimated profit.

\begin{example}[Optimizer's curse with two segmentation plans]
Suppose a firm chooses between two segmentation plans. Plan A generates profit
$V$ for each customer, and Plan B generates profit $V+\varepsilon_i$ for
customer $i$, where the profit shocks $\varepsilon_i$ are independent
$N(0,\sigma^2)$ draws with $\sigma>0$. Both plans therefore generate the same
expected profit per customer. Let $\bar{\varepsilon}_n$ denote the average shock
in a sample of $n$ customers. The firm selects Plan B if $\bar{\varepsilon}_n>0$
and Plan A otherwise. Its reported profit per customer is $\hat
V_n=V+\max\{0,\bar{\varepsilon}_n\}$. Because the firm selects Plan B only when
its estimated profit exceeds $V$, the reported profit exceeds $V$ on average by
$c/\sqrt n$, where $c=\sigma/\sqrt{2\pi}$. \footnote{To see this, we have
$\bar{\varepsilon}_n\sim N(0,\sigma^2/n)$ so the expected positive part of this
normal random variable gives
$
E[\hat V_n]-V
=E[\max\{0,\bar{\varepsilon}_n\}]
=\frac{\sigma}{\sqrt{2\pi n}}
=\frac{c}{\sqrt n}
$ so $c=\frac{\sigma}{\sqrt{2\pi}}$.
}

To apply our correction, we draw $m$ customers with replacement and again select
the plan with the higher estimated profit. Let $W_i$ denote customer $i$'s share
of the resample. Conditional on these weights, the estimated profit difference
between Plan B and Plan A is $\sum_i W_i\varepsilon_i$, which is normal with
mean zero and standard deviation $\sigma\|W\|_2$. The expected bootstrap
correction before scaling is therefore\footnote{The expectations are taken over
both the original sample and the bootstrap resampling.}
\[
  E[\hat b_m]
  = c\left(E\|W\|_2-\frac{1}{\sqrt n}\right).
\]
Since $\sum_i W_i=1$, we have $\|W\|_2\ge 1/\sqrt n$, so the expected unscaled
correction is nonnegative. For $n\ge2$ and $1\le m\le n$, subtracting
$\sqrt{m/n}\,\hat b_m$ strictly reduces the upward bias. To see how the
adjustment accounts for the resample size, we have that
\[
  1 \le \sqrt m\,E\|W\|_2 \le \sqrt{1+(m-1)/n}.
\]
The lower bound follows because the resample contains at most $m$
distinct customers. The upper bound follows from Jensen's inequality
and $E\|W\|_2^2=[1+(m-1)/n]/m$. Thus, as $m/n\to0$, the expected
unscaled correction is asymptotic to $c/\sqrt m$. Multiplying by
$\sqrt{m/n}$ recovers the leading bias $c/\sqrt n$ in the original
profit estimate.
\end{example}

The correction is intended to mitigate the optimizer's curse and may not
eliminate all bias. As \citet{Xu2025} note, bootstrap corrections can leave
residual bias in finite samples. In our application, the correction depends on
its scaling exponent and resample size. We therefore examine alternative values
for both.

First, replacing the exponent $0.5$ in $(m/n)^{0.5}$ with approximately $0.45$,
$0.55$, or $0.6$ preserves the ranking of the methods by estimated profit and
the qualitative results remain the same. As the exponent increases in our
setting, the scaling factor decreases, so we subtract a smaller bias correction.
Second, across alternative resample sizes, including $m=\lfloor n^{0.9}\rfloor$
and $m=\lfloor n^{0.85}\rfloor$, coarse personalization continues to generate
profits close to full personalization. The number of promotions that maximizes
estimated profits net of implementation costs is slightly sensitive to these
choices. For example, for $m=\lfloor n^{0.85}\rfloor$, it increases from $L=3$
to $L=5$. Thus, coarse personalization retains much of the profit from full
personalization, although the preferred number of promotions depends on the
resample size and scaling adjustment.

\section{Randomization checks for the experiment \label{sec:Randomization-check}}

To examine whether randomization was correctly implemented by the firm, we
compare the distributions of estimated propensity scores between promotion
recipients and the holdout group. The propensity score comparison checks for
overlap and covariate balance which should be satisfied by the randomization in
the field experiment \citep{Imbens2015}. We assume SUTVA holds because the
coupons are linked to each customer's account on the platform and we assume that
sharing of coupons among customers is unlikely. 

\begin{figure}[H]
\begin{centering}
\includegraphics[width=0.75\textwidth]{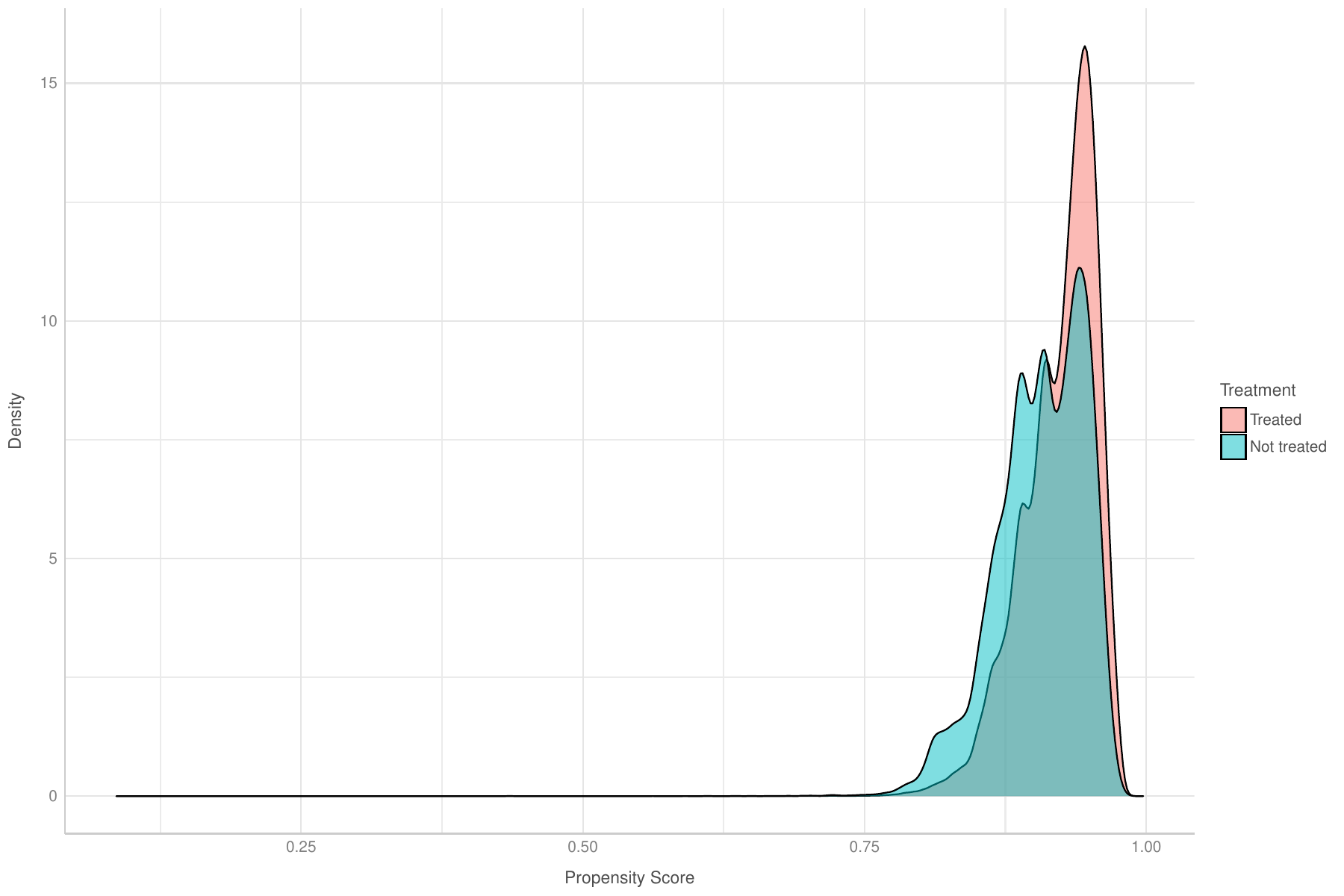}
\par\end{centering}
{\footnotesize Note: Propensity scores are obtained using the Lasso
with ten-fold cross-validation.}\caption{Estimated propensity score (any treated vs. not treated) \label{fig:Propensity-Score-Binary}}
\end{figure}

\begin{figure}[H]
\begin{centering}
\includegraphics[width=0.75\textwidth]{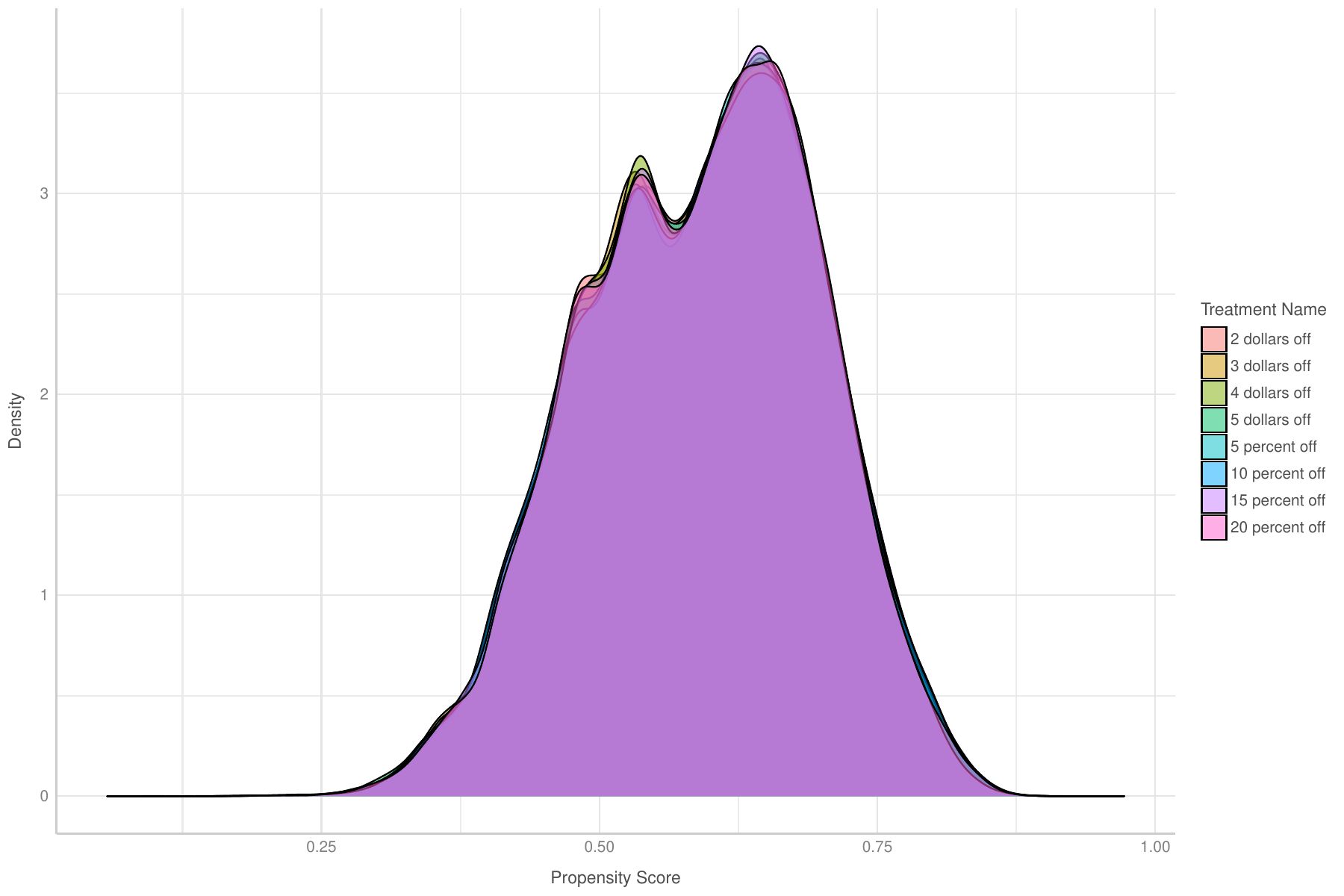}
\par\end{centering}
{\footnotesize Note: Propensity scores are obtained using the Lasso
with ten-fold cross-validation.}\caption{Estimated propensity score across treated arms \label{fig:Propensity-Score-Treatments}}
\end{figure}

We estimate a logistic regression for the probability of receiving any promotion
using our covariates and the Lasso. We then plot the distribution of the
predicted propensity score in Figure \ref{fig:Propensity-Score-Binary}
separately for treated and not treated individuals. We find that there is a
sizable difference in the propensity scores across the two types of individuals
which suggests the data might not be balanced. We then examine each treatment
arm separately and estimate a logistic regression for the propensity score of
each treatment arm to the no treatment arm using the Lasso and plot the
predicted propensity score for only the treated individuals for each treatment
arm in Figure \ref{fig:Propensity-Score-Treatments}. We find that there is
virtually no difference in predicted propensity scores for the treated
individuals in each treatment arm. These results suggest that randomization was
correctly implemented across treatment arms, but may be incorrectly performed
for the control group or no treatment sample. The company has not indicated that
the randomization was incorrectly implemented, but we adjust our CATE
estimations with the propensity score to account for possible mistakes or
stratification procedures during the firm's randomization procedure.

\section{Honest validation for CATE estimation\label{subsec:Honest-validation}}

We use the honest validation procedure introduced in \citet{Misra2021} to
evaluate the parametric form for the continuous CATEs (Equation
\ref{eq:CATE-Log-Form}) at held-out treatment levels against the non-parametric
CATE estimates for each treatment arm. The honest validation procedure uses the
fact that the treatment arms are discrete while the treatment variables are
themselves continuous to provide a comparison of the parameterized estimate to
the nonparametric estimates at the held-out treatment levels. We detail the
procedure and describe the evaluation in the remainder of this section. 

The honest validation procedure is implemented in four steps. First, we use the
Causal Forest to nonparametrically estimate the CATEs for each of the eight
treatment arms ($2,3,4,5$ dollars off and $5,10,15,20$ percent off).

Second, for each individual, we hold out the individual's actual treatment
assignment and use the predicted CATEs to estimate the individual's
parameterized CATE function (Equation \ref{eq:CATE-Log-Form}).\footnote{The
average $R^{2}$ of the restricted parameterized CATEs is $0.950$ for dollar-off
and $0.852$ for percentage-off promotions, which suggests the parameterization
fits the restricted set of data well.} We estimate the function only for the
type of promotion the individual received. For example, if the individual was
assigned to a two dollar off promotion in the field experiment, we would use the
customer's predicted CATEs for the other dollar off treatment arms ($0,3,4,5$
dollars off) to estimate the individual's parameterized CATE function. The two
dollar off promotion is held out in this step and forms the holdout set for the
parameterized estimate. 

Third, we use the parameterized CATE function to predict the CATE estimate for
the held-out treatment. In our running example, we would predict the
individual's CATE using the parameterized CATE function for two dollars off. The
two dollar off promotion was the individual's assigned treatment from the field
experiment so we have the nonparametric Causal Forest estimate for the
individual at that treatment. 

Lastly, we compare the predicted CATE in the held-out treatment to the
nonparametric Causal Forest estimate for the individual. In our running example,
we would compare the non-parametric estimate of the individual's CATE at two
dollars off to her predicted CATE estimate at two dollars off from the
parameterized CATE function. 

Using the nonparametric estimates as the "ground truth" benchmark, this
procedure evaluates how closely the chosen parameterization approximates the
estimated responses at held-out treatment levels. We perform this procedure
separately for dollar-off and percentage-off promotions.

\begin{figure}[H]
\begin{centering}
\caption{Honest validation of the parametric CATE estimates on
sales\label{fig:Honest-validation}} \subfloat[Dollar off
promotion]{\includegraphics[width=0.72\textwidth]{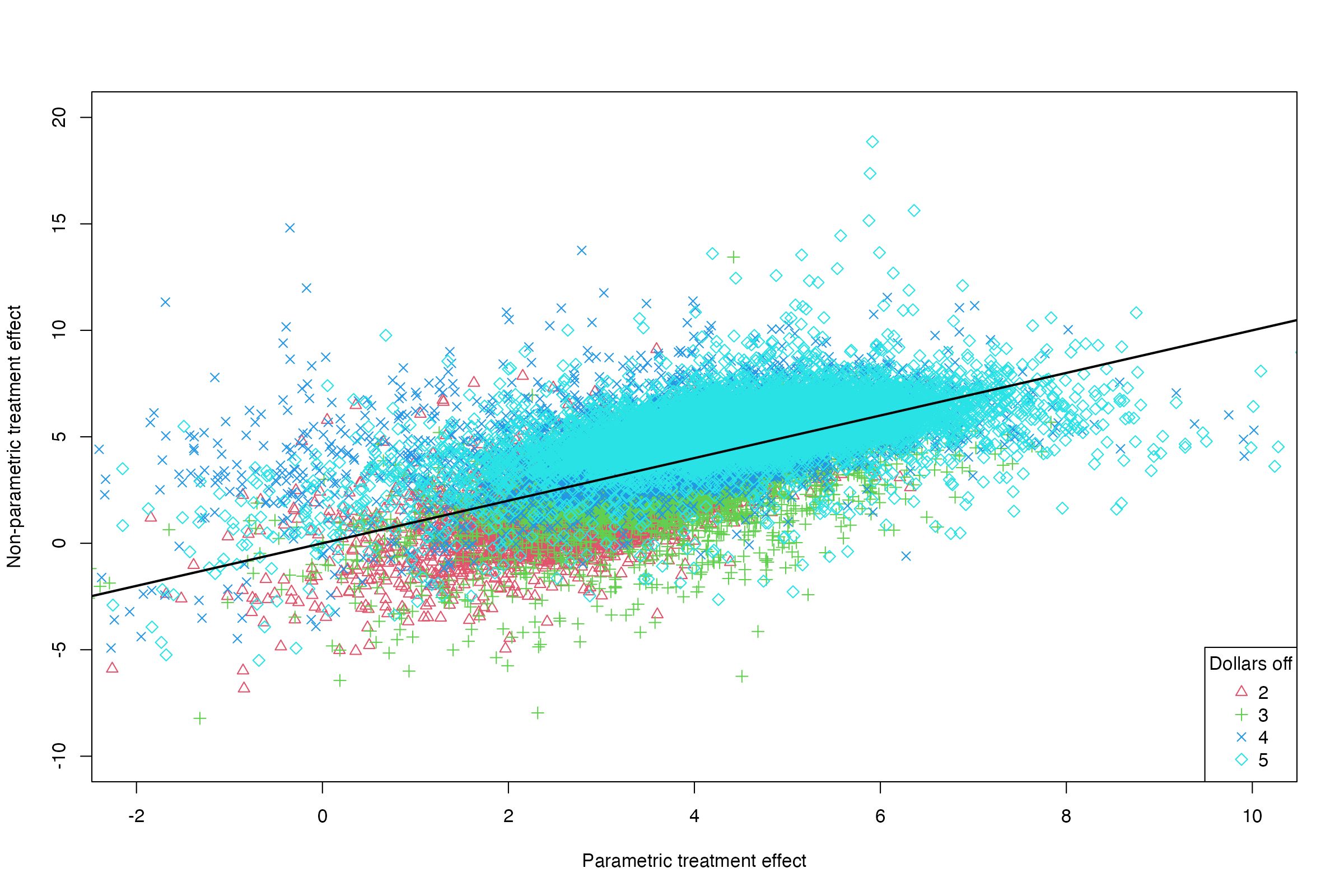}}
\par\end{centering}
\begin{centering}
\subfloat[Percentage off
promotion]{\includegraphics[width=0.72\textwidth]{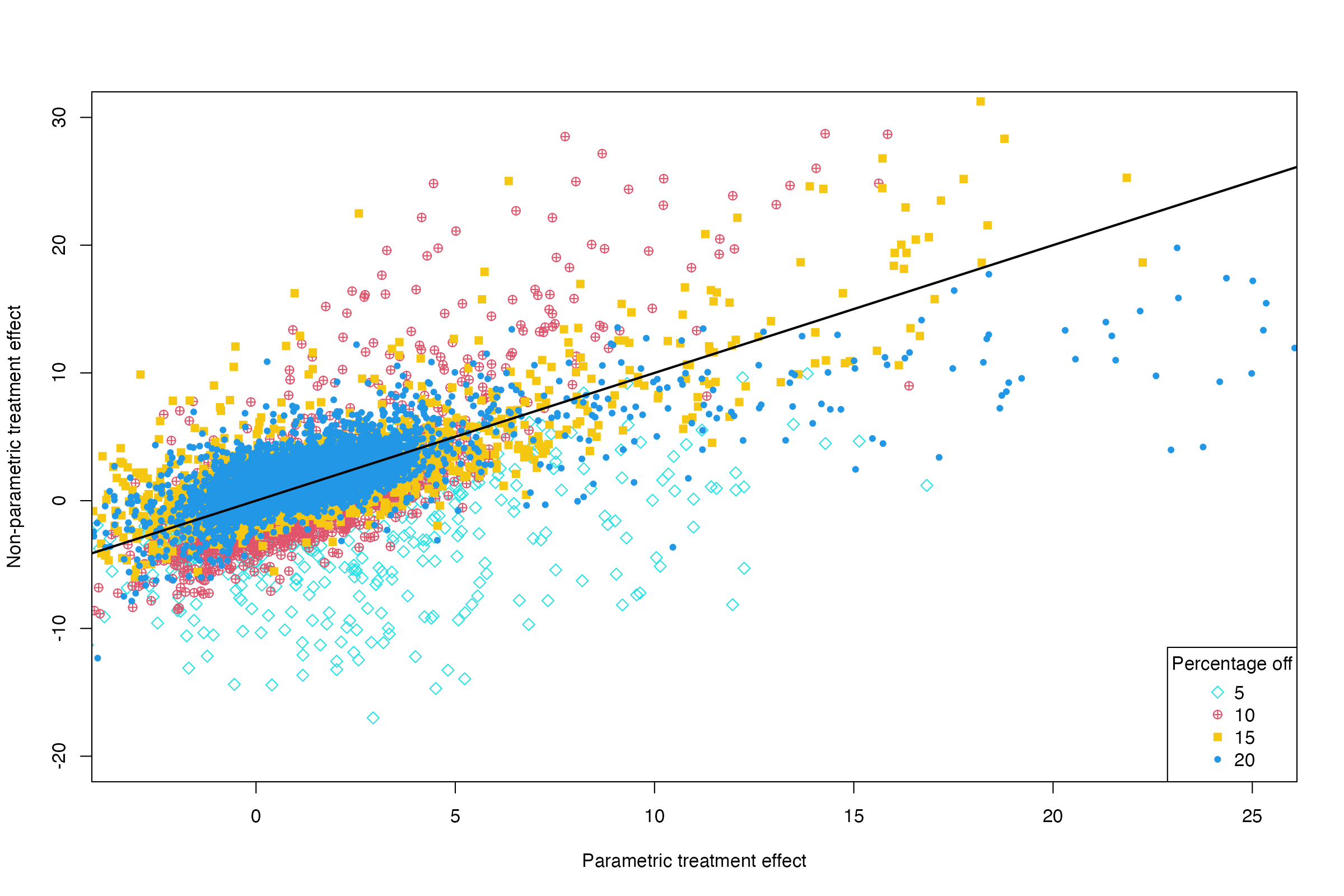}}\medskip{}
\par\end{centering} {\footnotesize Note: These figures provide comparisons of
the predicted treatment effect from the CATE parameterization (Equation
\ref{eq:CATE-Log-Form}) in the holdout treatment to its non-parametric estimate.
The dollar off and percentage off promotions are plotted separately in the two
panels. Each point represents an individual assigned a dollar off or percentage
off promotion in the field experiment. The shape and color of the points
represent which treatment arm was assigned to the individual. A $45$-degree
line is provided as a visual reference.}{\footnotesize\par}
\end{figure}

Figure \ref{fig:Honest-validation} plots the parametric estimates against the
non-parametric estimate for each individual on the sales scale. The horizontal
axis represents the parametric estimates in the holdout set and the vertical
axis represents the non-parametric estimates. The shapes and colors of the
points represent the assigned treatment arm in the field experiment. The
$45$-degree line indicates equality between the two estimates. We see that most
of the points lie scattered around the $45$-degree line which suggests the two
types of estimates overlap. Further, the correlation between the parametric and
nonparametric estimates is $0.850$ for dollar-off and $0.665$ for
percentage-off promotions. These results suggest that the parameterized CATEs
perform well in the holdout set.

Within treatment arms, the non-parametric estimates generally display more
dispersion than the parametric estimates in Figure \ref{fig:Honest-validation}.
This pattern is consistent with the smoothing imposed by the parametric form.
For both sets of estimates, dispersion is greater on average across treatment
arms for percentage-off promotions than for dollar-off promotions.

\section{Surplus calculation details\label{sec:Surplus-Appendix}}

With our coarse personalization framework, we can empirically examine how
surplus changes with the degree of personalization. We compare each coarse
personalization solution with the optimized blanket promotion, which assigns the
same treatment to all customers. As the firm expands the treatment set,
customers may receive promotions with higher or lower value than the blanket
offer. Although greater flexibility allows the firm to choose more profitable
promotions, deploying additional treatments also incurs implementation costs. We
therefore decompose changes in consumer, producer, and total surplus across
levels of personalization to assess whether the firm's net profit gains offset
any losses in consumer surplus.

Intuitively, the firm forms segments and chooses promotions to maximize profits,
rather than consumers' valuations. A customer who receives a smaller discount
than the blanket promotion may therefore generate more profit for the firm while
receiving less consumer surplus. Other customers may receive more generous
promotions when their responses make these offers profitable. Consequently,
consumer surplus may not move in the same direction as producer surplus as the
treatment set expands. The effect on total surplus depends on the balance
between these changes and the additional implementation costs. 

We formulate a simple surplus analysis using the estimated treatment responses
and assumptions on consumers' valuations of the promotions. Let
$\boldsymbol{t}^{\mathrm{bl}}$ denote the optimized blanket promotion, common to all
customers. Under coarse personalization with $L$ treatments, customer $i$
receives $\boldsymbol{t}_i^{(L)}$, which may differ from the blanket promotion
in both treatment type and level. Let $\Pi_i(\boldsymbol{t})$ denote the
expected incremental profit from assigning treatment $\boldsymbol{t}$ to
customer $i$, net of the promotion's monetary cost but before implementation
costs, and let $v_i(\boldsymbol{t})$ denote the customer's monetary valuation of
that treatment.

To complete the surplus calculation, we assume that consumers value dollar-off
promotions at face value and percentage-off promotions at the discount rate
times their average past promotional order value. Let $\bar{s}_i$ denote this average
order value for customer $i$. Since each promotion is nonzero in only one
treatment dimension, the valuation functions are
\[
v_i((t_1,0))=t_1,
\qquad
v_i((0,t_2))=\frac{t_2}{100}\bar{s}_i,
\]
where $t_2$ is expressed in percentage points. These valuations equal the
monetary cost of the promotions to the firm, so the promotion itself is a
transfer from the firm to the consumer. Implementation costs are separate costs
borne by the firm and do not enter the consumer's valuation.

At the individual level, we compare the valuation and expected profit under
coarse personalization with those under the optimized blanket promotion. The
change in consumer surplus for customer $i$ is
\begin{equation}
\Delta CS_i(L)
=v_i(\boldsymbol{t}_i^{(L)})-v_i(\boldsymbol{t}^{\mathrm{bl}}),
\end{equation}
and the change in expected profit before implementation costs is
\begin{equation}
\Delta \Pi_i(L)
=\Pi_i(\boldsymbol{t}_i^{(L)})-\Pi_i(\boldsymbol{t}^{\mathrm{bl}}).
\end{equation}
Positive values indicate gains relative to the blanket promotion. We incorporate
implementation costs at the aggregate level because they are incurred for
deploying the treatment set rather than for serving an individual customer.

Aggregating across customers, the changes in consumer and producer surplus are
\begin{align}
\Delta CS(L) &= \sum_{i=1}^{N}\Delta CS_i(L),\\
\Delta PS(L) &= \sum_{i=1}^{N}\Delta\Pi_i(L)
-\bigl[\psi(L)-\psi(1)\bigr].
\end{align}
The implementation cost in the producer surplus accounts for the costs of both
the coarse personalization solution and the blanket promotion. It reduces
producer and total surplus but leaves consumer surplus unchanged. The change in
total surplus is therefore
\begin{equation}
\Delta TS(L)=\Delta CS(L)+\Delta PS(L).
\label{eq:Total-Surplus}
\end{equation}
In our application, $\psi(L)=\$1{,}000L$, so the additional implementation cost
relative to the blanket promotion is $\$1{,}000(L-1)$.

\begin{example}[Dollar-off promotions]

To build intuition for Equation \ref{eq:Total-Surplus}, consider a setting with
only dollar-off promotions. Let $t_1^{\mathrm{bl}}$ denote the optimized blanket
discount and $t_{i,1}^{(L)}$ the discount assigned to customer $i$ under coarse
personalization. The customer values each coupon at face value, which also
equals its monetary cost to the firm. Since expected incremental profit is
$\Pi_i((t_1,0))=\tau_1(\boldsymbol{x}_i,t_1)-t_1$, the individual surplus
changes satisfy
\begin{align}
\Delta CS_i(L)+\Delta\Pi_i(L)
&=t_{i,1}^{(L)}-t_1^{\mathrm{bl}}
+\bigl[\tau_1(\boldsymbol{x}_i,t_{i,1}^{(L)})
-t_{i,1}^{(L)}\bigr]
-\bigl[\tau_1(\boldsymbol{x}_i,t_1^{\mathrm{bl}})-t_1^{\mathrm{bl}}\bigr]
\nonumber\\
&=\tau_1(\boldsymbol{x}_i,t_{i,1}^{(L)})
-\tau_1(\boldsymbol{x}_i,t_1^{\mathrm{bl}}).
\end{align}
The coupon transfer cancels, leaving the change in the expected incremental
outcome. Aggregating across customers and accounting for implementation costs
gives
\begin{equation}
\Delta TS(L)
=\sum_{i=1}^{N}
\bigl[\tau_1(\boldsymbol{x}_i,t_{i,1}^{(L)})
-\tau_1(\boldsymbol{x}_i,t_1^{\mathrm{bl}})\bigr]
-\bigl[\psi(L)-\psi(1)\bigr].
\end{equation}
Thus, under these valuation assumptions, expanding the treatment set increases
total surplus when the aggregate gain in expected outcomes exceeds the
additional implementation costs.

In this example, a larger discount directly benefits the customer but costs the
firm the same amount, so this transfer alone does not change total surplus.
Total surplus rises only to the extent that the promotion generates a higher
expected outcome, after accounting for additional implementation costs. However,
the firm chooses discounts based on the expected outcome net of the discount
cost. Its preferred treatment levels therefore do not necessarily maximize total
surplus under these valuation assumptions.

\end{example}

In our application, the optimized blanket promotion offers approximately \$1.69
off to all 1.2 million customers. Under the valuation assumptions, this policy
generates consumer surplus of \$2,041,543. After deducting its \$1,000
implementation cost, producer surplus is \$808,619 and total surplus is
\$2,850,161. Figure \ref{fig:Welfare-decomposition-bootstrap} and Online
Appendix Table \ref{tab:Welfare-decomposition-treatments} report changes
relative to this baseline, deducting the additional implementation costs of
\$1,000($L-1$) from producer and total surplus. All calculations use the
unrounded treatment levels and their profit estimates.

\begin{figure}[H]
\centering
\includegraphics[width=\textwidth]{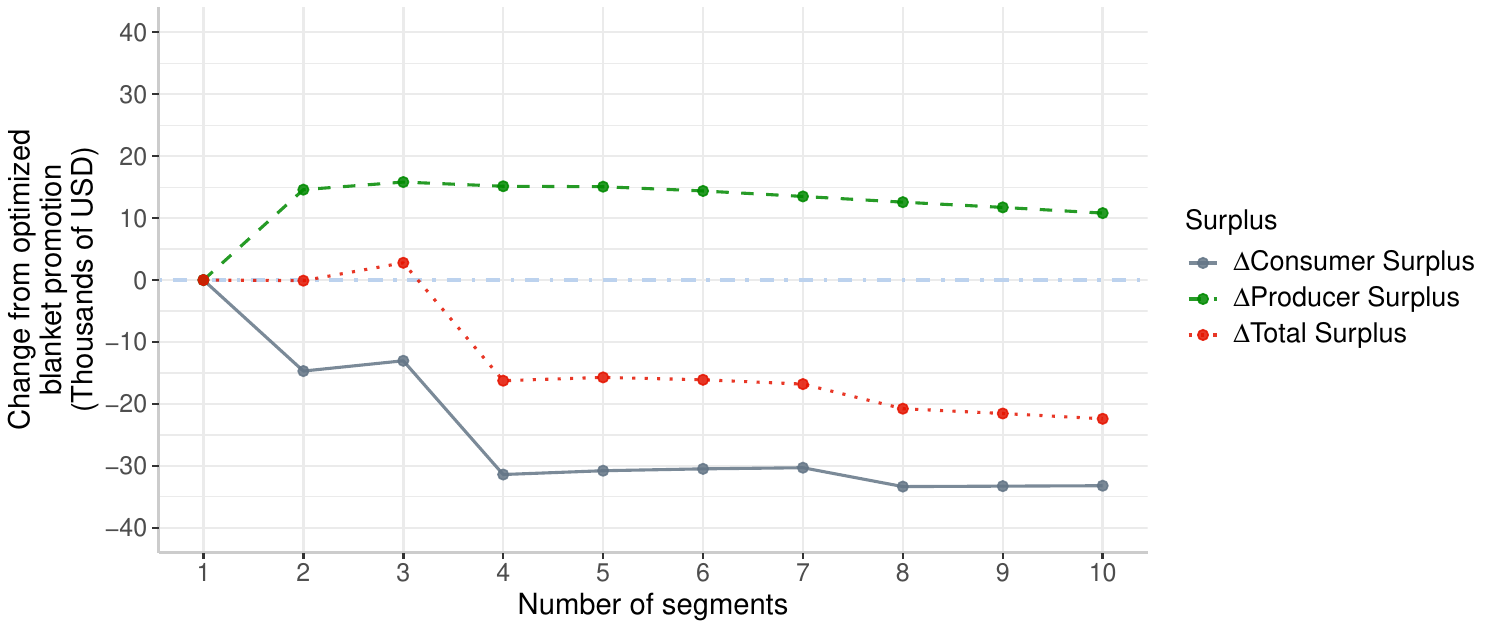}
\par
\raggedright{\footnotesize Note: Changes in surplus relative to the optimized
blanket promotion ($L=1$), incorporating implementation costs of \$1,000 per
treatment for both policies. Producer surplus uses bootstrap bias-corrected
profits. Total surplus adds consumer surplus and producer surplus.\par}
\caption{Surplus decomposition incorporating bias-corrected profits
\label{fig:Welfare-decomposition-corrected}}
\end{figure}

We also incorporate the bootstrap bias correction into producer profits for both
the coarse personalization solution and the blanket baseline. For the blanket
promotion, corrected total surplus is \$2,839,600. The corrected comparisons
appear in Table \ref{tab:Welfare-decomposition-treatments} and Figure
\ref{fig:Welfare-decomposition-corrected}. Consumer surplus remains uncorrected
and the reported total surplus combines consumer surplus with corrected producer
profits. 

\newpage{}

\section*{Algorithms}

\begin{algorithm}[!htbp]
\begin{spacing}{1.0}
\caption{Adapted Lloyd's Algorithm for the Empirical Application}
\label{alg:Empirical-Application}

\textbf{Notation:}
$L$ is the number of treatments. Dollar-off levels lie in $[0,5]$ and
percentage-off levels in $[0,20]$, measured in percentage points.
Customer $i$'s expected incremental profits are
\[
\pi_{i,1}(t)=\hat\alpha_{i,1}+\hat\beta_{i,1}\log(1+t)-t,
\qquad
\pi_{i,2}(t)=\hat\alpha_{i,2}+\hat\beta_{i,2}\log(1+t)-\frac{\bar{s}_i t}{100},
\]
where $\bar{s}_i>0$ is the customer's average past promotional order value.
Profit regret $C$ is defined in Equation \ref{eq:Assignment-Regret}.
Treatment $\ell$ at iteration $k$ has type $d_\ell^{\,k}\in\{1,2\}$
and level $v_\ell^{\,k}$, represented by
$\tilde{\boldsymbol t}_\ell^{\,k}=(v_\ell^{\,k},0)$ for dollars or
$(0,v_\ell^{\,k})$ for percentages. Its customer segment is $S_\ell^{\,k}$.
The tolerance is $\delta>0$.

\textbf{Step 0:}
Initialize $L$ feasible treatments.

\textbf{Step $\boldsymbol{k}$} ($k=0,1,\ldots$):
\begin{itemize}
\item (\emph{Form segments})
Assign each customer to the lowest-regret treatment, equivalently the
treatment yielding the highest expected incremental profit, and resolve
ties by the smallest treatment index:
\[
\ell_i^{\,k}\in
\arg\min_{j\in\{1,\ldots,L\}}
C(\boldsymbol x_i,\tilde{\boldsymbol t}_j^{\,k}),
\qquad
S_\ell^{\,k}=\{i:\ell_i^{\,k}=\ell\}.
\]
These segments are generalized Voronoi cells under profit regret.

\item (\emph{Optimize treatment levels})
For each nonempty segment, let $n_\ell=|S_\ell^{\,k}|$ and compute
the profit-maximizing level in each dimension:
\[
t'_{\ell,1}=\left[\frac{\sum_{i\in S_\ell^{\,k}}\hat\beta_{i,1}}
{n_\ell}-1\right]_{[0,5]},
\qquad
t'_{\ell,2}=\left[\frac{100\sum_{i\in S_\ell^{\,k}}\hat\beta_{i,2}}
{\sum_{i\in S_\ell^{\,k}}\bar{s}_i}-1\right]_{[0,20]},
\]
where $[z]_{[a,b]}=\min\{b,\max\{a,z\}\}$ clips a treatment level to its
domain.

\item (\emph{Choose treatment dimensions})
Compute $\Pi_{\ell,d}=\sum_{i\in S_\ell^{\,k}}
\pi_{i,d}(t'_{\ell,d})$ for $d=1,2$, and set
\[
\tilde{\boldsymbol t}_\ell^{\,k+1}=
\begin{cases}
(t'_{\ell,1},0), & \Pi_{\ell,1}\ge\Pi_{\ell,2},\\
(0,t'_{\ell,2}), & \text{otherwise}.
\end{cases}
\]
If $S_\ell^{\,k}$ is empty, retain its previous treatment.
\end{itemize}

\textbf{Terminate} when $\max_{\ell=1,\ldots,L}\|\tilde{\boldsymbol
t}_\ell^{\,k+1} -\tilde{\boldsymbol t}_\ell^{\,k}\|_\infty<\delta$ or the
iteration limit is reached.

\textbf{Output:}
Reassign customers to the final treatments and return the segmentation
plan.
\end{spacing}
\end{algorithm}

\begin{algorithm}[!htbp]
\begin{spacing}{1.0}
\caption{Adapted Lloyd's Algorithm with Discrete and Continuous Treatments}
\label{alg:Mixed-Treatments}

\textbf{Notation:}
Let $\mathcal T_d=\{\boldsymbol e_d(t):t\in[0,\bar t_d]\}$
be the feasible treatments in continuous dimension $d$, where
$\boldsymbol e_d(t)$ has level $t$ in dimension $d$ and zero elsewhere.
Let $\mathcal A$ be a finite set of discrete actions.
The feasible set is $\mathcal T=(\bigcup_{d=1}^{D}\mathcal T_d)\cup\mathcal A$.
Customer $i$'s expected incremental profit from treatment $a$ is $\pi_i(a)$,
net of its treatment cost. For a discrete action,
$\pi_i(a)=\hat\gamma_a(\boldsymbol x_i)-c_a(\boldsymbol x_i)$.
Profit regret is $C(\boldsymbol x_i,a) =\max_{b\in\mathcal T}\pi_i(b)-\pi_i(a),$
where the reference profit is maximized over both continuous treatments and
discrete actions.
$L$ is the number of treatments. The tolerance is $\delta>0$.

\textbf{Step 0:}
Initialize $L$ distinct feasible treatments
$\tilde{\mathbf t}^{\,0}=(\tilde t_1^{\,0},\ldots,\tilde t_L^{\,0})$.

\textbf{Step $\boldsymbol{k}$} ($k=0,1,\ldots$):
\begin{itemize}
\item (\emph{Form segments})
Assign each customer to the lowest-regret treatment, equivalently the
treatment yielding the highest expected incremental profit, and resolve
ties by the smallest treatment index:
\[
\ell_i^{\,k}\in
\arg\min_{j\in\{1,\ldots,L\}}
C(\boldsymbol x_i,\tilde t_j^{\,k}),
\qquad
S_\ell^{\,k}=\{i:\ell_i^{\,k}=\ell\}.
\]
These segments are generalized Voronoi cells under profit regret.

\item (\emph{Update segment treatments})
Set a working treatment set $\mathbf u=\tilde{\mathbf t}^{\,k}$.
Visit segments in order $\ell=1,\ldots,L$, holding their memberships fixed.
For each nonempty segment, find the best treatment in each continuous
dimension:
\[
t'_{\ell,d}\in\arg\max_{t\in[0,\bar t_d]}
\sum_{i\in S_\ell^{\,k}}\pi_i(\boldsymbol e_d(t)),
\qquad d=1,\ldots,D.
\]
Compare these $D$ candidates with every discrete action not already used
in another slot of the working set,
$\mathcal A_\ell=\mathcal A\setminus\{u_j:j\ne\ell\}$.
Replace $u_\ell$ with the candidate giving the highest aggregate profit
for $S_\ell^{\,k}$ and use a fixed candidate order to resolve ties.
Use this updated working set when visiting the next segment so that
each discrete action occupies at most one slot. Leave an empty segment's
treatment unchanged. After all updates, set
$\tilde{\mathbf t}^{\,k+1}=\mathbf u$.

\end{itemize}

\textbf{Terminate} when every slot retains its treatment type (continuous
dimension or discrete action) and every continuous level changes by at
most $\delta$ or the iteration limit is reached.

\textbf{Output:}
Reassign customers to their final treatments and return the customer
assignments.
\end{spacing}
\end{algorithm}

\begin{algorithm}[!htbp]
\begin{spacing}{1.0}
\caption{Adapted Lloyd's Algorithm with Rounded Treatment Levels}
\label{alg:Rounded-Treatments}

\textbf{Notation:}
$G_d\subseteq[0,\bar t_d]$ is the finite set of permitted treatment levels in
dimension $d$. In the application, increments of $h\in\{0.25,0.5,1\}$ give the
grids $G_1=\{0,h,\ldots,5\}$ dollars and $G_2=\{0,h,\ldots,20\}$ percentage
points. Let $\pi_{i,d}(t)$ be customer $i$'s expected incremental profit from
level $t$ in dimension $d$, net of treatment costs. Profit regret $C$ is defined
in Equation \ref{eq:Assignment-Regret}. Treatment $\ell$ at iteration $k$ is
$\tilde{\boldsymbol t}_\ell^{\,k}$, with dimension $d_\ell^{\,k}$, level
$v_\ell^{\,k}$, and segment $S_\ell^{\,k}$. The treatment budget is $L$ and the
tolerance is $\delta>0$.

\textbf{Step 0:}
Initialize $L$ feasible treatments with levels on their respective grids.

\textbf{Step $\boldsymbol{k}$} ($k=0,1,\ldots$):
\begin{itemize}
\item (\emph{Form segments})
Assign each customer to the lowest-regret treatment, equivalently the
treatment yielding the highest expected incremental profit, and resolve
ties by the smallest treatment index:
\[
\ell_i^{\,k}\in
\arg\min_{j\in\{1,\ldots,L\}}
C(\boldsymbol x_i,\tilde{\boldsymbol t}_j^{\,k}),
\qquad
S_\ell^{\,k}=\{i:\ell_i^{\,k}=\ell\}.
\]
These segments are generalized Voronoi cells under profit regret.

\item (\emph{Optimize treatment levels on the grid})
For each nonempty segment and dimension, define
$\Pi_{\ell,d}(t)=\sum_{i\in S_\ell^{\,k}}\pi_{i,d}(t)$ and select
\[
t'_{\ell,d}\in\arg\max_{t\in G_d}\Pi_{\ell,d}(t),
\qquad d=1,\ldots,D.
\]
Under the concave profit specification, evaluate only the grid points
immediately below and above the continuous maximizer of $\Pi_{\ell,d}$.
If that maximizer lies outside the grid's range, use the nearest endpoint.
If it lies on the grid, use that point. Choose the candidate with higher
segment profit, resolving ties in favor of the smaller level.

\item (\emph{Choose treatment dimensions})
Compare the maximized segment profits across dimensions and resolve ties by 
the smallest dimension index:
\[
d_\ell^{\,k+1}\in\arg\max_{d\in\{1,\ldots,D\}}
\Pi_{\ell,d}(t'_{\ell,d}).
\]
Set $\tilde{\boldsymbol t}_\ell^{\,k+1}$ to the vector with level
$t'_{\ell,d_\ell^{\,k+1}}$ in dimension $d_\ell^{\,k+1}$ and zero
elsewhere. If $S_\ell^{\,k}$ is empty, retain its previous treatment.
\end{itemize}

\textbf{Terminate} when
$\max_{\ell=1,\ldots,L}\|\tilde{\boldsymbol t}_\ell^{\,k+1}
-\tilde{\boldsymbol t}_\ell^{\,k}\|_\infty<\delta$
or the iteration limit is reached.

\textbf{Output:}
Reassign customers to the final treatments and return the segmentation.
\end{spacing}
\end{algorithm}

\newpage{}

\section*{Figures}

\begin{spacing}{1.0}

\begin{figure}[H]
\centering
\includegraphics[width=0.75\textwidth]{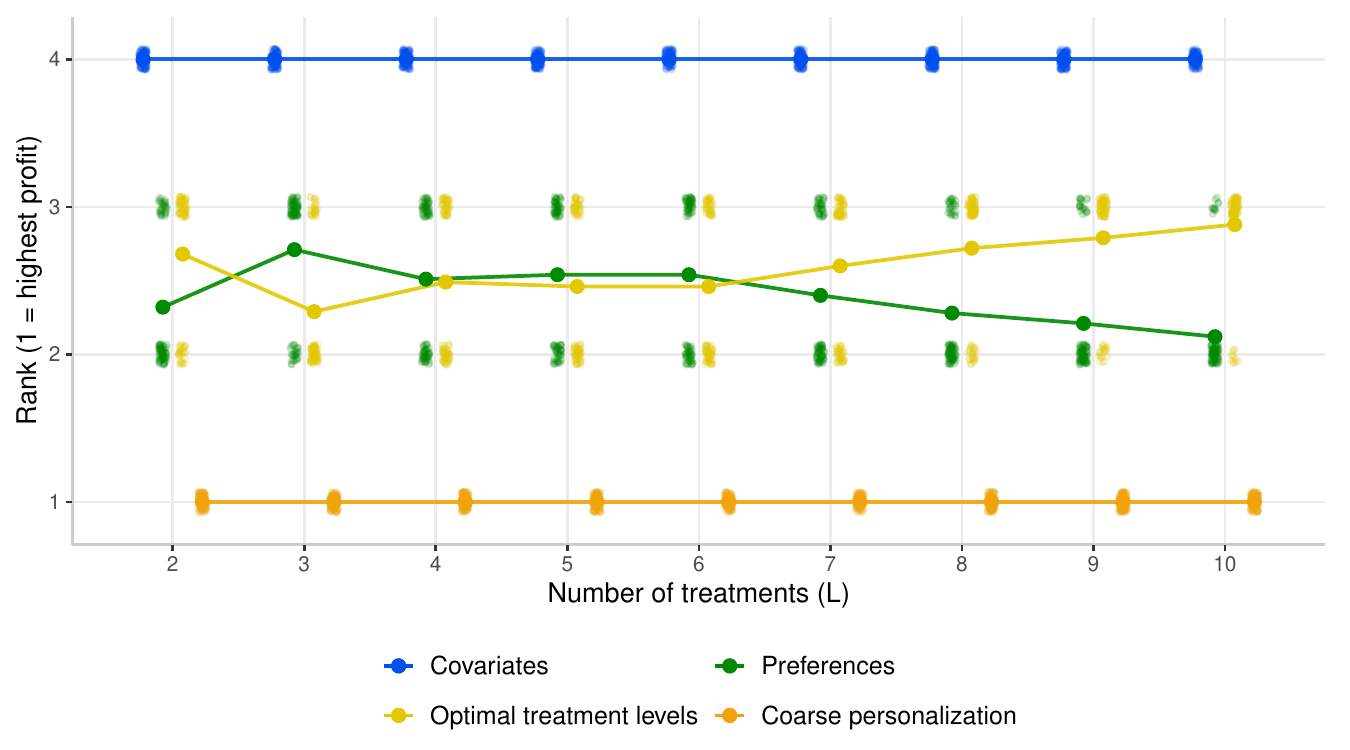}
\caption{Segmentation-method ranks across subsamples}
\label{fig:Segmentation-Ranks}
\par\medskip {\footnotesize\raggedright The figure ranks the four segmentation
methods by uncorrected in-sample expected incremental profit at each $L$. Each
of 100 replications draws 80\% of customers without replacement. We reestimate
the continuous CATE functions and reconstruct the policies in each replication.
Rank 1 indicates the highest profit. The faint points show individual
replications and are jittered to make overlapping points visible. Solid points
and lines show mean ranks. We omit $L=1$ because all four methods select the
same blanket treatment. \par}
\end{figure}

\begin{figure}[H]
\centering
\includegraphics[width=0.9\textwidth]{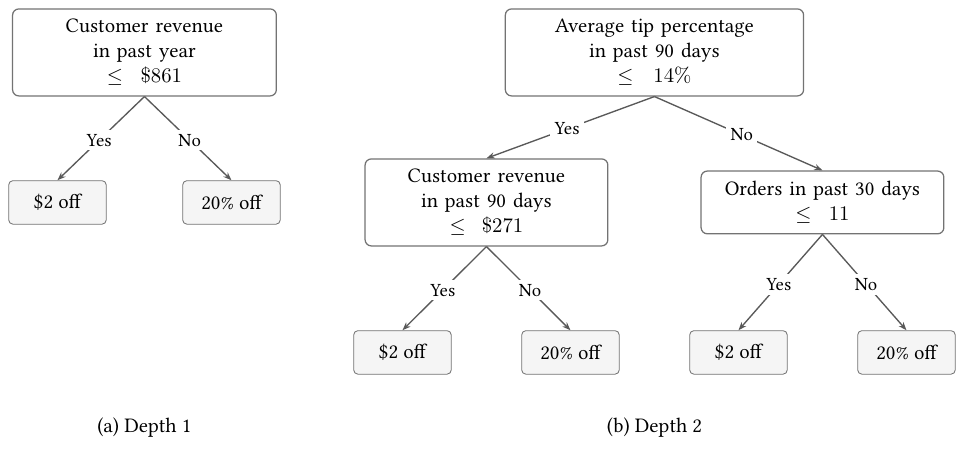}
\caption{Estimated policy trees}
\label{fig:Policy-Trees}
\par\medskip {\footnotesize\raggedright The figure shows the baseline policy
trees used in the profit comparisons. Splits use pre-treatment customer
characteristics. We fit the trees by minimizing profit regret over the eight
experimental promotions. Even though the depth 2 tree has four terminal nodes, both trees
assign two distinct promotions. Split thresholds are rounded to the nearest
integer for the figure and reported profits use the unrounded thresholds.\par}
\end{figure}

\begin{figure}[H]
\begin{centering}
\includegraphics[width=0.7\textwidth]{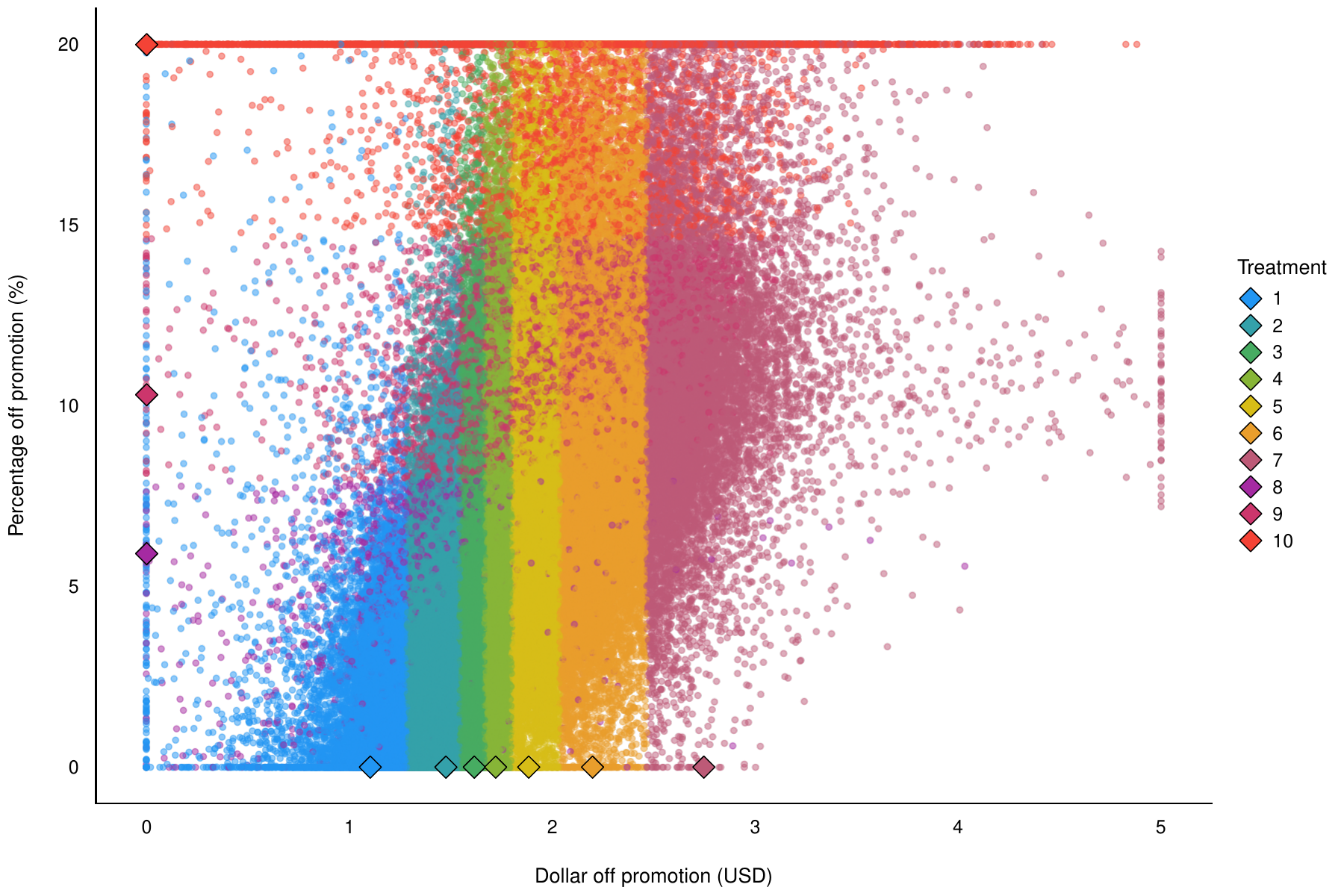}
\par\end{centering} \raggedright{}{\footnotesize Note: This figure provides a
top-down view of Figure \ref{fig:Treatment-assignments-10} and visualizes the
coarse personalization solution with $L=10$ segments. The optimal treatment
levels ($t^{*}_{i,1},t^{*}_{i,2}$) for each individual are plotted for the
dollar off and the percentage off promotions in the scatterplot. Segment
membership is indicated by color and each segment's assigned treatment is
represented by a diamond. These assigned treatments are in Table
\ref{tab:Treatment-assignments-10}. }\caption{Treatment assignments for the
coarse personalization solution ($L=10$
treatments)\label{fig:Treatment-assignments-10-2D}}
\end{figure}

\begin{figure}[H]
\begin{centering}
\includegraphics[width=0.7\textwidth]{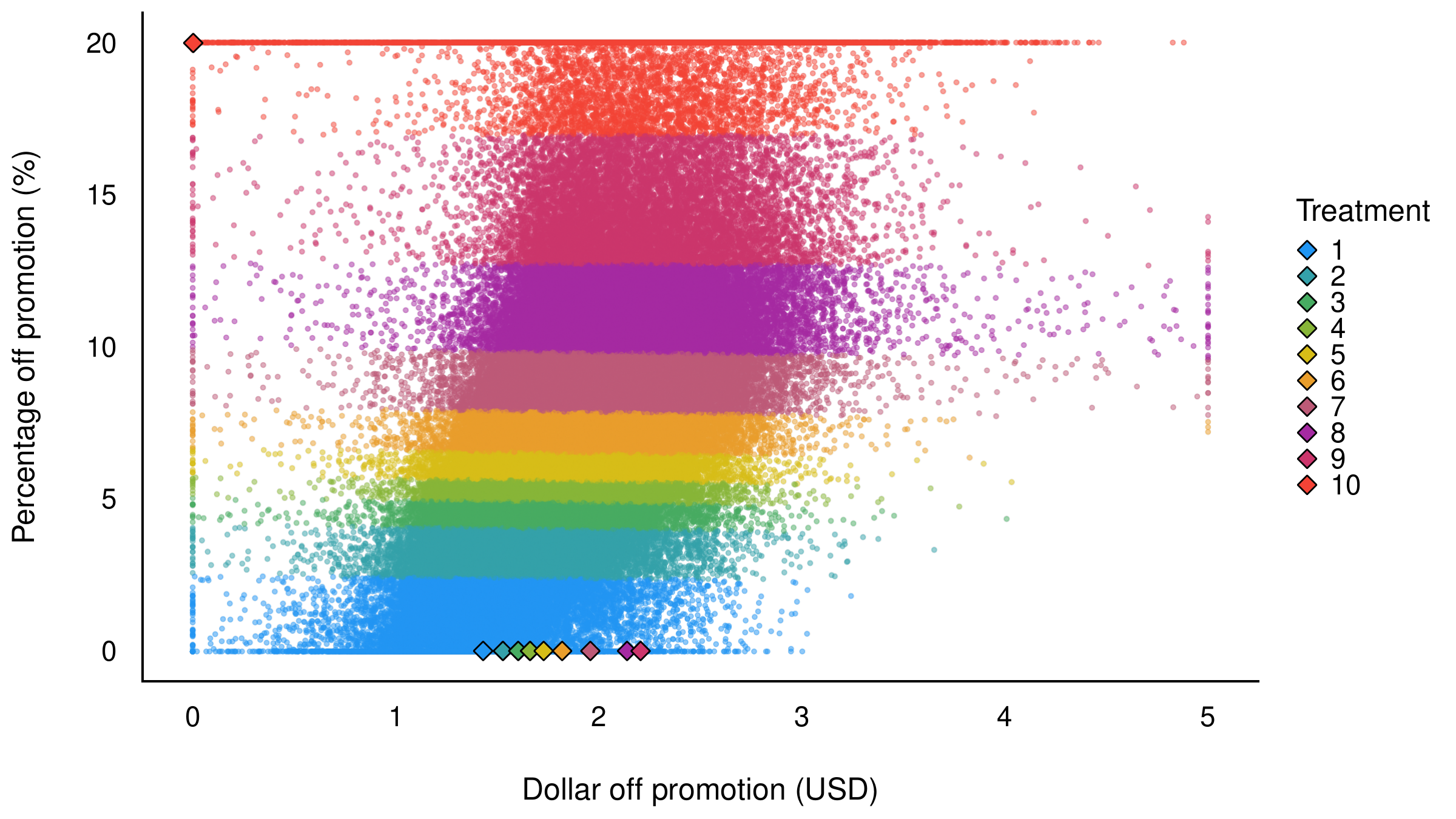}
\par\end{centering} \raggedright{}{\footnotesize Note: This figure visualizes
the results from forming $L=10$ segments based on optimal treatment levels. The
construction of segments uses Euclidean distance between the optimal treatment
levels. The optimal treatment levels ($t^{*}_{i,1},t^{*}_{i,2}$) for each
individual are plotted for the dollar off and the percentage off promotions in
the scatterplot. Segment membership is indicated by color and each segment's
assigned treatment is represented by a diamond. }\caption{Treatment assignments
for segmenting on optimal treatment levels ($L=10$
treatments)\label{fig:Treatment-assignments-10-2D-opt-t-levels}}
\end{figure}

\end{spacing}

\newpage 
\section*{Tables}

\begin{spacing}{1.0}

\begin{table}[H]
\centering
\caption{Computational performance on the full-population treatment grid}
\label{tab:Solver-Performance}
\begin{tabular}{rrrrrr}
 & & \multicolumn{2}{c}{Algorithm 1} & \multicolumn{2}{c}{Gurobi} \\
\cline{3-4}\cline{5-6}
$L$ & \shortstack{Gap to bound\\(\%)} & \shortstack{Time\\(seconds)} &
\shortstack{Memory Usage\\(GB)} & \shortstack{Time\\(hours)} &
\shortstack{Memory Usage \\(GB)} \\
\hline
 5 & 0.00043 & 4.16 & 0.94 &  5.83 & 395.0 \\
 6 & 0.02205 & 4.35 & 0.92 & 10.89 & 440.6 \\
 7 & 0.00010 & 4.94 & 0.92 & 13.76 & 451.1 \\
 8 & 0.00046 & 5.11 & 0.90 & 12.41 & 438.3 \\
 9 & 0.00003 & 5.17 & 0.90 & 28.04 & 439.5 \\
10 & 0.00803 & 5.33 & 0.89 & 15.48 & 442.4 \\
\end{tabular}

\medskip
\begin{minipage}{\textwidth}
\footnotesize
Note: Both methods use the same 102-treatment grid, with spacing 0.25 in each
dimension, for 1.2 million customers. The gap is the difference between Gurobi's
final upper bound and Algorithm 1's best profit expressed as a percentage of the
bound. Algorithm 1 times include initialization and five starts run
in parallel on five workers. Both Algorithm 1 and Gurobi use five CPU cores and
ran on the same AMD EPYC 9374F CPU node. Gurobi solves to a relative MIP gap of
$10^{-5}$ or smaller. Memory Usage is reported as peak RAM of the job. We use
Gurobi version 12.0.3.

Gurobi obtains its upper bound by relaxing the integer treatment-selection
constraints and tightening these relaxations through branching. The optimal
profit on this grid lies between Algorithm 1's achieved profit and Gurobi's
upper bound. A small gap thus establishes that little additional profit can
be obtained by improving the optimization on this grid.
\end{minipage}
\end{table}

\begin{table}[H]
\centering
\caption{Comparison with continuous-treatment optimizers}
\label{tab:Solver-Heuristics}
{\small
\setlength{\tabcolsep}{4pt}
\begin{tabular}{lrrrrrr}
& \multicolumn{3}{c}{$L=5$} & \multicolumn{3}{c}{$L=10$} \\
\cline{2-4}\cline{5-7} Method & Gap (\$) & Evals. & Sec. & Gap (\$) & Evals. &
Sec. \\
\hline
Algorithm 1, five starts & 0 & 395 & 10 & 0 & 1,289 & 26 \\
L-BFGS-B, five starts & 0 & 11,364 & 74 & 0 & 128,647 & 1,308 \\
Simulated annealing& 5 & 39,501 & 251 & 64 & 128,901 & 1,308 \\
Differential evolution & 13 & 39,500 & 253 & 44 & 128,800 & 1,306
\\
Genetic algorithm & 45 & 32,328 & 203 & 33 & 105,419 & 1,081 \\
\end{tabular}
}

\medskip
\begin{minipage}{\textwidth}
\footnotesize
Note: Results use approximately 1.2 million customers and continuous treatment
levels. Algorithm \ref{alg:Adapted-Lloyd's-Algorithm} used five starts and
L-BFGS-B used five starts for each mix of dollar-off and percentage-off
treatments. The other three methods, simulated annealing, differential
evolution, and the genetic algorithm, were each allowed up to 100 times
Algorithm 1's objective-function evaluations. Actual evaluations and runtimes
are reported. Entries are medians across three solver seeds, with all methods
run on one core. Gap measures expected-profit difference from the best solution
found across methods for the same seed and not from a certified global optimum.
We used R's \texttt{optim} for L-BFGS-B and simulated annealing (\texttt{SANN}),
the \texttt{DEoptim} package for differential evolution, and the \texttt{GA}
package for the genetic algorithm.
\end{minipage}
\end{table}

\begin{table}[H]
\centering
\caption{Uncorrected profits, standard errors, and bootstrap corrections}
\label{tab:Profits-Uncertainty-Correction}
\begingroup
\setstretch{1}
\footnotesize
\setlength{\tabcolsep}{2pt}
\renewcommand{\arraystretch}{1.05}
\begin{tabular*}{\textwidth}{@{\extracolsep{\fill}}l*{10}{r}@{}}
Method / statistic & \multicolumn{10}{c}{Number of treatments ($L$)} \\
 & 1 & 2 & 3 & 4 & 5 & 6 & 7 & 8 & 9 & 10 \\
\hline
Discrete treatments & 790.1 & 810.3 & 812.2 & 813.2 & 813.9 & 814.0 & 814.0 & 814.0 & \textemdash & \textemdash \\
\quad SE & (30.4) & (33.2) & (33.5) & (33.8) & (33.8) & (33.9) & (34.0) & (34.0) & \textemdash & \textemdash \\
\quad Correction (\%) & 1.44 & 1.94 & 2.18 & 2.30 & 2.34 & 2.38 & 2.38 & 2.39 & \textemdash & \textemdash \\
\addlinespace[5pt]
Covariates & 809.6 & 809.6 & 809.7 & 809.7 & 809.7 & 809.8 & 809.8 & 810.1 & 810.0 & 809.9 \\
\quad SE & (27.9) & (27.9) & (27.9) & (28.0) & (28.0) & (28.0) & (28.1) & (28.1) & (28.1) & (28.2) \\
\quad Correction (\%) & 1.30 & 1.31 & 1.31 & 1.32 & 1.32 & 1.33 & 1.34 & 1.32 & 1.33 & 1.34 \\
\addlinespace[5pt]
Preferences & 809.6 & 814.4 & 827.0 & 828.1 & 828.6 & 828.7 & 829.0 & 831.2 & 831.5 & 831.4 \\
\quad SE & (27.9) & (30.2) & (30.9) & (30.9) & (31.1) & (31.2) & (31.2) & (31.1) & (31.2) & (31.3) \\
\quad Correction (\%) & 1.30 & 1.93 & 1.33 & 1.56 & 1.68 & 1.77 & 1.83 & 1.72 & 1.77 & 1.84 \\
\addlinespace[5pt]
Optimal treatment levels & 809.6 & 814.7 & 826.6 & 828.4 & 829.7 & 830.2 & 830.4 & 830.6 & 830.8 & 830.8 \\
\quad SE & (27.9) & (30.3) & (31.4) & (31.2) & (31.3) & (31.4) & (31.4) & (31.4) & (31.4) & (31.5) \\
\quad Correction (\%) & 1.30 & 1.96 & 1.57 & 1.62 & 1.63 & 1.64 & 1.65 & 1.65 & 1.65 & 1.65 \\
\addlinespace[5pt]
Coarse personalization & 809.6 & 830.1 & 834.4 & 836.1 & 837.8 & 838.6 & 839.1 & 839.4 & 839.7 & 839.9 \\
\quad SE & (27.9) & (31.2) & (32.1) & (32.2) & (32.3) & (32.3) & (32.3) & (32.3) & (32.3) & (32.4) \\
\quad Correction (\%) & 1.30 & 1.86 & 2.10 & 2.27 & 2.34 & 2.41 & 2.45 & 2.48 & 2.49 & 2.51 \\
\end{tabular*}
\par\medskip
\begin{tabular*}{\textwidth}{@{\extracolsep{\fill}}lrrr@{}}
Benchmark & $L$ & Uncorrected profits (SE) & Correction (\%) \\
\hline
Full personalization & \textemdash & 841.1 (32.4) & 2.57 \\
Policy tree, depth 1 & 2 & 804.1 (32.5) & 1.54 \\
Policy tree, depth 2 & 2 & 804.9 (32.8) & 1.63 \\
\end{tabular*}
\par\medskip
\begin{minipage}{\textwidth}
\footnotesize\raggedright
Note: For each method, the first row reports uncorrected profits, the second reports their bootstrap standard errors in parentheses, and the third reports the bias correction as a percentage of uncorrected profits. Profits and SEs are in thousands of dollars for over 1.2 million customers for 28 days, net of promotion costs and before implementation costs. SEs and bias corrections use 500 $m$-out-of-$n$ bootstrap replications with $m=\lfloor n^{0.95}\rfloor$. SEs measure uncertainty in uncorrected profits. Corrected profits equal uncorrected profits $ \times (1-\text{Correction \%}/100)$.
Discrete treatments are restricted to eight experimental offers. Both policy trees use two distinct offers and Figure \ref{fig:Policy-Trees} shows their assignment rules. Full personalization is unrestricted by $L$ and entails 2,084 distinct promotions at three significant figures.
\end{minipage}
\par\endgroup
\end{table}

\begin{table}[H]
\centering
\caption{Paired profit differences between coarse personalization and benchmarks}
\label{tab:Profits-Paired-Comparisons}
\begingroup
\setstretch{1}
\footnotesize
\setlength{\tabcolsep}{2pt}
\renewcommand{\arraystretch}{1.08}
\begin{tabular*}{\textwidth}{@{\extracolsep{\fill}}l*{10}{r}@{}}
Coarse personalization & \multicolumn{10}{c}{Number of treatments ($L$)} \\
minus & 1 & 2 & 3 & 4 & 5 & 6 & 7 & 8 & 9 & 10 \\
\hline
\hspace{0.5em}Discrete treatments & 19.5$^{***}$ & 19.8$^{***}$ & 22.2$^{***}$ & 23.0$^{***}$ & 23.9$^{***}$ & 24.6$^{***}$ & 25.1$^{***}$ & 25.4$^{***}$ & \textemdash & \textemdash \\
\hspace{1.5em}SE & (2.7) & (2.5) & (1.9) & (2.0) & (2.0) & (2.1) & (2.1) & (2.1) & \textemdash & \textemdash \\
\addlinespace[5pt]
\hspace{0.5em}Covariates & 0.0 & 20.5$^{**}$ & 24.7$^{***}$ & 26.5$^{***}$ & 28.1$^{***}$ & 28.9$^{***}$ & 29.3$^{***}$ & 29.3$^{***}$ & 29.7$^{***}$ & 30.0$^{***}$ \\
\hspace{1.5em}SE & (0.0) & (9.3) & (9.5) & (9.8) & (9.9) & (9.9) & (10.0) & (10.0) & (10.0) & (9.9) \\
\addlinespace[5pt]
\hspace{0.5em}Preferences & 0.0 & 15.7$^{**}$ & 7.4 & 8.1 & 9.1$^{*}$ & 9.9$^{**}$ & 10.1$^{**}$ & 8.2$^{*}$ & 8.2$^{*}$ & 8.6$^{**}$ \\
\hspace{1.5em}SE & (0.0) & (6.8) & (5.4) & (5.0) & (4.8) & (4.6) & (4.5) & (4.3) & (4.2) & (4.0) \\
\addlinespace[5pt]
\hspace{0.5em}Optimal treatment levels & 0.0 & 15.4$^{***}$ & 7.8$^{***}$ & 7.8$^{***}$ & 8.1$^{***}$ & 8.4$^{***}$ & 8.7$^{***}$ & 8.8$^{***}$ & 9.0$^{***}$ & 9.1$^{***}$ \\
\hspace{1.5em}SE & (0.0) & (3.8) & (2.2) & (2.2) & (2.2) & (2.2) & (2.3) & (2.3) & (2.3) & (2.3) \\
\midrule
\multicolumn{11}{l}{\textit{Benchmarks held fixed across $L$}} \\
\addlinespace[3pt]
\hspace{0.5em}Policy tree, depth 1 & 5.6 & 26.0$^{***}$ & 30.3$^{***}$ & 32.1$^{***}$ & 33.7$^{***}$ & 34.6$^{***}$ & 35.0$^{***}$ & 35.4$^{***}$ & 35.7$^{***}$ & 35.9$^{***}$ \\
\hspace{1.5em}SE & (8.9) & (3.6) & (3.2) & (3.5) & (3.6) & (3.7) & (3.8) & (3.8) & (3.8) & (3.8) \\
\addlinespace[5pt]
\hspace{0.5em}Policy tree, depth 2 & 4.7 & 25.2$^{***}$ & 29.5$^{***}$ & 31.2$^{***}$ & 32.9$^{***}$ & 33.7$^{***}$ & 34.2$^{***}$ & 34.5$^{***}$ & 34.8$^{***}$ & 35.0$^{***}$ \\
\hspace{1.5em}SE & (9.3) & (3.5) & (2.9) & (3.1) & (3.2) & (3.3) & (3.3) & (3.3) & (3.4) & (3.4) \\
\addlinespace[5pt]
\hspace{0.5em}Full personalization & -31.5$^{***}$ & -11.0$^{***}$ & -6.7$^{***}$ & -4.9$^{***}$ & -3.3$^{***}$ & -2.5$^{***}$ & -2.0$^{***}$ & -1.7$^{***}$ & -1.4$^{***}$ & -1.1$^{***}$ \\
\hspace{1.5em}SE & (10.2) & (2.5) & (1.3) & (0.9) & (0.6) & (0.4) & (0.3) & (0.3) & (0.2) & (0.2) \\
\end{tabular*}
\par\medskip
\begin{minipage}{\textwidth}
\footnotesize\raggedright
Note: $^{*}p<0.10$, $^{**}p<0.05$, $^{***}p<0.01$. Entries report coarse
personalization's uncorrected expected incremental profit minus each benchmark's
profit. Differences and paired standard errors in parentheses are in thousands
of dollars over 1.2 million customers for 28 days, net of promotion costs and
before implementation costs. Positive differences favor coarse personalization.
Standard errors use 500 $m$-out-of-$n$ bootstrap replications with
$m=\lfloor n^{0.95}\rfloor$. $p$-values are from a two-sided $t$-test between
coarse personalization and each method. The first four benchmarks use the same
$L$ as coarse personalization. Discrete treatments are limited to eight
experimental offers. In the lower block, the benchmarks remain fixed as CP's $L$
changes. Both policy trees use two offers and full personalization is
unrestricted. At $L=1$, the four continuous treatment segmentation methods
choose the same optimized blanket treatment, so their differences and SEs are
zero. Discrete offers cannot optimize treatment levels, so their profits can
differ even at $L=1$.
\end{minipage}
\par\endgroup
\end{table}

\begin{table}[H]
\begin{centering}
\caption{Treatment assignments ($L=10$) \label{tab:Treatment-assignments-10}}
\centering\begin{tabular}{r|rr|rr}
Treatment ($\ell$) & Dimension ($d$) & Value ($\tilde{t}_{\ell,d_\ell}$) & $N$ & Share (\%) \\
\hline
1 & Dollar & 1.10 & 11,576 & 0.96 \\
2 & Dollar & 1.48 & 133,451 & 11.07 \\
3 & Dollar & 1.61 & 482,762 & 40.06 \\
4 & Dollar & 1.72 & 357,301 & 29.65 \\
5 & Dollar & 1.88 & 137,297 & 11.39 \\
6 & Dollar & 2.20 & 40,584 & 3.37 \\
7 & Dollar & 2.75 & 13,118 & 1.09 \\
8 & Percentage & 5.91 & 9,660 & 0.80 \\
9 & Percentage & 10.31 & 11,319 & 0.94 \\
10 & Percentage & 20.00 & 8,101 & 0.67 \\
\end{tabular}
{\small\medskip{}
}{\small\par}
\par\end{centering}
{\footnotesize Note: The table reports the ten treatments and segment
sizes for the coarse personalization solution in Figure
\ref{fig:Treatment-assignments-10}. Treatment values are expressed
in dollars for dollar-off promotions and percentage points for
percentage-off promotions, rounded to two decimal places.
$N$ denotes the number of customers assigned to each treatment,
and Share denotes the percentage of customers assigned.\par}
\end{table}

\begin{table}[H]
\centering
\caption{Expected incremental profits with integer treatment levels}
\label{tab:Profits-table-by-treatments-discrete}
\begin{spacing}{1.0}
\small
\setlength{\tabcolsep}{7pt}
\renewcommand{\arraystretch}{1.05}
\begin{tabular}{@{}lrrr@{}}
$L$ & \shortstack{Uncorrected\\profit} & \shortstack{Bias-corrected\\profit} & \shortstack{Percent of integer-grid\\full personalization (\%)} \\
\hline
1 & 790.1 & 778.8 & 96.70 \\
2 & 810.3 & 794.5 & 99.16 \\
3 & 813.3 & 794.5 & 99.53 \\
4 & 815.2 & 794.9 & 99.77 \\
5 & 816.2 & 795.0 & 99.89 \\
6 & 816.6 & 795.0 & 99.94 \\
7 & 816.7 & 794.8 & 99.95 \\
8 & 816.9 & 794.9 & 99.97 \\
9 & 816.9 & 794.9 & 99.97 \\
10 & 817.0 & 794.9 & 99.98 \\
\hline
\shortstack[l]{Full personalization\\(26 unique treatments)} & 817.1 & 794.9 & --- \\
\end{tabular}
\par\medskip
\begin{minipage}{\textwidth}
\footnotesize\raggedright Note: Profits are in thousands of dollars over 1.2
million customers, net of promotion costs and before implementation costs.
Treatments are restricted to whole-dollar and whole-percentage-point increments,
with the profit-maximizing grid value selected within each segment update.
Bias-corrected profits subtract the global-target correction from 500
$m$-out-of-$n$ bootstrap replications with $m=\lfloor n^{0.95}\rfloor$. The
final column expresses uncorrected profit as a percentage of uncorrected
full-personalization profit on the same integer grid.
\end{minipage}
\end{spacing}
\end{table}

\begin{table}[H]
\centering
\caption{Surplus decomposition by number of segments}
\label{tab:Welfare-decomposition-treatments}
\begingroup
\setstretch{1}
\footnotesize
\setlength{\tabcolsep}{2pt}
\renewcommand{\arraystretch}{1.10}
\begin{tabular*}{\textwidth}{@{\extracolsep{\fill}}l*{10}{r}@{}}
 & \multicolumn{10}{c}{Number of segments ($L$)} \\
 & 1 & 2 & 3 & 4 & 5 & 6 & 7 & 8 & 9 & 10 \\
\hline
Consumer surplus & 0.0 & -14.7 & -13.0 & -31.4 & -30.8 & -30.5 & -30.3 & -33.3 & -33.3 & -33.2 \\
\addlinespace[8pt]
\multicolumn{11}{l}{\textit{Uncorrected profits}} \\
\quad Producer surplus & 0.0 & 19.5 & 22.8 & 23.5 & 24.1 & 24.0 & 23.5 & 22.8 & 22.1 & 21.3 \\
\quad Total surplus & 0.0 & 4.8 & 9.7 & -7.9 & -6.6 & -6.5 & -6.8 & -10.5 & -11.2 & -11.9 \\
\addlinespace[8pt]
\multicolumn{11}{l}{\textit{Bias-corrected profits}} \\
\quad Producer surplus & 0.0 & 14.6 & 15.8 & 15.1 & 15.1 & 14.4 & 13.5 & 12.6 & 11.7 & 10.8 \\
\quad Total surplus & 0.0 & -0.1 & 2.8 & -16.3 & -15.7 & -16.1 & -16.8 & -20.8 & -21.5 & -22.4 \\
\end{tabular*}
\par\medskip
\begin{minipage}{\textwidth}
\footnotesize\raggedright Note: Values are changes relative to the optimized
blanket promotion ($L=1$), in thousands of dollars. Producer and total surplus
deduct implementation costs of \$1,000 per treatment, which gives a cost
difference of $\$1{,}000(L-1)$ to the baseline. Corrected producer surplus uses
bootstrap bias-corrected profits for both policies. Total surplus with corrected
profits is the sum of consumer surplus and corrected producer surplus.
\end{minipage}
\par\endgroup
\end{table}

\end{spacing}

\begin{spacing}{1.0}
\small
\renewcommand{\refname}{Online Appendix References}
\putbib[bib]
\end{spacing}
\end{bibunit}


\begin{thebibliography}{63}
\newcommand{\enquote}[1]{``#1''}
\expandafter\ifx\csname natexlab\endcsname\relax\def\natexlab#1{#1}\fi

\bibitem[\protect\citeauthoryear{Ansari and Mela}{Ansari and
  Mela}{2003}]{Ansari2003}
\textsc{Ansari, A. and C.~F. Mela} (2003): \enquote{{E-Customization},}
  \emph{Journal of Marketing Research}, 40, 131--145.

\bibitem[\protect\citeauthoryear{Aouad, Elmachtoub, Ferreira, and
  McNellis}{Aouad et~al.}{2023}]{Aouad2023}
\textsc{Aouad, A., A.~N. Elmachtoub, K.~J. Ferreira, and R.~McNellis} (2023):
  \enquote{{Market Segmentation Trees},} \emph{Manufacturing \& Service
  Operations Management}, 25, 648--667.

\bibitem[\protect\citeauthoryear{Arora, Dreze, Ghose, Hess, Iyengar, Jing,
  Joshi, Kumar, Lurie, Neslin, Sajeesh, Su, Syam, Thomas, and Zhang}{Arora
  et~al.}{2008}]{Arora2008}
\textsc{Arora, N., X.~Dreze, A.~Ghose, J.~D. Hess, R.~Iyengar, B.~Jing,
  Y.~Joshi, V.~Kumar, N.~Lurie, S.~Neslin, S.~Sajeesh, M.~Su, N.~Syam,
  J.~Thomas, and Z.~J. Zhang} (2008): \enquote{{Putting one-to-one marketing to
  work: Personalization, customization, and choice},} \emph{Marketing Letters},
  19, 305--321.

\bibitem[\protect\citeauthoryear{Ascarza}{Ascarza}{2018}]{Ascarza2018}
\textsc{Ascarza, E.} (2018): \enquote{{Retention Futility: Targeting High-Risk
  Customers Might be Ineffective},} \emph{Journal of Marketing Research}, 55,
  80--98.

\bibitem[\protect\citeauthoryear{Athey and Imbens}{Athey and
  Imbens}{2016}]{Athey2016}
\textsc{Athey, S. and G.~Imbens} (2016): \enquote{{Recursive partitioning for
  heterogeneous causal effects},} \emph{Proceedings of the National Academy of
  Sciences of the United States of America}, 113, 7353--7360.

\bibitem[\protect\citeauthoryear{Athey, Tibshirani, and Wager}{Athey
  et~al.}{2019}]{Athey2019}
\textsc{Athey, S., J.~Tibshirani, and S.~Wager} (2019): \enquote{{Generalized
  random forests},} \emph{Annals of Statistics}, 47, 1148--1178.

\bibitem[\protect\citeauthoryear{Athey and Wager}{Athey and
  Wager}{2021}]{Athey2021}
\textsc{Athey, S. and S.~Wager} (2021): \enquote{{Policy Learning With
  Observational Data},} \emph{Econometrica}, 89, 133--161.

\bibitem[\protect\citeauthoryear{Bergemann and Bonatti}{Bergemann and
  Bonatti}{2011}]{Bergemann2011}
\textsc{Bergemann, D. and A.~Bonatti} (2011): \enquote{{Targeting in
  advertising markets: Implications for offline versus online media},}
  \emph{RAND Journal of Economics}, 42, 417--443.

\bibitem[\protect\citeauthoryear{Bickel, G{\"{o}}tze, and {Van Zwet}}{Bickel
  et~al.}{1997}]{Bickel1997}
\textsc{Bickel, P.~J., F.~G{\"{o}}tze, and W.~R. {Van Zwet}} (1997):
  \enquote{{Resampling fewer than n observations: Gains, losses, and remedies
  for losses},} \emph{Statistica Sinica}, 7, 1--31.

\bibitem[\protect\citeauthoryear{Briesch}{Briesch}{1997}]{Briesch1997}
\textsc{Briesch, R.~A.} (1997): \enquote{{Does it matter how price promotions
  are operationalized?}} \emph{Marketing Letters}, 8, 167--181.

\bibitem[\protect\citeauthoryear{Bucklin and Gupta}{Bucklin and
  Gupta}{1992}]{Bucklin1992}
\textsc{Bucklin, R.~E. and S.~Gupta} (1992): \enquote{{Brand Choice, Purchase
  Incidence, and Segmentation: An Integrated Modeling Approach},} \emph{Journal
  of Marketing Research}, 29.

\bibitem[\protect\citeauthoryear{Bucklin, Gupta, and Siddarth}{Bucklin
  et~al.}{1998}]{Bucklin1998a}
\textsc{Bucklin, R.~E., S.~Gupta, and S.~Siddarth} (1998):
  \enquote{{Determining Segmentation in Sales Response across Consumer Purchase
  Behaviors},} \emph{Journal of Marketing Research}, 35, 189--197.

\bibitem[\protect\citeauthoryear{Cui and Hamilton}{Cui and
  Hamilton}{2026}]{Cui2022}
\textsc{Cui, T. and M.~L. Hamilton} (2026): \enquote{{Optimal Feature-Based
  Market Segmentation and Pricing},} \emph{Operations Research}, 74,
  1354--1370.

\bibitem[\protect\citeauthoryear{{De Vos}, Bockel-Rickermann, Lessmann, and
  Verbeke}{{De Vos} et~al.}{2026}]{DeVos2026}
\textsc{{De Vos}, S., C.~Bockel-Rickermann, S.~Lessmann, and W.~Verbeke}
  (2026): \enquote{{Uplift modeling with continuous treatments: A
  predict-then-optimize approach},} \emph{European Journal of Operational
  Research}, 330, 230--244.

\bibitem[\protect\citeauthoryear{DoorDash}{DoorDash}{2022}]{doordash2026promotion}
\textsc{DoorDash} (2022): \enquote{{Building a Promo Code DevOps Tool for a
  Better Understanding of Campaigns},}
  \href{https://careersatdoordash.com/blog/doordash-2022-summer-intern-survey-post/#promo}{Building
  a Promo Code DevOps Tool}.

\bibitem[\protect\citeauthoryear{Dube and Misra}{Dube and
  Misra}{2022}]{Dube2017}
\textsc{Dube, J.-P. and S.~Misra} (2022): \enquote{{Personalized Pricing and
  Consumer Welfare},} \emph{Journal of Political Economy}.

\bibitem[\protect\citeauthoryear{Ellickson, Kar, and Reeder}{Ellickson
  et~al.}{2022}]{Ellickson2022}
\textsc{Ellickson, P.~B., W.~Kar, and J.~C. Reeder} (2022):
  \enquote{{Estimating Marketing Component Effects: Double Machine Learning
  from Targeted Digital Promotions},} \emph{Marketing Science}.

\bibitem[\protect\citeauthoryear{Ellickson, Kar, Reeder, and Zeng}{Ellickson
  et~al.}{2026}]{Ellickson2026}
\textsc{Ellickson, P.~B., W.~Kar, J.~C. Reeder, and G.~Zeng} (2026):
  \enquote{{Evaluating Novel Unstructured Treatments with Generative AI: A
  Causal Prediction Framework},} \emph{Journal of Marketing Research}.

\bibitem[\protect\citeauthoryear{Farrell, Liang, and Misra}{Farrell
  et~al.}{2020}]{Farrell2020}
\textsc{Farrell, M.~H., T.~Liang, and S.~Misra} (2020): \enquote{{Deep Learning
  for Individual Heterogeneity},} .

\bibitem[\protect\citeauthoryear{Farrell, Liang, and Misra}{Farrell
  et~al.}{2021}]{Farrell2021}
---\hspace{-.1pt}---\hspace{-.1pt}--- (2021): \enquote{{Deep Neural Networks
  for Estimation and Inference},} \emph{Econometrica}, 89, 181--213.

\bibitem[\protect\citeauthoryear{Galichon}{Galichon}{2016}]{Galichon2016}
\textsc{Galichon, A.} (2016): \emph{{Optimal Transport Methods in Economics}},
  Princeton University Press.

\bibitem[\protect\citeauthoryear{Gilmore and Pine}{Gilmore and
  Pine}{1997}]{Gilmore1997}
\textsc{Gilmore, J.~H. and B.~J.~I. Pine} (1997): \enquote{{The Four Faces of
  Mass Customization},} \emph{Harvard Business Review}.

\bibitem[\protect\citeauthoryear{Grover and Srinivasan}{Grover and
  Srinivasan}{1987}]{Grover1987}
\textsc{Grover, R. and V.~Srinivasan} (1987): \enquote{{A Simultaneous Approach
  to Market Segmentation and Market Structuring},} \emph{Journal of Marketing
  Research}, 24, 139.

\bibitem[\protect\citeauthoryear{Gupta and Chintagunta}{Gupta and
  Chintagunta}{1994}]{Gupta1994}
\textsc{Gupta, S. and P.~K. Chintagunta} (1994): \enquote{{On Using Demographic
  Variables to Determine Segment Membership in Logit Mixture Models},}
  \emph{Journal of Marketing Research}, 31, 128.

\bibitem[\protect\citeauthoryear{{Harvard Business Review Analytic
  Services}}{{Harvard Business Review Analytic
  Services}}{2018}]{HarvardBusinessReviewAnalyticServices2018}
\textsc{{Harvard Business Review Analytic Services}} (2018): \enquote{{The Age
  of Personalization},} \emph{Harvard Business Review}.

\bibitem[\protect\citeauthoryear{Hitsch, Misra, and Zhang}{Hitsch
  et~al.}{2024}]{Hitsch2024}
\textsc{Hitsch, G.~J., S.~Misra, and W.~W. Zhang} (2024):
  \enquote{{Heterogeneous treatment effects and optimal targeting policy
  evaluation},} \emph{Quantitative Marketing and Economics}.

\bibitem[\protect\citeauthoryear{Huang, Ascarza, and Israeli}{Huang
  et~al.}{2024}]{Huang2024}
\textsc{Huang, T.-W., E.~Ascarza, and A.~Israeli} (2024):
  \enquote{{Incrementality Representation Learning: Synergizing Past
  Experiments for Intervention Personalization},} \emph{SSRN Electronic
  Journal}.

\bibitem[\protect\citeauthoryear{Imbens and Rubin}{Imbens and
  Rubin}{2015}]{Imbens2015}
\textsc{Imbens, G.~W. and D.~B. Rubin} (2015): \emph{{Causal Inference for
  Statistics, Social, and Biomedical Sciences}}, Cambridge University Press.

\bibitem[\protect\citeauthoryear{Jain, Bass, and Chen}{Jain
  et~al.}{1990}]{Jain1990}
\textsc{Jain, D., F.~M. Bass, and Y.-M. Chen} (1990): \enquote{{Estimation of
  Latent Class Models with Heterogeneous Choice Probabilities: An Application
  to Market Structuring},} \emph{Journal of Marketing Research}, 27, 94.

\bibitem[\protect\citeauthoryear{Kallus and Zhou}{Kallus and
  Zhou}{2018}]{pmlr-v84-kallus18a}
\textsc{Kallus, N. and A.~Zhou} (2018): \enquote{{Policy Evaluation and
  Optimization with Continuous Treatments},} in \emph{Proceedings of the
  Twenty-First International Conference on Artificial Intelligence and
  Statistics}, ed. by A.~Storkey and F.~Perez-Cruz, PMLR, vol.~84 of
  \emph{Proceedings of Machine Learning Research}, 1243--1251.

\bibitem[\protect\citeauthoryear{Kallus and Zhou}{Kallus and
  Zhou}{2021}]{Kallus2021}
---\hspace{-.1pt}---\hspace{-.1pt}--- (2021): \enquote{{Fairness, Welfare, and
  Equity in Personalized Pricing},} in \emph{Proceedings of the 2021 ACM
  Conference on Fairness, Accountability, and Transparency}, New York, NY, USA:
  ACM, 296--314.

\bibitem[\protect\citeauthoryear{Kamakura and Russell}{Kamakura and
  Russell}{1989}]{Kamakura1989a}
\textsc{Kamakura, W.~A. and G.~J. Russell} (1989): \enquote{{A Probabilistic
  Choice Model for Market Segmentation and Elasticity Structure},}
  \emph{Journal of Marketing Research}, 26, 379.

\bibitem[\protect\citeauthoryear{Khan, Lewis, and Singh}{Khan
  et~al.}{2009}]{Khan2009}
\textsc{Khan, R., M.~Lewis, and V.~Singh} (2009): \enquote{{Dynamic Customer
  Management and the Value of One-to-One Marketing},} \emph{Marketing Science},
  28, 1063--1079.

\bibitem[\protect\citeauthoryear{Kitagawa and Tetenov}{Kitagawa and
  Tetenov}{2018}]{Kitagawa2018}
\textsc{Kitagawa, T. and A.~Tetenov} (2018): \enquote{{Who Should Be Treated?
  Empirical Welfare Maximization Methods for Treatment Choice},}
  \emph{Econometrica}, 86, 591--616.

\bibitem[\protect\citeauthoryear{Kotler and Keller}{Kotler and
  Keller}{2014}]{KotlerP.andKeller2014}
\textsc{Kotler, P. and K.~L. Keller} (2014): \emph{{Marketing Management}},
  Pearson, 15 ed.

\bibitem[\protect\citeauthoryear{Krieger and Green}{Krieger and
  Green}{1996}]{Krieger1996}
\textsc{Krieger, A.~M. and P.~E. Green} (1996): \enquote{{Modifying
  Cluster-Based Segments to Enhance Agreement with an Exogenous Response
  Variable},} \emph{Journal of Marketing Research}, 33, 351.

\bibitem[\protect\citeauthoryear{Laber and Zhao}{Laber and
  Zhao}{2015}]{Laber2015}
\textsc{Laber, E.~B. and Y.~Q. Zhao} (2015): \enquote{{Tree-based methods for
  individualized treatment regimes},} \emph{Biometrika}, 102, 501--514.

\bibitem[\protect\citeauthoryear{Lloyd}{Lloyd}{1982}]{Lloyd1982}
\textsc{Lloyd, S.~P.} (1982): \enquote{{Least Squares Quantization in PCM},}
  \emph{IEEE Transactions on Information Theory}, 28, 129--137.

\bibitem[\protect\citeauthoryear{Lu, Simester, and Zhu}{Lu
  et~al.}{2025}]{Lu2025}
\textsc{Lu, H., D.~Simester, and Y.~Zhu} (2025): \enquote{{Optimizing Scalable
  Targeted Marketing Policies with Constraints},} \emph{Marketing Science}, 44,
  1082--1103.

\bibitem[\protect\citeauthoryear{Peyr{\'{e}} and Cuturi}{Peyr{\'{e}} and
  Cuturi}{2019}]{Peyre2019}
\textsc{Peyr{\'{e}}, G. and M.~Cuturi} (2019): \enquote{{Computational Optimal
  Transport},} \emph{Foundations and Trends in Machine Learning}.

\bibitem[\protect\citeauthoryear{Politis, Romano, and Wolf}{Politis
  et~al.}{1999}]{Politis1999}
\textsc{Politis, D.~N., J.~P. Romano, and M.~Wolf} (1999):
  \emph{{Subsampling}}, Springer Series in Statistics, New York, NY: Springer
  New York.

\bibitem[\protect\citeauthoryear{Pollard}{Pollard}{1982}]{Pollard1982}
\textsc{Pollard, D.} (1982): \enquote{{Quantization and the Method of
  k-Means},} \emph{IEEE Transactions on Information Theory}, 28, 199--205.

\bibitem[\protect\citeauthoryear{Rafieian and Yoganarasimhan}{Rafieian and
  Yoganarasimhan}{2021}]{Rafieian2021}
\textsc{Rafieian, O. and H.~Yoganarasimhan} (2021): \enquote{{Targeting and
  Privacy in Mobile Advertising},} \emph{Marketing Science}, 40, 193--218.

\bibitem[\protect\citeauthoryear{Rossi, McCulloch, and Allenby}{Rossi
  et~al.}{1996}]{Rossi1996}
\textsc{Rossi, P.~E., R.~E. McCulloch, and G.~M. Allenby} (1996): \enquote{{The
  value of purchase history data in target marketing},} \emph{Marketing
  Science}, 15, 321--340.

\bibitem[\protect\citeauthoryear{Rust and Verhoef}{Rust and
  Verhoef}{2005}]{Rust2005}
\textsc{Rust, R.~T. and P.~C. Verhoef} (2005): \enquote{{Optimizing the
  Marketing Interventions Mix in Intermediate-Term CRM},} \emph{Marketing
  Science}, 24, 477--489.

\bibitem[\protect\citeauthoryear{Shaffer and Zhang}{Shaffer and
  Zhang}{1995}]{Shaffer1995}
\textsc{Shaffer, G. and Z.~J. Zhang} (1995): \enquote{{Competitive Coupon
  Targeting},} \emph{Marketing Science}, 14, 395--416.

\bibitem[\protect\citeauthoryear{Simester, Timoshenko, and Zoumpoulis}{Simester
  et~al.}{2020}]{Simester2020}
\textsc{Simester, D., A.~Timoshenko, and S.~I. Zoumpoulis} (2020):
  \enquote{{Efficiently evaluating targeting policies: Improving on champion
  vs. Challenger experiments},} \emph{Management Science}, 66, 3412--3424.

\bibitem[\protect\citeauthoryear{Smith and Winkler}{Smith and
  Winkler}{2006}]{Smith2006}
\textsc{Smith, J.~E. and R.~L. Winkler} (2006): \enquote{{The Optimizer's
  Curse: Skepticism and Postdecision Surprise in Decision Analysis},}
  \emph{Management Science}, 52, 311--322.

\bibitem[\protect\citeauthoryear{Smith}{Smith}{1956}]{Smith1956}
\textsc{Smith, W.~R.} (1956): \enquote{{Product Differentiation and Market
  Segmentation as Alternative Marketing Strategies},} \emph{Journal of
  Marketing}, 21, 3.

\bibitem[\protect\citeauthoryear{Stremersch and Tellis}{Stremersch and
  Tellis}{2002}]{Stremersch2002}
\textsc{Stremersch, S. and G.~J. Tellis} (2002): \enquote{{Strategic Bundling
  of Products and Prices: A New Synthesis for Marketing},} \emph{Journal of
  Marketing}, 66, 55--72.

\bibitem[\protect\citeauthoryear{Swaminathan and Joachims}{Swaminathan and
  Joachims}{2015}]{Swaminathan-Joachims-2015}
\textsc{Swaminathan, A. and T.~Joachims} (2015): \enquote{{Batch Learning from
  Logged Bandit Feedback through Counterfactual Risk Minimization},}
  \emph{Journal of Machine Learning Research}, 1731--1755.

\bibitem[\protect\citeauthoryear{{Uber Technologies}}{{Uber
  Technologies}}{2026}]{uber2026marketing}
\textsc{{Uber Technologies}} (2026): \enquote{{Managing Promotions through Uber
  Marketing Manager},}
  \href{https://help.uber.com/en/merchants-and-restaurants/article/managing-promotions-through-uber-marketing-manager?nodeId=58381469-f12a-4f89-9851-6ca9709f37ae}{Managing
  Promotions through Uber Marketing Manager}.

\bibitem[\protect\citeauthoryear{Wager and Athey}{Wager and
  Athey}{2018}]{Wager2018}
\textsc{Wager, S. and S.~Athey} (2018): \enquote{{Estimation and Inference of
  Heterogeneous Treatment Effects using Random Forests},} \emph{Journal of the
  American Statistical Association}, 113, 1228--1242.

\bibitem[\protect\citeauthoryear{Wernerfelt}{Wernerfelt}{2008}]{Wernerfelt2008}
\textsc{Wernerfelt, B.} (2008): \enquote{{Class Pricing},} \emph{Marketing
  Science}, 27, 755--763.

\bibitem[\protect\citeauthoryear{Wind}{Wind}{1978}]{Wind1978}
\textsc{Wind, Y.} (1978): \enquote{{Issues and Advances in Segmentation
  Research},} \emph{Journal of Marketing Research}, 15, 317--337.

\bibitem[\protect\citeauthoryear{Xu, Thomadsen, and Zhang}{Xu
  et~al.}{2025}]{Xu2025}
\textsc{Xu, S., R.~Thomadsen, and D.~J. Zhang} (2025): \enquote{{The Winner's
  Curse in Data-Driven Decision Making: Evidence and Solutions},} \emph{SSRN
  Electronic Journal}.

\bibitem[\protect\citeauthoryear{Yao, Lo, Nir, Tan, Evnine, Lerer, and
  Peysakhovich}{Yao et~al.}{2022}]{Yao2022}
\textsc{Yao, L., C.~Lo, I.~Nir, S.~Tan, A.~Evnine, A.~Lerer, and
  A.~Peysakhovich} (2022): \enquote{{Efficient Heterogeneous Treatment Effect
  Estimation With Multiple Experiments and Multiple Outcomes},} .

\bibitem[\protect\citeauthoryear{Yoganarasimhan, Barzegary, and
  Pani}{Yoganarasimhan et~al.}{2022}]{Yoganarasimhan2020}
\textsc{Yoganarasimhan, H., E.~Barzegary, and A.~Pani} (2022): \enquote{{Design
  and Evaluation of Optimal Free Trials},} \emph{Management Science}.

\bibitem[\protect\citeauthoryear{Zhang and Krishnamurthi}{Zhang and
  Krishnamurthi}{2004}]{Zhang2004}
\textsc{Zhang, J. and L.~Krishnamurthi} (2004): \enquote{{Customizing
  Promotions in Online Stores},} \emph{Marketing Science}, 23, 561--578.

\bibitem[\protect\citeauthoryear{Zhang and Wedel}{Zhang and
  Wedel}{2009}]{Zhang2009}
\textsc{Zhang, J. and M.~Wedel} (2009): \enquote{{The Effectiveness of
  Customized Promotions in Online and Offline Stores},} \emph{Journal of
  Marketing Research}, 46, 190--206.

\bibitem[\protect\citeauthoryear{Zhang}{Zhang}{2023}]{Zhang2023}
\textsc{Zhang, W.~W.} (2023): \enquote{{Optimal Comprehensible Targeting},} .

\bibitem[\protect\citeauthoryear{Zhang, Zeng, Zhan, and Zhang}{Zhang
  et~al.}{2026}]{Zhang2026}
\textsc{Zhang, Z., Z.~Zeng, R.~Zhan, and D.~Zhang} (2026):
  \enquote{{Personalized Policy Learning through Discrete Experimentation:
  Theory and Empirical Evidence},} .

\bibitem[\protect\citeauthoryear{Zhou, Athey, and Wager}{Zhou
  et~al.}{2022}]{Zhou2022}
\textsc{Zhou, Z., S.~Athey, and S.~Wager} (2022): \enquote{{Offline
  Multi-Action Policy Learning: Generalization and Optimization},}
  \emph{Operations Research}.

\end{thebibliography}


\begin{thebibliography}{10}
\newcommand{\enquote}[1]{``#1''}
\expandafter\ifx\csname natexlab\endcsname\relax\def\natexlab#1{#1}\fi

\bibitem[\protect\citeauthoryear{Efron and Tibshirani}{Efron and
  Tibshirani}{1994}]{Efron1994}
\textsc{Efron, B. and R.~Tibshirani} (1994): \emph{{An Introduction to the
  Bootstrap}}, Chapman and Hall/CRC.

\bibitem[\protect\citeauthoryear{Galichon}{Galichon}{2016}]{Galichon2016}
\textsc{Galichon, A.} (2016): \emph{{Optimal Transport Methods in Economics}},
  Princeton University Press.

\bibitem[\protect\citeauthoryear{Imbens and Rubin}{Imbens and
  Rubin}{2015}]{Imbens2015}
\textsc{Imbens, G.~W. and D.~B. Rubin} (2015): \emph{{Causal Inference for
  Statistics, Social, and Biomedical Sciences}}, Cambridge University Press.

\bibitem[\protect\citeauthoryear{Misra}{Misra}{2021}]{Misra2021}
\textsc{Misra, S.} (2021): \enquote{{Algorithmic Nudges},} .

\bibitem[\protect\citeauthoryear{Nemhauser, Wolsey, and Fisher}{Nemhauser
  et~al.}{1978}]{Nemhauser1978}
\textsc{Nemhauser, G.~L., L.~A. Wolsey, and M.~L. Fisher} (1978): \enquote{{An
  analysis of approximations for maximizing submodular set functions-I},}
  \emph{Mathematical Programming}, 14, 265--294.

\bibitem[\protect\citeauthoryear{Politis, Romano, and Wolf}{Politis
  et~al.}{1999}]{Politis1999}
\textsc{Politis, D.~N., J.~P. Romano, and M.~Wolf} (1999):
  \emph{{Subsampling}}, Springer Series in Statistics, New York, NY: Springer
  New York.

\bibitem[\protect\citeauthoryear{Santambrogio}{Santambrogio}{2015}]{Santambrogio2015}
\textsc{Santambrogio, F.} (2015): \emph{{Optimal Transport for Applied
  Mathematicians}}, vol.~87.

\bibitem[\protect\citeauthoryear{Villani}{Villani}{2009}]{Villani2009}
\textsc{Villani, C.} (2009): \emph{{Optimal Transport}}, vol. 338 of
  \emph{Grundlehren der mathematischen Wissenschaften}, Berlin, Heidelberg:
  Springer Berlin Heidelberg.

\bibitem[\protect\citeauthoryear{Xu, Thomadsen, and Zhang}{Xu
  et~al.}{2025}]{Xu2025}
\textsc{Xu, S., R.~Thomadsen, and D.~J. Zhang} (2025): \enquote{{The Winner's
  Curse in Data-Driven Decision Making: Evidence and Solutions},} \emph{SSRN
  Electronic Journal}.

\bibitem[\protect\citeauthoryear{Xu, Thomadsen, Zhang, and Zhang}{Xu
  et~al.}{2026}]{Xu2026}
\textsc{Xu, S., R.~Thomadsen, D.~J. Zhang, and H.~Zhang} (2026): \enquote{{SOTA
  or Luck? The Winner's Curse in LLM Leaderboards},} .

\end{thebibliography}
\end{document}